\documentclass[longauth]{aa}

\usepackage{graphicx}
\usepackage{txfonts}
\usepackage{hyperref}
\hypersetup{
    colorlinks=true,
    linkcolor=blue,
    filecolor=blue,      
    urlcolor=blue,
    citecolor=blue,
    pdftitle={A study of Gaia18ajz, A camdidate Black Hole Revealed by Microlensing},
    pdfpagemode=FullScreen, 
    }
\usepackage{multirow}
\usepackage{amsmath}
\usepackage{booktabs}

\def\msun{M_\odot}

\def\I0st{{I_{\mathrm 0}^{\mathrm{st}}}}
\def\V0{V_{\mathrm 0}}

\def\tE{t_{\mathrm E}}

\def\t0{t_{\mathrm 0}}

\def\u0{u_{\mathrm 0}}

\def\piE{\pi_{\mathrm{E}}}

\def\thetaE{\theta_{\mathrm{E}}}

\newcommand{\gaia}{{\it{Gaia}}\xspace}
\newcommand{\Gaia}{{\it{Gaia}}\xspace}

\newcommand{\DLC}{{\texttt{DarkLensCode}}\space}

\newcommand{\orcid}[1]{\protect\href{https://orcid.org/#1}{\protect\includegraphics[width=8pt]{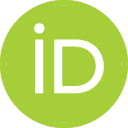}}}

\begin{document} 

  \title{Uncovering the Invisible: A Study of Gaia18ajz, a Candidate Black Hole Revealed by Microlensing}

\titlerunning{A Study of Gaia18ajz, a Candidate Black Hole Revealed by Microlensing}
\authorrunning{Kornel Howil et al.}

   \author{
   K.~Howil \orcid{0000-0002-4085-935X} \inst{\ref{oauw}, \ref{mii}}\fnmsep\thanks{\email{k.howil@student.uw.edu.pl}},
   {\L}.~Wyrzykowski \orcid{0000-0002-9658-6151} \inst{\ref{oauw}},
   K. Kruszy{\'n}ska \inst{\ref{oauw},\ref{lco}}, %
   P. Zieli\'{n}ski \orcid{0000-0001-6434-9429} \inst{\ref{umk}}, %
   E. Bachelet \inst{\ref{ipac}}, %
   M.~Gromadzki \orcid{0000-0002-1650-1518} \inst{\ref{oauw}}, %
   P.J.~Miko{\l}ajczyk \orcid{0000-0001-8916-8050} \inst{\ref{oauw},\ref{wroc}}, %
   K.~Kotysz
   \inst{\ref{oauw},\ref{wroc}}, %
   M.~Jab{\l}o{\'n}ska \orcid{https://orcid.org/0000-0001-6962-4979} \inst{\ref{oauw},\ref{anu}}, %
   Z.~Kaczmarek \inst{\ref{heidelberg}}, %
   P.~Mr\'oz \orcid{https://orcid.org/0000-0001-7016-1692} \inst{\ref{oauw}}, %
   N.~Ihanec \inst{\ref{oauw}}, %
   M.~Ratajczak \orcid{0000-0002-3218-2684} \inst{\ref{oauw}}, %
   U.~Pylypenko \inst{\ref{oauw}}, %
   K.~Rybicki \inst{\ref{wis}}, %
   D.~Sweeney \orcid{0000-0002-7528-1463} \inst{\ref{swee1},\ref{swee2}}, %
   S.T.~Hodgkin \inst{\ref{cam}}, %
   M.~Larma \inst{\ref{martina}} %
   J.~M.~Carrasco \orcid{0000-0002-3029-5853} \inst{\ref{iccub},\ref{fqa},\ref{ieec}}, %
   U.~Burgaz \orcid{0000-0003-0126-3999} \inst{\ref{tcd}}, %
   V.~Godunova \orcid{0000-0001-7668-7994} \inst{\ref{icamer}},  %
   A.~Simon \orcid{0000-0003-0404-5559} \inst{\ref{taras}, \ref{jasu}},  %
   F.~Cusano \inst{\ref{inaf}}, %
   M.~Jelinek \inst{\ref{czk}}, %
   J.~\v{S}trobl\inst{\ref{czk}}, %
   R.~Hudec \inst{\ref{czk},\ref{czk2}}, %
   J.~Merc \inst{\ref{jaro}}, %
   H.~Ku\v{c}\'akov\'a \orcid{0000-0002-1330-1318} \inst{\ref{czk},\ref{czk3},\ref{jaro}}, %
   O.~Erece \orcid{0000-0002-9723-6823} \inst{\ref{trk1}, \ref{trk2}}, %
   Y.~Kilic \orcid{0000-0001-8641-0796} \inst{\ref{trk1}, \ref{trk2}}, %
   F.~Olivares \inst{\ref{filipe}}, %
   M.~Morrell \inst{\ref{ou}}, %
   M.~Wicker \inst{\ref{oauw}} %
   }

\institute{Astronomical Observatory, University of Warsaw, Al. Ujazdowskie 4, 00-478 Warszawa, Poland
    \label{oauw}
    \and
    Faculty of Mathematics and Computer Science, Jagiellonian University, {\L}ojasiewicza 6, 30-348 Krak{\'o}w, Poland
    \label{mii}
    \and
    Las Cumbres Observatory, 6740 Cortona Drive, Suite 102, Goleta, CA 93117, USA
    \label{lco}
    \and
    Institute of Astronomy, Faculty of Physics, Astronomy and Informatics, Nicolaus Copernicus University in Toru{\'n}, Grudzi\k{a}dzka 5, 87-100 Toru{\'n}, Poland
    \label{umk}
    \and
    IPAC, Mail Code 100-22, Caltech, 1200 E. California Blvd., Pasadena, CA 91125, USA
    \label{ipac}
    \and
    Research School of Astronomy and Astrophysics, Australian National University, Mount Stromlo Observatory, Cotter Road Weston Creek, ACT 2611, Australia
    \label{anu}
    \and
    Zentrum f{\"u}r Astronomie der Universit{\"a}t Heidelberg, Astronomisches Rechen-Institut, M{\"o}nchhofstr. 12-14, 69120 Heidelberg, Germany
    \label{heidelberg}
    \and
    Department of Particle Physics and Astrophysics, Weizmann Institute of Science, Rehovot 76100, Israel
    \label{wis}
    \and
    Astronomical Institute, University of Wroc{\l}aw, ul. Miko{\l}aja Kopernika 11, 51-622 Wroc{\l}aw, Poland
    \label{wroc}
    \and 
    Sydney Institute for Astronomy (SIfA), The University of Sydney, Physics Road, Sydney 2050, Australia\label{swee1}
    \and
    Donald Bren School of Information and Computer Sciences, University of California, Irvine, CA 92697, USA\label{swee2}
    \and
    Institute of Astronomy, University of Cambridge, Madingley Road, Cambridge, CB3 0HA, United Kingdom
    \label{cam}
    \and
    University of Bonn, Argelander-Institut für Astronomie, Auf dem Hügel 71, 53121 Bonn, Germany
    \label{martina}
    \and
    School of Physics, Trinity College Dublin, College Green, Dublin 2, Ireland
    \label{tcd}
    \and
    Institut de Ci\`encies del Cosmos (ICCUB), Universitat de Barcelona (UB), Mart\'{i} i Franqu\`es 1, E-08028 Barcelona, Spain
    \label{iccub}
    \and
    Departament de F\'isica Qu\`antica i Astrof\'{i}sica (FQA), Universitat de Barcelona (UB), Mart\'{i} i Franqu\`es 1, E-08028 Barcelona, Spain
    \label{fqa}
    \and
    Institut d'Estudis Espacials de Catalunya (IEEC), Esteve Terradas, 1, Edifici RDIT, Campus PMT-UPC, 08860 Castelldefels (Barcelona), Spain
    \label{ieec}
    \and
    ICAMER Observatory of National Academy of Sciences of Ukraine, 27 Acad. Zabolotnoho str., Kyiv, 03143 Ukraine
    \label{icamer}
    \and
    Astronomy and Space Physics Department, Taras Shevchenko National University of Kyiv, 4 Glushkova ave., Kyiv, 03022 Ukraine
    \label{taras}
    \and
    National Center "Junior Academy of Sciences of Ukraine", 38-44 Dehtiarivska St., Kyiv, 04119 Ukraine
    \label{jasu}
    \and
    INAF-Osservatorio di Astrofisica e Scienza dello Spazio, Via Gobetti $93/3$, I-40129 Bologna, Italy
    \label{inaf}
    \and
    Astronomical Institute of the Academy of Sciences of the Czech Republic (ASU CAS), 25165 Ondrejov, Czech Republic
    \label{czk}
    \and
    Czech Technical University in Prague, Faculty of Electrical Engineering, Prague, Czech Republic
    \label{czk2}
    \and 
Astronomical Institute, Faculty of Mathematics and Physics, Charles University, V Hole\v{s}ovi\v{c}k{\'a}ch 2, 180 00 Prague, Czech Republic    \label{jaro}
    \and
    Research Centre for Theoretical Physics and Astrophysics, Institute of Physics, Silesian University, Bezru\v{c}ovo n\'{a}m. 13, 746 01 Opava, Czechia
    \label{czk3}
    \and
    Department of Space Sciences and Technologies, Akdeniz University, Campus, Antalya, 07058, Turkey
    \label{trk1}
    \and
    T\"UB\.{I}TAK National Observatory, Akdeniz University Campus, Antalya, 07058, Turkey
    \label{trk2}
    \and
    Instituto de Astronom\'{i}a y Ciencias Planetarias Universidad de Atacama, 
    Av. Copayapu 485 1531772, Copiapo, Chile
    \label{filipe}
    \and
    School of Physical Sciences, The Open University, Walton Hall, Milton Keynes MK7 6AA, UK
    \label{ou}
}

   \date{October 2024}

  \abstract  
  {Identifying black holes is essential for comprehending the development of stars and uncovering novel principles of physics. Gravitational microlensing provides an exceptional opportunity to examine an undetectable population of black holes in the Milky Way. In particular, long-lasting events are likely to be associated with massive lenses, including black holes.
  }
  {We present an analysis of the Gaia18ajz microlensing event, reported by the Gaia Science Alerts system, which has exhibited a long timescale and features indicative of the annual microlensing parallax effect. Our objective is to estimate the lens parameters based on the best-fitting model.
  }
  {We utilized photometric data obtained from the Gaia satellite and terrestrial observatories to investigate a variety of microlensing models and calculate the most probable mass and distance to the lens, taking into consideration a Galactic model as a prior. Subsequently, we applied a mass-brightness relation to evaluate the likelihood that the lens is a main-sequence star.
  We also describe the \DLC (DLC), an open-source routine which computes the distribution of probable lens mass, distance, and luminosity employing the Galaxy priors on stellar density and velocity for microlensing events with detected microlensing parallax. 
  }
  {We modelled Gaia18ajz event and found its two possible models with most likely Einstein timescale of $316^{+36}_{-30}$ days and $299^{+25}_{-22}$ days. Applying Galaxy priors for stellar density and motion, we calculated the most probable lens mass of $4.9^{+5.4}_{-2.3} M_\odot$ located at $1.14^{+0.75}_{-0.57}\,\text{kpc}$ or less likely $11.1^{+10.3}_{-4.7} M_\odot$ located at $1.31^{+0.80}_{-0.60}\,\text{kpc}$. Our analysis of the blended light suggests that the lens is likely a dark remnant of stellar evolution, rather than a main-sequence star.}


   \maketitle
\section{Introduction}
In recent years, the field of black hole astronomy has witnessed significant advancements in the detection and characterisation of stellar-mass black holes (sBHs) \citep{Askar2023IMBH} in our Galaxy. Historically, the identification of sBHs was achieved primarily through X-ray observations of accreting binary systems, for example, \cite{1992Natur.355..614C}, \cite{2017MNRAS.467.2199B},  \cite{Corral-Santana2016}. 
More recently, novel techniques for the detection of sBH have emerged, which have expanded our understanding of these enigmatic objects. The direct observation of gravitational waves, as demonstrated in landmark events like the detection of binary black hole mergers \citep{Abbott2016, Abbott2017GW50MSUN, Abbott2019}, has opened new avenues for probing the universe's most massive and elusive black holes. Additionally, binary systems with luminous companions, as exemplified by the works of \cite{Gomel2023-Gaia-DR3-ellipsoidal-compact-binaries}, \cite{Shahaf2023} and, in particular, by the discoveries of GaiaBH1 \citep{GaiaBH1, Chakrabarti2023}, GaiaBH2 \citep{GaiaBH2, Ataru2022} and GaiaBH3 \citep{GaiaBH3}, have further enriched our knowledge of sBHs with masses around 10 $\msun$.

Nevertheless, our comprehension of the sBHs population and their evolution has predominantly been investigated through binary systems. To construct a comprehensive picture of the mass spectrum and spatial distribution of sBHs, it has become increasingly crucial to explore and investigate also isolated black holes. These solitary black holes can have diverse origins, including the remnants of massive single stars \citep{Belczynski2020, Pejcha2015}, disruptions of binary systems \citep{Wiktorowicz2019}, or ejections from globular clusters \citep{Giersz2022, Leveque2023}.

One promising avenue for the study of solitary sBHs is gravitational microlensing. This gravitational phenomenon, a consequence of Einstein's General Theory of Relativity, occurs when a massive object passes in front of a distant star within the Milky Way or its vicinity \citep{Einstein1936, Paczynski1986}. Unlike strong gravitational lensing, where multiple, distorted images of a source can be observed, images in microlensing events are typically almost impossible to resolve, only with the very first successful attempts using optical interferometry \citep{Dong2019, Cassan2023}. Instead, they manifest as temporary brightening of the background source, coupled with a positional shift in the source's centroid, a phenomenon referred to as astrometric microlensing \citep{DominikSahu2000, BelokurovEvans2002}.

Astrometric microlensing, although challenging to observe, has been successfully employed with precise observatories such as the Hubble Space Telescope (HST) \citep{SahuWD2017, McGill2023}.
The combination of astrometric and photometric microlensing effects enables determination of the mass of the lensing object ($M_\mathrm{L}$) \citep{Gould2000b}. The formula used for this purpose is derived from the angular Einstein radius ($\theta_\mathrm{E}$) and the microlensing parallax ($\pi_\mathrm{E}$). Remarkably, this technique has recently led to the direct measurement of the mass of a solitary sBH for the first time \citep{Sahu_2022, Lam_2022, Mroz_2022, Lam2023}.
This has opened a new channel for studies of stellar evolution and black hole production, e.g. \citep{Andrews2022, Horvath2023, Vigna-Gomez2023}.

European Space Agency's Gaia mission \citep{Gaia}, is going to be a valuable asset for astrometric microlensing studies due to its long-term all-sky submilliarcsecond accuracy of epoch astrometric measurements for nearly 2 billion stars \citep{Rybicki2018}.
Gaia has proven to be capable of detecting microlensing events in its time-domain photometric data. In its latest data release, Gaia DR3 contained a catalogue of microlensing events from all over the sky \citep{Wyrzykowski2023}. Another source of microlensing events is the near-real-time system of Gaia Science Alerts (GSA) \citep{Wyrzykowski2012, Hodgkin_2013, Hodgkin_2021}, which detects ongoing events and alerts the astronomical community.

Here we present an analysis of one of the GSA alerts, Gaia18ajz, which was one of the longest microlensing events ever studied.
Long-lasting microlensing events are the prime suspects of being caused by massive lenses, as their large timescale could be related to a large Einstein radius, with prior examples presented in \cite{Mao2002} and \cite{Kruszynska2022}. However, the lack of Einstein radius measurement prevents mass derivation, but such events typically exhibit the microlensing parallax effect \citep{Gould2000b}. The effect, combined with information on the Galaxy and the possible kinematics of the lenses, allows us to estimate the most likely mass and distance of the lens (e.g. \citealt{Wyrzykowski2016, Wyrzykowski2020, MrozWyrzykowski2021}). In this work, we present the \DLC (DLC), an open-source software for generating posterior probability distributions for the physical parameters of the lens using microlensing parallax posteriors obtained from the light curve.

\section{Discovery and ground based observations of Gaia18ajz}

\begin{figure}
   \centering
   \includegraphics[width=6cm]{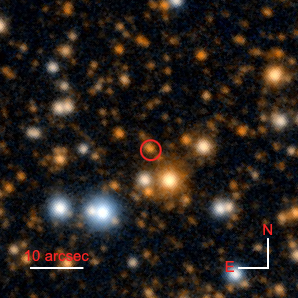}
   \caption{Location of Gaia18ajz and its neighbourhood during the event. The image comes from Pan-STARRS DR1 via Aladin Tool implemented in BHTOM.}
   \label{fig:fchart}
\end{figure}

\quad Gaia18ajz (AT2018uh according to the IAU transient name server) was discovered by \textit{Gaia} Science Alerts on February 9 2018 (HJD’ = HJD - 2450000.0 = 8159.09). On 14 February it was posted on the GSA website \footnote{\href{http://gsaweb.ast.cam.ac.uk/alerts/alert/Gaia18ajz/}{http://gsaweb.ast.cam.ac.uk/alerts/alert/Gaia18ajz/}}. The event was located at equatorial coordinates RA$_J2000$ = 18h30m14.46s, Dec$_J2000$ = -08$^{\circ}$13'12.76'' and galactic coordinates $l = 23.20506^{\circ},\, b = +0.925751^{\circ}$. The finding chart with the location of the event in the sky is presented in Figure \ref{fig:fchart}.

\textit{Gaia} Data Release 2 (GDR2) \citep{GaiaDR2} and \textit{Gaia}  Data Release 3 (GDR3) \citep{GaiaDR3} contain an entry for this source under \textit{Gaia} Source ID = 4156664130700362752. Astrometric parameters for this object from both GDR2 and GDR3 are presented in Table \ref{tab:gdrsVals}.

\begin{table}
\centering
\caption{\label{tab:gdrsVals}Gaia astrometric parameters for the source star in Gaia18ajz.}
     \centering
        \begin{tabular}{c c c}
        \hline
        \noalign{\smallskip}
             Parameter &  GDR2 & GDR3 \\
             \noalign{\smallskip}
        \hline
        \hline
        \noalign{\smallskip}
             $\varpi$ [mas] & $3.24\pm0.59$ &  $1.52\pm0.54$ \\
             $\mu_{\alpha^*}$ [$\mathrm{mas} \, \mathrm{yr}^{-1}$] & $-7.76\pm1.36$ & $-5.37\pm0.59$ \\
             $\mu_{\delta}$ [$\mathrm{mas}\, \mathrm{yr}^{-1}$] & $-4.27\pm1.46$ & $-6.69\pm0.51$ \\
             RUWE & 1.49 & 1.53\\
        \noalign{\smallskip}
        \hline
        \end{tabular}
\tablefoot{Parallax $\varpi$, proper motions $\mu_\alpha, \mu_\delta$ and Renormalised Unit Weight Error (RUWE) come from \textit{Gaia} Data Release 2 \citep{GaiaDR2} and \textit{Gaia} Data Release 3 \citep{GaiaDR3}.}
\end{table}

\subsection{Gaia photometry} \label{sec:phot}
\quad The \textit{Gaia} photometic data is collected in a wide $G$-band \citep{2010JordiGaiaPhot}. As of March 2024, \textit{Gaia} has collected 86 measurements for Gaia18ajz and the event has returned to its baseline. Light curve from \textit{Gaia} has one peak reaching about $17.6$ mag in the $G$ band and a baseline around $19.3$ mag in the same band. 

\textit{Gaia} Science Alerts does not provide errors of magnitude for published events. In order to estimate these error bars, we followed a method described previously in \cite{Kruszynska2022}.
The error bars for this event should vary from $0.02$ mag for $G = 17.6$ at peak to $0.06$ mag for $G = 19.3$ at baseline.

\subsection{Ground-based photometric observations}
Consistent photometric follow-up is essential to obtain accurate values for the best-fitting microlensing model, particularly blending parameters. This involves observing the event not just at its peak brightness but also during the baseline using the same telescope. Due to the fact that the event was faint, especially at its baseline with G = 19.3 mag, the follow-up data have a large measurement error. 

\begin{figure*}
   \centering
   \includegraphics[width=\hsize]{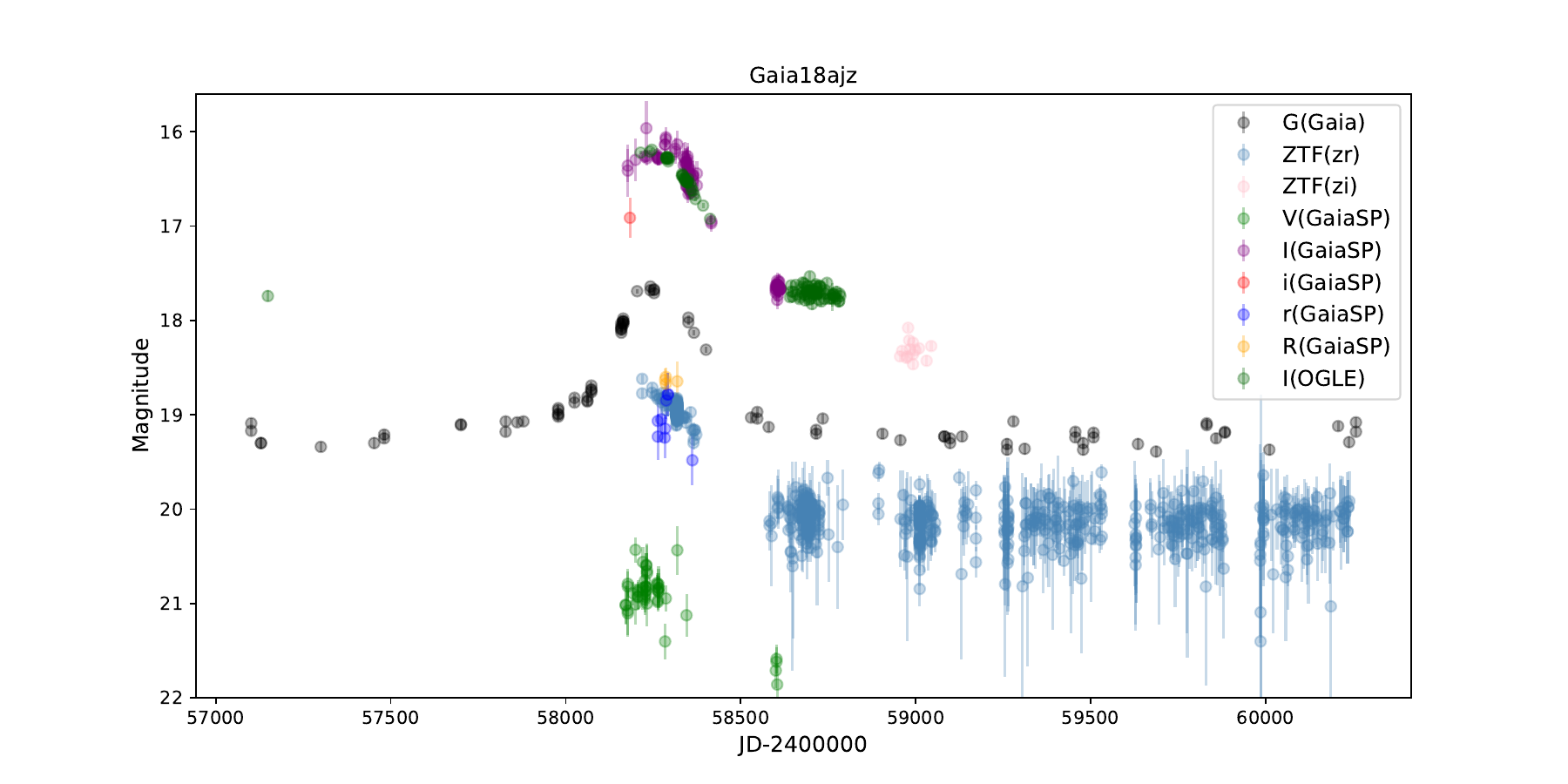}
   \caption{Photometric data collected for Gaia18ajz. The different colours represent different bands.}
   \label{fig:data}
\end{figure*}

Following the event announcement on the Gaia Science Alerts website, the earliest follow-up began 13 days later. The first data point was taken on 22 February 2018, by Pawe{\l} Zieli{\'n}ski 
using a 1.3-meter SMARTS telescope equipped with the ANDICAM instrument at Cerro Tololo Inter-American Observatory, Chile. 
Additional observations were also carried out with: 
the 0.8 m Joan Or\'o Telescope using MEIA2 CCD at Montsec astronomical observatory, Spain (ObsMontsec);
1.52m Loiano telescope with BFOSC, Italy (Loiano);
0.6m PIRATE (Physics Innovations Robotic Astronomical Telescope Explorer) telescope of the Open University in the UK, with FLI ProLine KAF-16803 camera, on Tenerife, Spain;  
2m Terskol with the FLI PL4301 camera and 0.6m Terskol with the SBIG STL1001E camera in the North Caucasus operated by the NAS of Ukraine; 
2m Ondrejov Telescope in Czechia (Ondrejov);
0.6m T60 robotic telescope with FLI Proline3041 at TÜBİTAK National Observatory, Turkey (TUG T60);
and 1.54m Danish Telescope in La Silla, Chile.
The number of data points observed by each observatory is presented in Table \ref{tab:photometry}. All data points are visible in the Figure \ref{fig:data}. Ground-based observations were reduced using the automatic tool for time-domain data - the Black Hole TOM (BHTOM) tool\footnote{\url{http://bhtom.space}}. The bias-, dark-, and flat-field-corrected images were processed to obtain instrumental photometry for all stars in the image. Then, the photometry was calibrated as described in \citep{Zielinski2019, Zielinski2020} to the Gaia Synthetic Photometry magnitudes \citep{Montegriffo2022}.

The Zwicky Transient Facility (ZTF) is an astronomical survey that aims to detect transient and variable objects in the sky through repeated observations using a wide-field camera mounted on the Samuel Oschin Telescope at the Palomar Observatory (CITE).
ZTF observed this target serendipitously while scanning the Northern sky. ZTF provides data in r- and g-band openly, however, only r-band data were collected during the actual event in 2018.

The event was also observed by the Optical Gravitational Lensing Experiment (OGLE; \citealt{Udalski2015}) as part of its Galaxy Variability Survey \citep{Mroz2020} from May 2015 to October 2019. OGLE uses a 1.3-m Warsaw Telescope located at Las Campanas Observatory, Chile. All OGLE observations were collected through an $I$-band filter and reduced using the standard OGLE pipeline \citep{Udalski2015}.

\begin{table*}
\centering
\begin{scriptsize}
\centering
\caption{Photometric data collected for Gaia18ajz from different observatories.}
\footnotesize
\label{tab:photometry}
\begin{tabular}{lllll}
\hline
Observatory  &  Filters (Observed or Standardised to) & Data points & Min.MJD & Max.MJD  \\
\hline
\hline
ZTF &ZTF(zr), ZTF(zi)&1095 &58218.47&60356.55\\
OGLE & I(OGLE) &104 & 57148.89 & 58784.54\\
SMARTS1.3$^B$ &V(GaiaSP), I(GaiaSP), i(GaiaSP)&103&58171.38&58613.32\\%
Gaia &G(Gaia)&86&57101.37&60258.08\\%
Ondrejov$^B$&R(GaiaSP), I(GaiaSP), U(GaiaSP) &49&58358.82&58422.72\\%
ObsMontsec$^B$ &I(GaiaSP)&22&58336.98&58374.84\\%
Loiano$^B$ &I(GaiaSP), R(GaiaSP), V(GaiaSP), r(GaiaSP)&12&58283.99&58312.87\\%
PIRATE$^B$ &r(GaiaSP)&8&58263.12&58361.95\\%
Terskol-2$^B$ &R(GaiaSP), V(GaiaSP), I(GaiaSP)&9&58322.91&58755.73\\%
Terskol-0.6$^B$ &R(GaiaSP), V(GaiaSP), &4&58345.81&58346.74\\%
Danish$^B$&B(GaiaSP), V(GaiaSP), R(GaiaSP), I(GaiaSP)&4&60378.38&60378.39\\%
\hline
\end{tabular}
\end{scriptsize}
\\
$^B$ Data from that observatory has been processed and calibrated using BHTOM
\end{table*}

\subsection{Spectroscopic follow-up}
The object was also observed spectroscopically with the X-Shooter instrument \citep{Vernet2011} mounted on the ESO Very Large Telescope (VLT). The first spectrum was obtained on 25 March and the second on 8 August 2018\footnote{ESO Programme ID: 0101.D-0035(A), PI: {\L}. Wyrzykowski}. The following setup was applied: the slit width 1.0 arcsec,  0.7 arcsec and 0.6 arcsec for UVB ($300-559.5$~nm), VIS ($559.5-1024$~nm) and NIR ($1024-2480$~nm) channels, respectively, and exposure times of 1872 s (UVB), 2104 s (VIS) and 2400 s (NIR) for both spectra. This gave us the resolution of $R\sim4300$ and the median signal-to-noise ratio ${\rm SNR}=0.2$ in UVB, $R\sim10500$ and ${\rm SNR}=29$ in VIS and $R\sim7900$ and ${\rm SNR}=188$ in NIR for the first spectrum, while for the second spectrum we have got $R\sim4300$ and ${\rm SNR}=0.3$ in UVB, $R\sim10500$ and ${\rm SNR}=37$ in VIS and $R\sim7900$ and ${\rm SNR}=243$ in the NIR part.

 X-Shooter data were reduced in a standard way using the EsoReflex\footnote{\url{https://www.eso.org/sci/software/esoreflex/}} pipeline. The ThAr comparison lamp was used for the calibration of UVB and VIS wavelengths, and the Ar, Hg, Ne, and Xe lamps were used for NIR wavelengths.

The first spectrum was used to classify this target as a microlensing event that was published in \citet{Kruszynska2018ATel}. The spectrum did not present emission features and only absorption lines, and the continuum shape indicated a late-type star with significant reddening. It ensured us to continue ground-based photometric follow-up of the Gaia18ajz event.

\section{Photometric Microlensing Model}

In this paper, following the procedure described previously in \cite{Kruszynska2022}, we fit two single source and single lens models: with and without taking into account the annual parallax effect (e.g. \cite{Wyrzykowski2016}, \cite{Kaczmarek2022}, \cite{Jablonska2022}). 
\subsection{Standard model}
The first is the standard Paczy\'nski light curve model described in \cite{Paczynski1986}. It is defined by the following parameters:

\begin{itemize}
    \item $t_0$: the time of the peak of brightness,
    \item $u_0$: impact parameter, defined as the source-lens separation in units of the Einstein radius at time $t_0$,
    \item $t_E$: time scale, defined as $\theta_E / \mu_{LS}$, where $\mu_{LS}$ is the relative proper motion of the source and the lens,
    \item $mag_0$: baseline magnitude, calculated separately in each of the observing bands, defined as $mag_0 = -2.5 \log(F_S+F_B)$, where $F_S$ is the baseline flux of the source and $F_B$ is the baseline flux of the blend,
    \item $f_S$: blending parameter, defined as part of the flux coming from the source star (separately in each of the observing bands), defined as $f_S = \frac{F_S}{F_S + F_B}$.
\end{itemize}
\subsection{Parallax model}

Since the event studied was extremely long, we have to consider the Earth's movement around the Sun. The second model considered, described in \cite{Gould2000b} and \cite{Gould2004}, is an extension of the Paczy\'nski model that takes into account the annual parallax effect. In this model, in addition to the parameters described previously, we have to include two additional parameters: $\pi_{EN}$ and $\pi_{EE}$. These parameters are, respectively the (equatorial) North and East components of the microlensing parallax $\mathbf{\pi_E}$, projected in the direction as the relative proper motion of the lens and the source.

\subsection{Modelling and results}

Due to the relatively low brightness of the event and the resulting large scatter in the ground-based data, we decided to fit three sets of models. The first set was trained using only Gaia data, and the second set was trained using both the \Gaia data and ground-based data. We decided to use data only from ZTF, OGLE, and SMARTS facilities. In addition, we trained the third set using photometric data from \Gaia and OGLE, which are of the best quality.

We used a Markov chain Monte Carlo (MCMC) method to explore the parameter space of the microlensing models described earlier. The models were trained using \texttt{MulensModel} \citep{MulensModel} with \texttt{emcee} package \citep{2013Emcee}. During modelling, we scaled the error bars separately for each model and each band so that $\chi^2$ per degree of freedom for every band is equal to one. In order to ensure that models are physical, we introduced a gaussian prior on negative flux:
\begin{equation}
    \log{(\text{Prior})} = -\frac{1}{2}\biggl(\frac{F_b}{4}\biggr)^2 
\end{equation}
for $F_b < 0$ where, $F_b$ is the blended flux. Flux equal to 1 corresponds to 22 mag. 

To ensure that all possible solutions were uncovered and the entire parameter space was explored, we visually examined the MCMC results. It was found that in all sets of models, all posterior distributions were bimodal with classic degeneracy with respect to the impact parameter. This degeneracy is caused by the fact that the lens can pass in front of the source from both sides, resulting in a change of sign in the impact parameter.
We have decided to divide bimodal solutions into two separate models one with a positive and one with negative impact parameter $u_0$ and model them separately. Finally, our results contain nine different models:
\begin{itemize}
    \item G0 : Gaia-only, no-parallax
    \item G+ : Gaia-only, parallax, positive $u_0$
    \item G- : Gaia-only, parallax, negative $u_0$
    \item GG0 : Gaia+ground, no-parallax
    \item GG+ : Gaia+ground, parallax, positive $u_0$
    \item GG- : Gaia+ground, parallax, negative $u_0$
    \item GO0 : Gaia+OGLE, no-parallax
    \item GO+ : Gaia+OGLE, parallax, positive $u_0$
    \item GO- : Gaia+OGLE, parallax, negative $u_0$
\end{itemize}
In the case of the standard model without taking into account the parallax effect, we decided to report only one solution with a positive impact parameter. Due to the lack of the parallax effect, the solution with a negative impact parameter is symmetrical. 

Additionally, to test which of the two modes of posterior distribution is more likely, we analysed the event using nested sampling with \textit{nested\_ulens\_parallax} code \footnote{\url{https://github.com/zofiakaczmarek/nested_ulens_parallax/}} \citep{Kaczmarek2022, 2020Speagle}.

Table \ref{tab:valSolutionsGaia} displays the optimal parameter values obtained using only Gaia data, while Table \ref{tab:valSolutionsFup} presents the optimal parameter values obtained using both Gaia and all ground-based follow-up photometric data. Table \ref{tab:valSolutionsOGLE} shows the results of the modelling for GO models. For each band, the baseline magnitude and blending parameters were determined independently. As a reference for calculating $\chi^2/dof$, we decided to use the data set used to model the GG+ model, with error bars rescaled according to this model.

\begin{table*}
    \centering
    \caption{\label{tab:valSolutionsGaia} The optimal parameter values for fitting Gaia18ajz using only Gaia photometric data.}
    \begin{tabular}{c c c c}
            \hline
            \noalign{\smallskip}
            Parameter & G0 & G+ & G- \\
            \noalign{\smallskip}
            \hline
            \hline
            \noalign{\smallskip}
            $t_{0, par}-2450000$ [days]  & -- & \multicolumn{2}{c}{$8231$} \\
            $t_{0}-2450000$ [days] & $8256.2^{+3.1}_{-3.2}$ &  $ 8229.4^{+3.3}_{-3.6}$ & $8227.3^{+4.1}_{-4.0}$\\
            \noalign{\smallskip}
            $u_0$ & $0.362^{+0.034}_{-0.034}$ & $0.239^{+0.051}_{-0.050}$ & $-0.224^{+0.053}_{-0.052}$ \\
            \noalign{\smallskip}
            $t_E$ [days] & $232^{+17}_{-15}$ & $347^{+64}_{-51}$ & $353^{+81}_{-50}$ \\
            \noalign{\smallskip}
            $\pi_{EN}$ & -- & $0.055^{+0.054}_{-0.035}$ & $-0.011^{+0.023}_{-0.023}$ \\
            \noalign{\smallskip}
            $\pi_{EE}$ & -- & $0.128^{+0.030}_{-0.035}$ & $0.098^{+0.011}_{-0.011}$\\
            \noalign{\smallskip}
            $G_0$ [mag] & $19.2744^{+0.0087}_{-0.0085}$ & $19.2479^{+0.0083}_{-0.0078}$ & $19.2466^{+0.0083}_{-0.0077}$ \\
            \noalign{\smallskip}
            $f_{s, G}$ & $1.90^{+0.24}_{-0.23}$ & $1.02^{+0.28}_{-0.25}$ & $0.95^{+0.28}_{-0.26}$ \\
            \noalign{\smallskip}
            $\chi^2/dof$ & $2.81$ & $0.98$ & $0.98$ \\
            \noalign{\smallskip}
            \noalign{\smallskip}
            \hline
    \end{tabular}
    \tablefoot{The model G0 represents a point source-point lens configuration that does not take into account the microlensing parallax effect. On the other hand, models G+ and G- incorporate the microlensing parallax effect into the point source-point lens configuration. In the case of G0, the model with $u_0>0$ was selected, as the posterior distribution for $u_0$ is symmetrical.}
\end{table*}

\begin{table*}
    \centering
    \caption{\label{tab:valSolutionsFup} The optimal parameter values for fitting Gaia18ajz, using both Gaia and ground-based follow-up photometric data.}
    \begin{tabular}{c c c c}
            \hline
            \noalign{\smallskip}
            Parameter & GG0 & GG+ & GG- \\
            \noalign{\smallskip}
            \hline
            \hline
            \noalign{\smallskip}
            $t_{0, par}-2450000$ [days]  & -- & \multicolumn{2}{c}{$8231$} \\
            $t_{0}$ [days] & $8262.0^{+1.4}_{-1.4}$ &  $ 8230.8^{+1.0}_{-1.0}$ & $8231.4^{+1.6}_{-1.6}$\\
            \noalign{\smallskip}
            $u_0$ & $0.2476^{+0.0093}_{-0.0094}$ & $0.204^{+0.023}_{-0.021}$ & $-0.223^{+0.015}_{-0.012}$ \\
            \noalign{\smallskip}
            $t_E$ [days] & $345^{+12}_{-12}$ & $299^{+25}_{-22}$ & $336^{+18}_{-13}$ \\
            \noalign{\smallskip}
            $\pi_{EN}$ & -- & $0.159^{+0.020}_{-0.023}$ & $-0.024^{+0.006}_{-0.006}$ \\
            \noalign{\smallskip}
            $\pi_{EE}$ & -- & $0.063^{+0.011}_{-0.008}$ & $0.0929^{+0.0058}_{-0.0056}$\\
            \noalign{\smallskip}
            $G_0(\text{Gaia})$ [mag] & $19.3260^{+0.0053}_{-0.0053}$ & $19.2427^{+0.0061}_{-0.0060}$ & $19.2447^{+0.0051}_{-0.0050}$ \\
            \noalign{\smallskip}
            $f_{s, G(\text{Gaia})}$ & $1.173^{+0.054}_{-0.053}$ & $0.83^{+0.11}_{-0.10}$ & $0.99^{+0.06}_{-0.07}$ \\
            \noalign{\smallskip}
            $I_0(\text{GaiaSP})$ [mag] & $18.080^{+0.020}_{-0.019}$ & $17.783^{+0.013}_{-0.012}$ & $17.812^{+0.012}_{-0.011}$ \\
            \noalign{\smallskip}
            $f_{s, I(\text{GaiaSP})}$ & $1.528^{+0.056}_{-0.056}$ & $0.82^{+0.11}_{-0.09}$ & $1.018^{+0.059}_{-0.072}$ \\
            \noalign{\smallskip}
            $V_0(\text{GaiaSP})$ [mag] & $21.741^{+0.013}_{-0.013}$ & $21.7264^{+0.0053}_{-0.0050}$ & $21.7359^{+0.0047}_{-0.0046}$ \\
            \noalign{\smallskip}
            $f_{s, V(\text{GaiaSP})}$ & $0.454^{+0.005}_{-0.005}$ & $0.315^{+0.043}_{-0.037}$ & $0.382^{+0.023}_{-0.028}$ \\
            \noalign{\smallskip}
            $r_0(\text{ZTF})$ [mag] & $20.2112^{+0.0048}_{-0.0036}$ & $20.1697^{+0.0040}_{-0.0035}$ & $20.1732^{+0.0029}_{-0.0026}$ \\
            \noalign{\smallskip}
            $f_{s, r(\text{ZTF})}$ & $0.875^{+0.039}_{-0.039}$ & $0.656^{+0.090}_{-0.078}$ & $0.788^{+0.049}_{-0.059}$ \\
            \noalign{\smallskip}
            $I_0(\text{OGLE})$ [mag] & $17.928^{+0.012}_{-0.012}$ & $17.810^{+0.012}_{-0.010}$ & $17.8185^{+0.0085}_{-0.0078}$ \\
            \noalign{\smallskip}
            $f_{s, I(\text{OGLE})}$ & $1.239^{+0.051}_{-0.050}$ & $0.86^{+0.11}_{-0.10}$ & $1.042^{+0.061}_{-0.074}$ \\
            \noalign{\smallskip}
            $\chi^2/dof$ & $1.38$ & $0.97$ & $0.98$ \\
            \noalign{\smallskip}
            \hline
    \end{tabular}
   \tablefoot{The model GG0 represents a point source-point lens configuration that does not take into account the microlensing parallax effect. On the other hand, models GG+ and GG- incorporate the microlensing parallax effect into the point source-point lens configuration. In the case of GG0, the model with $u_0>0$ was selected, as the posterior distribution for $u_0$ is symmetrical.}
\end{table*}

\begin{table*}
    \centering
    \caption{\label{tab:valSolutionsOGLE} The optimal parameter values for fitting Gaia18ajz, using Gaia and OGLE photometric data.}
    \begin{tabular}{c c c c}
            \hline
            \noalign{\smallskip}
            Parameter & GO0 & GO+ & GO- \\
            \noalign{\smallskip}
            \hline
            \hline
            \noalign{\smallskip}
            $t_{0, par}-2450000$ [days]  & -- & \multicolumn{2}{c}{$8231$} \\
            \noalign{\smallskip}
            $t_{0}-2450000$ [days] & $8259.4^{+1.9}_{-1.9}$ &  $ 8231.5^{+1.2}_{-1.1}$ & $8232.7^{+1.8}_{-1.8}$ \\
            \noalign{\smallskip}
            $u_0$ & $0.241^{+0.015}_{-0.015}$ & $0.196^{+0.030}_{-0.026}$ & $-0.228^{+0.023}_{-0.017}$ \\
            \noalign{\smallskip}
            $t_E$ [days] & $375^{+23}_{-20}$ & $316^{+36}_{-30}$ & $346^{+30}_{-20}$ \\
            \noalign{\smallskip}
            $\pi_{EN}$ & -- & $0.145^{+0.022}_{-0.026}$ & $-0.0232^{+0.0069}_{-0.0069}$ \\
            \noalign{\smallskip}
            $\pi_{EE}$ & -- & $0.060^{+0.015}_{-0.010}$ & $0.0865^{+0.0066}_{-0.0063}$\\
            \noalign{\smallskip}
            $G_0(\text{Gaia})$ [mag] & $19.3451^{+0.0083}_{-0.0079}$ & $19.2484^{+0.0075}_{-0.0071}$ & $19.2497^{+0.0063}_{-0.0060}$ \\
            \noalign{\smallskip}
            $f_{s, G(\text{Gaia})}$ & $1.106^{+0.083}_{-0.082}$ & $0.79^{+0.15}_{-0.12}$ & $0.96^{+0.09}_{-0.11}$ \\
            \noalign{\smallskip}
            $I_0(\text{OGLE})$ [mag] & $17.972^{+0.022}_{-0.020}$ & $17.819^{+0.017}_{-0.014}$ & $17.826^{+0.011}_{-0.010}$ \\
            \noalign{\smallskip}
            $f_{s, I(\text{OGLE})}$ & $1.278^{+0.076}_{-0.076}$ & $0.82^{+0.15}_{-0.12}$ & $1.02^{+0.09}_{-0.12}$ \\
            \noalign{\smallskip}
            $\chi^2/dof$ & $3.12$ & $0.93$ & $0.97$ \\
            \noalign{\smallskip}
            \hline
    \end{tabular}
   \tablefoot{The model GO0 represents a point source-point lens configuration that does not take into account the microlensing parallax effect. On the other hand, models GO+ and GO- incorporate the microlensing parallax effect into the point source-point lens configuration. In the case of GO0, the model with $u_0>0$ was selected, as the posterior distribution for $u_0$ is symmetrical.}
\end{table*}

\begin{figure*}
   \centering
   \includegraphics[width=\hsize]{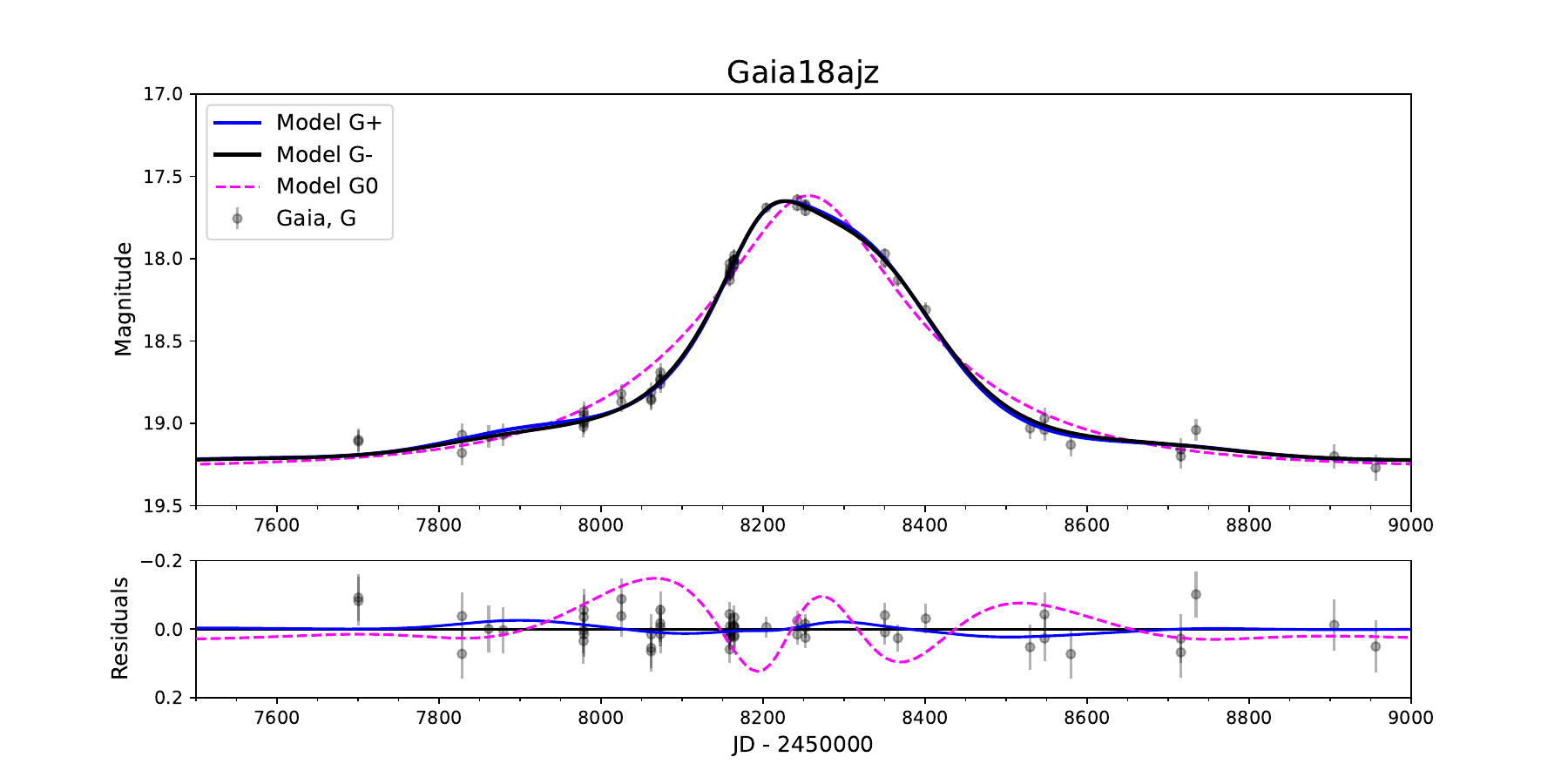}
   \caption{The light curve of the Gaia18azj event and the microlensing model fit using only the Gaia data. The dashed magenta line represents the model without parallax, while the blue and black solid lines show the positive and negative solutions for the parallax model, respectively. The bottom panel shows the residuals with respect to the negative solution for the parallax model.}
   \label{fig:lc1}
\end{figure*}

\begin{figure*}
   \centering
   \includegraphics[width=\hsize]{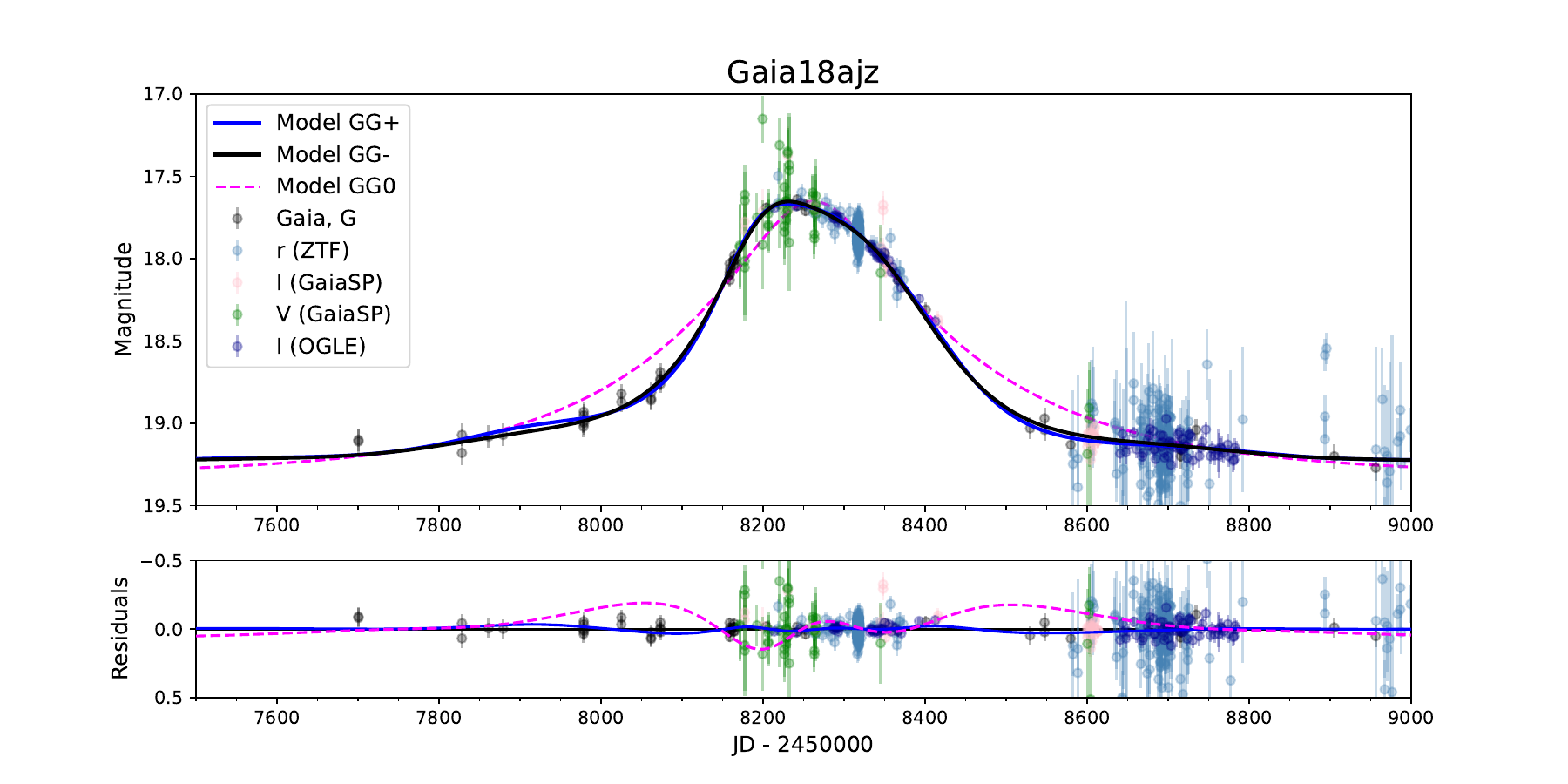}
   \caption{The light curve of the Gaia18azj event and the microlensing model fit using the Gaia data as well as the ground-based survey data from ZTF, SMARTS, and OGLE. The dashed magenta line represents the model without parallax, while the blue and black solid lines show the positive and negative solutions for the parallax model, respectively. The bottom panel shows the residuals with respect to the negative solution for the parallax model.). 
  }
   \label{fig:lc2}
\end{figure*}

\begin{figure*}
   \centering
   \includegraphics[width=\hsize]{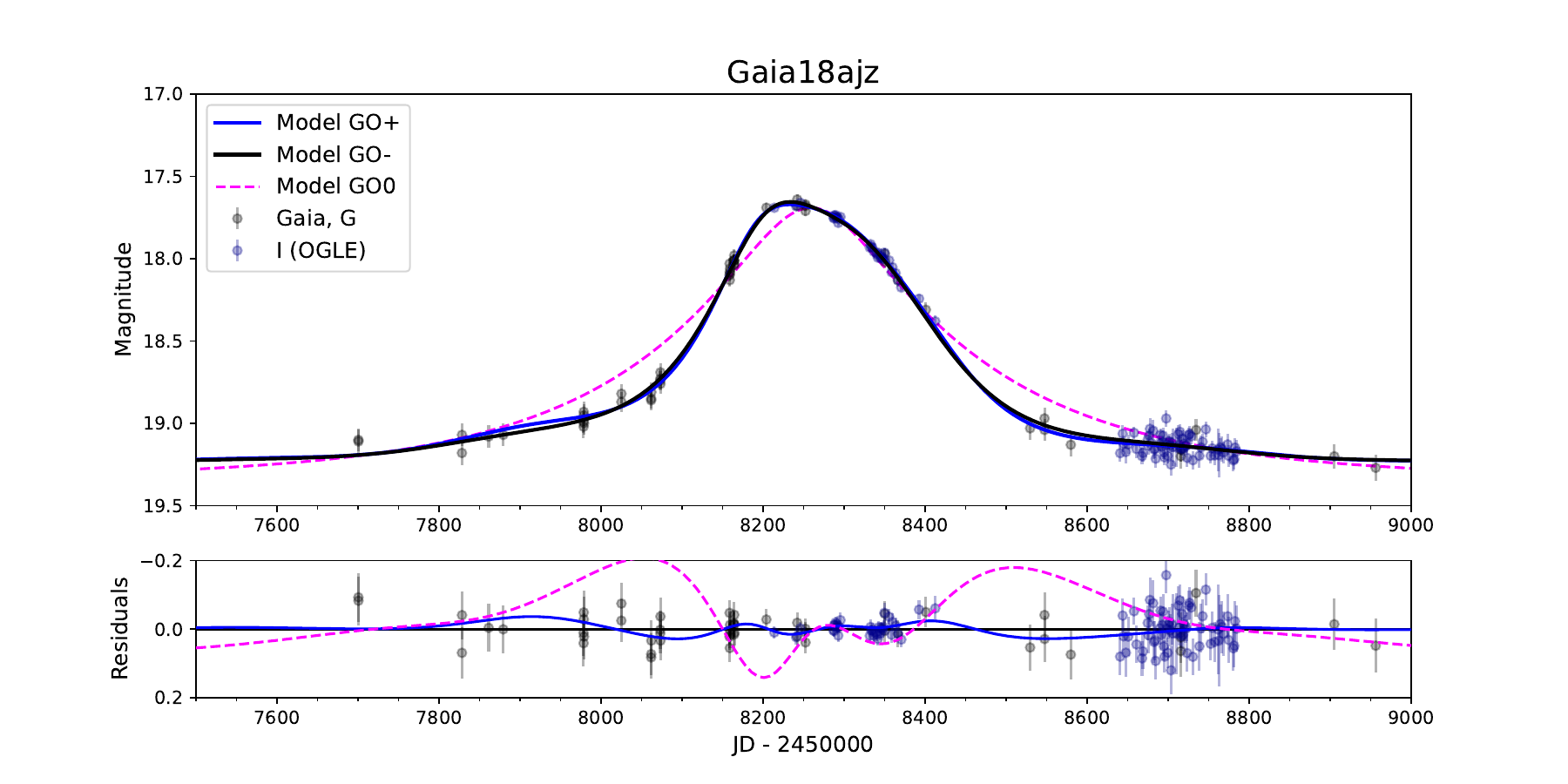}
   \caption{The light curve of the Gaia18azj event and the microlensing model fit using the Gaia data as well as the ground-based data from OGLE. The dashed magenta line represents the model without parallax, while the blue and black solid lines show the positive and negative solutions for the parallax model, respectively. The bottom panel shows the residuals with respect to the negative solution for the parallax model.}
   \label{fig:lc3}
\end{figure*}

\begin{figure*}
   \centering
   \includegraphics[width=\hsize]{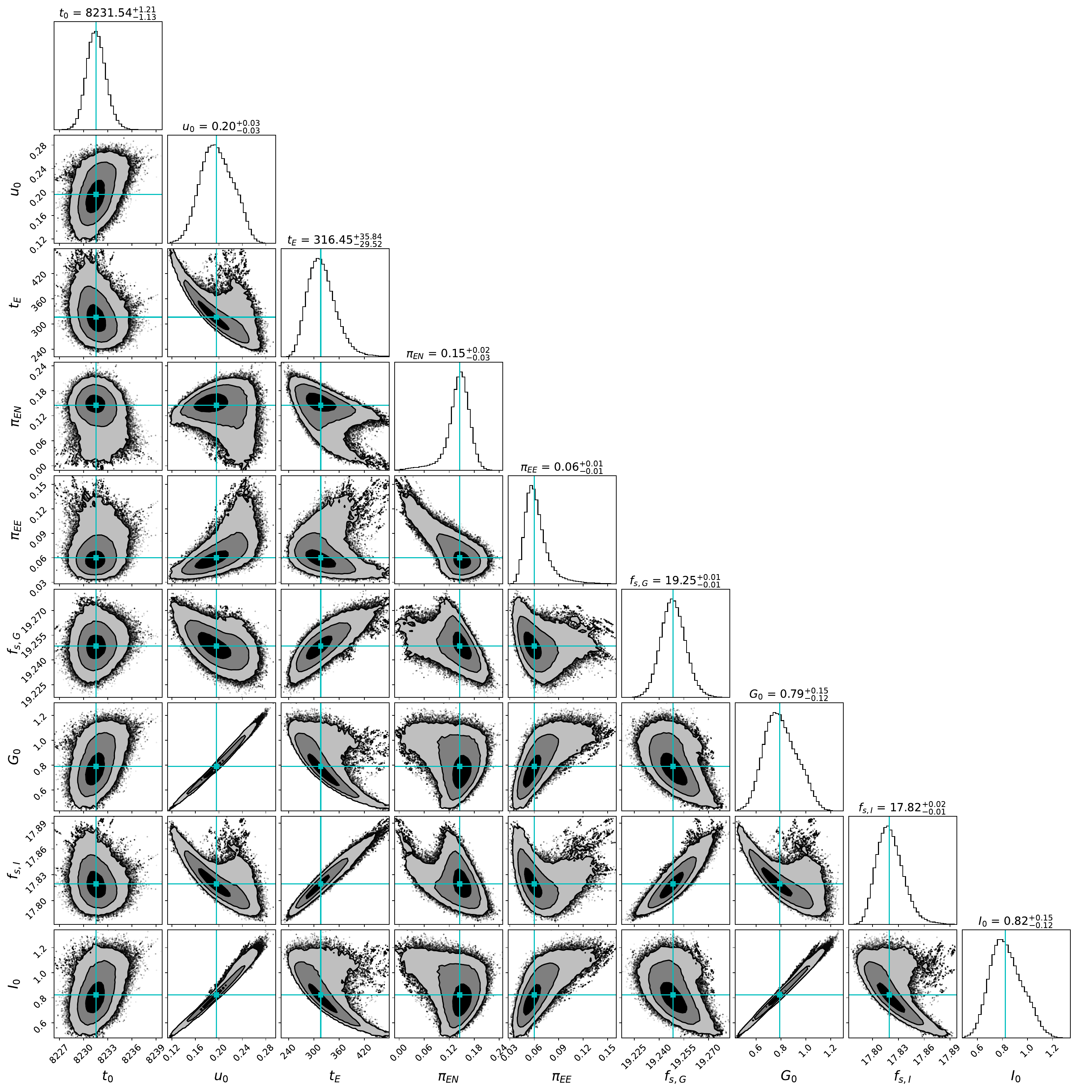}
   \caption{Corner plot of the microlensing parameters from the MCMC fit of the GO+ solution. The plot exhibits 1$\sigma$, 2$\sigma$, and 3$\sigma$ confidence regions with solid black, dark grey, and light grey colors, respectively. Any solutions outside of the 3$\sigma$ confidence level are represented by black dots.}
   \label{fig:cplot1}
\end{figure*}

Figures \ref{fig:lc1}, \ref{fig:lc2} and \ref{fig:lc3} show light curves for all models for the Gaia-only, Gaia+follow-up  and Gaia+OGLE datasets, respectively. The corner plot for the GO+ solution is shown in Figure \ref{fig:cplot1}. The plot was produced utilizing the \texttt{corner} python package developed by \cite{corner}.

\begin{figure*}
   \centering
   \includegraphics[width=\hsize]{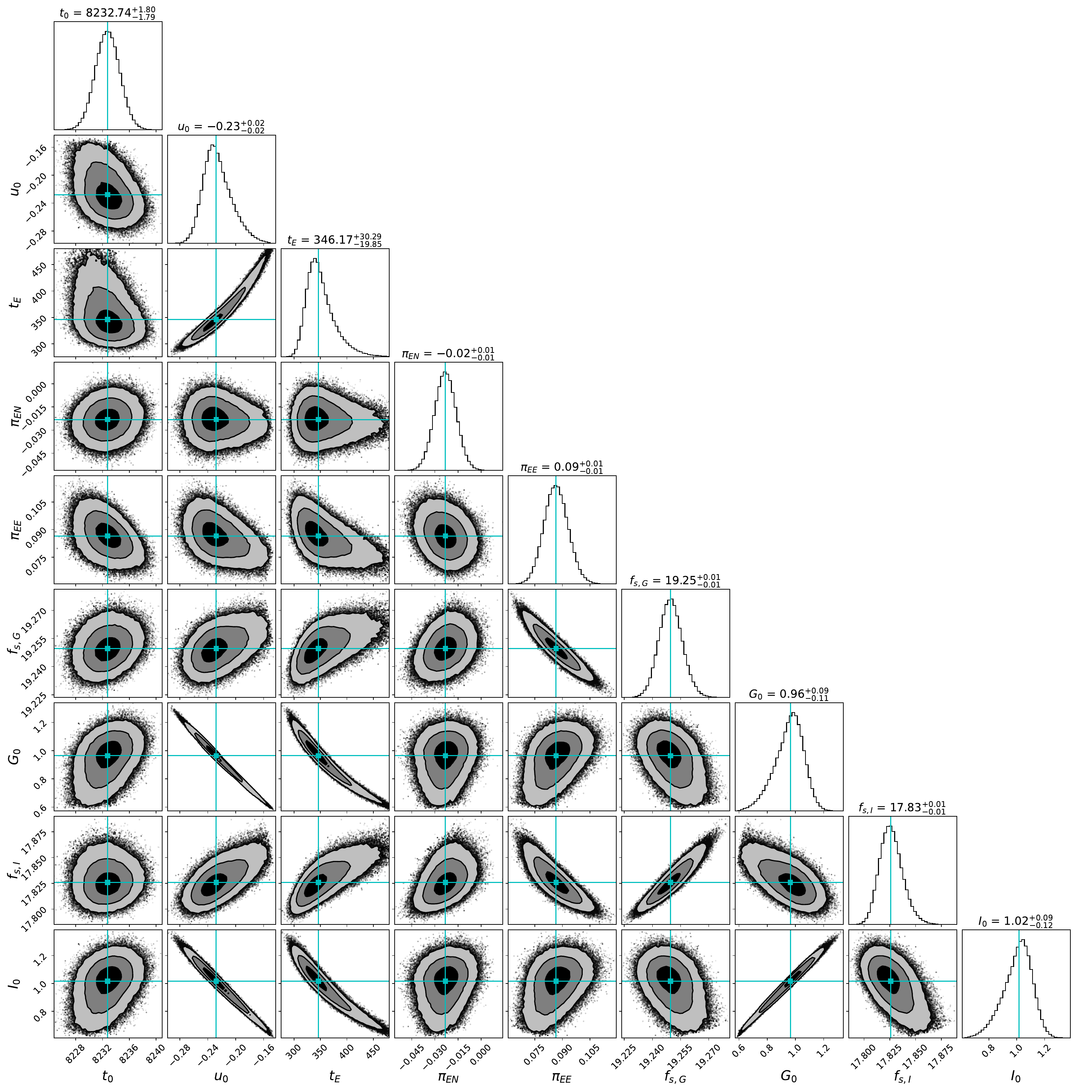}
   \caption{Corner plot of the microlensing parameters from the MCMC fit of the GO- solution. The plot exhibits 1$\sigma$, 2$\sigma$, and 3$\sigma$ confidence regions with solid black, dark grey, and light grey colors, respectively. Any solutions outside of the 3$\sigma$ confidence level are represented by black dots.}
   \label{fig:cplot2
}
\end{figure*}

Table \ref{tab:nested_sampling} shows equal-weight sample ratio of solution with positive to solution with negative impact parameter $u_0$ for Gaia-only and Gaia + OGLE models. Ratios were obtained using nested sampling.
\begin{table}
    \centering
    \caption{\label{tab:nested_sampling} Equal-weight sample ratios of solution with positive to solution with negative impact parameter $u_0$.}
    \begin{tabular}{c c}
            \hline
            \noalign{\smallskip}
            Model & $N(u_0+)/N(u_0-)$ \\
            \noalign{\smallskip}
            \hline
            \hline
            \noalign{\smallskip}
            Gaia-only   &  1.90 \\
            \noalign{\smallskip}
            Gaia + OGLE & 24.68  \\
            \hline
    \end{tabular}
\end{table}

\section{Source star}

\subsection{Atmospheric parameters}
\begin{figure}
   \centering
   \includegraphics[width=\hsize]{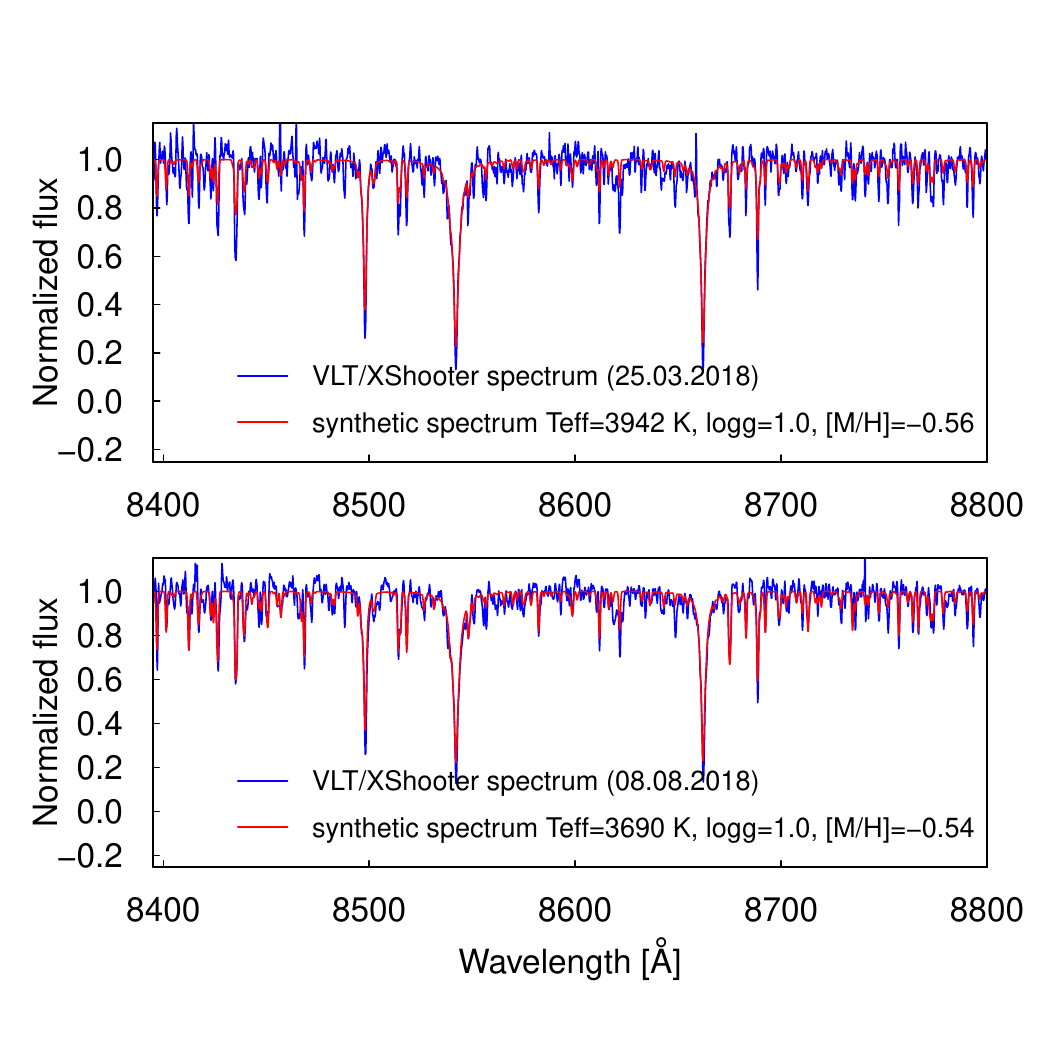}
   \caption{The result of the synthetic spectrum fitting (red spectra) to the X-Shooter spectroscopic data (blue spectra) generated for the best-matching atmospheric parameters. The Ca~II triplet region is presented for two spectra obtained in 2018, March 25 (top panel) and August 8 (bottom panel) by using X-Shooter.}
   \label{fig:specfit}
\end{figure}

In order to obtain the atmospheric parameters of the source star of Gaia18ajz event, the spectral analysis was performed with the \textit{iSpec} \footnote{\url{https://www.blancocuaresma.com/s/iSpec}} framework of various radiative transfer codes \citep{BlancoCuaresma2014, BlancoCuaresma2019}. The effective temperature $T_{\rm eff}$, surface gravity $\log g$ and metallicity [M/H] were determined by using the SPECTRUM\footnote{\url{http://www.appstate.edu/~grayro/spectrum/spectrum.html}} code, a grid of MARCS atmospheric models \citep{Gustafsson2008}, solar abundances from \citet{Grevesse2007} and atomic line list from Gaia-ESO Survey (GESv6; \citealt{Heiter2021}). The GESv6 line list covers the wavelength range from 420 to 920~nm, therefore, only UVB and VIS part of the X-Shooter spectra could be used. But due to the poor quality and low SNR of the UVB region in both spectra we decided to fit the synthetic spectra for VIS part only. In consequence, we have obtained the $T_{\rm eff}$, $\log g$ and [M/H] of the source star in both cases. The final solutions of atmospheric parameters resulted from the fitting procedure of both X-Shooter spectra are presented in Tab.~\ref{tab:params} and in Fig.~\ref{fig:specfit}.

\begin{figure}[h]
   \centering
   \includegraphics[width=\hsize]{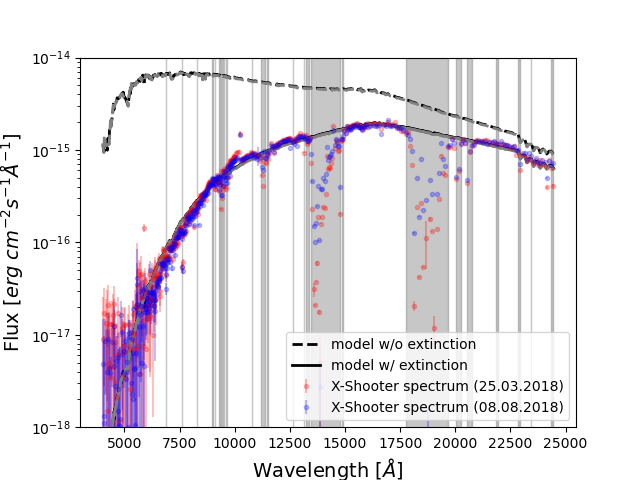}
   \caption{The two spectra from VLT/X-Shooter obtained in 2018, March 25 (red) and August 8 (blue), as well as the best models with (solid lines) and without (dashed lines) extinction correction are visible. The gray vertical lines and regions indicate the telluric bands, where the data were not used for the modelling.}
   \label{fig:templatematch}
\end{figure}

In addition to the absorption line analysis, we applied a template matching method using Spyctres \footnote{\url{https://github.com/ebachelet/Spyctres}}, similarly to \citet{2022Bachelet}. We modeled both X-Shooter spectra using the stellar template library from \citet{Kurucz1993} as well as the Spectral Energy Distribution (SED) at the time of spectra acquisition, including the source magnification $A(\mathrm{t})\sim5$ (for the first spectrum) and $A(\mathrm{t})\sim4$ (for the second spectrum). We used Markov Chain Monte Carlo (MCMC, \citet{2013Emcee}) sampling in order to generate the chains for final values of the effective temperature $T_{\mathrm{eff}}$, the surface gravity $log\mathrm{g}$, the metallicity $\mathrm{[Fe/H]}$, the radial velocity of the source star, the angular source radius $\theta_*$ and the line-of-sight extinction $A_V$. We have constrained MCMC chains by using the mean values of the parameters obtained in synthetic spectra fitting as input values. As a result, we obtained the stellar parameters and the line-of-sight extinction $A_V$ thanks to an accurate flux calibration in wide wavelength range of X-Shooter data (UVB+VIS+NIR parts) and the updated extinction law taken from \citet{Wang2019}. The extinction parameter $A_V$ was additionally constrained by taking into account the unblended apparent V magnitude. The final values of parameters obtained in template matching are presented in Tab.~\ref{tab:params} (last column) and in Fig.~\ref{fig:templatematch}.

Based on the parameters determined in these two methods, we state that the Gaia18ajz source is reddened K5-type supergiant or bright giant (luminosity class I or II) located in the Galactic disk. \citet{Schlegel98} and \citet{Schlafly2011} estimated the value of $A_V$ for as high as 11.76~mag and 10.11~mag, respectively, assuming a visual extinction to reddening ratio $A_V / E(B-V) = 3.1$\footnote{\url{https://irsa.ipac.caltech.edu/applications/DUST/}}. High reddening for this location in the Galaxy is confirmed by the Galactic dust distribution which is in agreement with the value obtained by us $A_V=7.3$~mag.

\begin{table*}
    \centering
    \caption{\label{tab:params} The parameters of the source star determined in synthetic spectrum fitting and template matching.}
    \begin{tabular}{l c c c}
            \hline
            \noalign{\smallskip}
            Parameter & \multicolumn{2}{c}{Synthetic spectrum fitting} & Template matching \\
             & 1st spectrum & 2nd spectrum & both spectra \\
             & 25 Mar 2018 & 8 Aug 2018 &  \\
            \noalign{\smallskip}
            \hline
            \hline
            \noalign{\smallskip}
            $T_\mathrm{eff}$ [K] & $3942\pm182$ & $3690\pm100$ & $3887\pm33$ \\ 
            $\log g$ (cgs) & $1.00\pm0.29$ & $1.00\pm0.75$ & $1.00\pm0.06$ \\
            $[\mathrm{M}/\mathrm{H}]$ [dex] & $-0.56\pm0.21$ & $-0.54\pm0.61$ &$-0.46\pm0.22$ \\
            $A_V$ [mag] & - & - & $7.3\pm0.1$ \\
            \hline
    \end{tabular}
\end{table*}

\subsection{Distance}
The atmospheric parameters and exctinction $A_V$ in the direction to the Gaia18ajz event were used to calculate the distance to the source from the equation:
\begin{equation}
5\log D_S = V - M_V + 5 - A_V,
\end{equation}
where $D_S$ is the distance to source star, $V$ is the apparent magnitude, $M_V$ is the absolute
magnitude and $A_V$ is the interstellar line-of-sight extinction. 
Assuming the typical absolute magnitude $M_V=-1.4\pm0.2$~mag for metal-poor K5~I/II star estimated by using CMD 3.7\footnote{\url{http://stev.oapd.inaf.it/cmd}} with PARSEC (v1.2S) isochrones \citep{Bressan2012} as well as apparent brightness $V=21.82\pm0.05$~mag, the distance is $D_S=15.26\pm2.46$~kpc. 

\begin{figure*}
   \centering
   \includegraphics[width=1.0\textwidth]{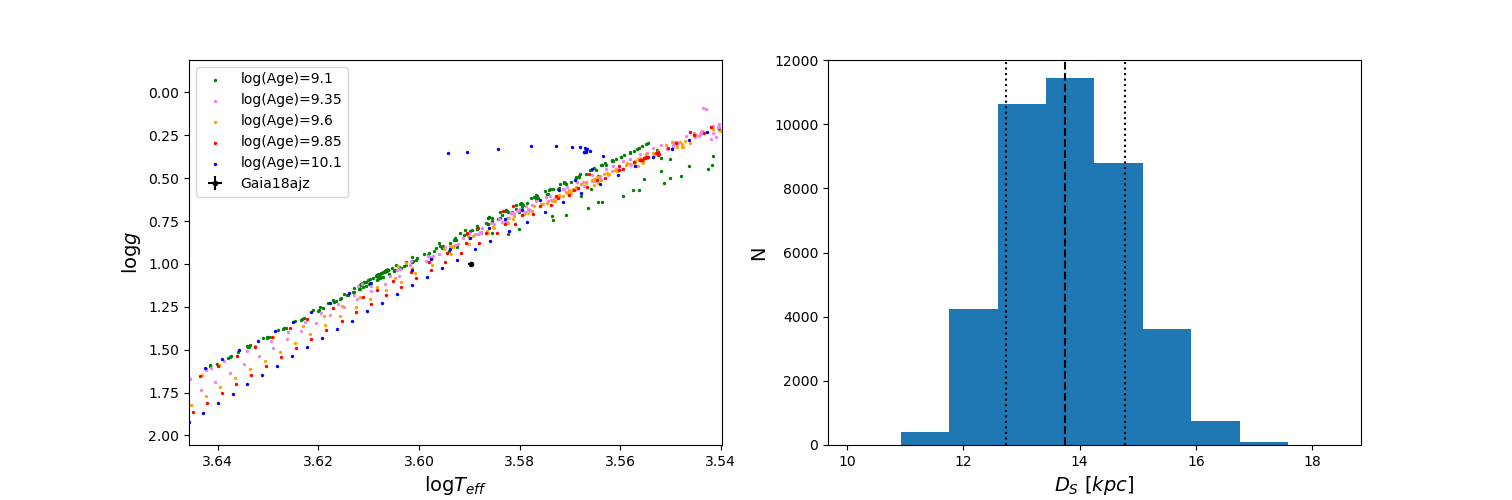}
   \caption{(Left) PARSEC stellar isochrones for 1.3, 2.2, 4.0, 7.1 and 12.6 Gyr with a fixed metallicity of -0.5. The source is most likely an old bright giant or supergiant.
   (Right) Source distance distribution based on PARSEC isochrones and the extinction estimated from the spectra modelling. Black dashed and dotted lines indicate the median distance value and the 1-$\sigma$ confident region, respectively.}
   \label{fig:kiel-dist}
\end{figure*}

Moreover, the template matching analysis yields to the source distance distribution based on PARSEC isochrones that is presented in Fig.~\ref{fig:kiel-dist}. According to this distribution, the average distance to the source of Gaia18ajz event is $13.75\pm1.02~$kpc. This value is in agreement with the spectroscopic distance calculation within 1-$\sigma$ uncertainty. The plot also shows that the source is located in the region of Kiel diagram ($\log\mathrm{g}-\log{T}_{\mathrm{eff}}$) where isochrones for ages~$>4$~Gyr are dominant, assuming metal-poor stars.

The spectroscopic distances obtained according to the above-mentioned approach are inconsistent with distances based on the Gaia parallax measurements. \citep{2021Bailer-Jones} inferred the distance to Gaia18ajz taking into account astrometric (geometric model) and astrometric + photometric measurements (photogeometric model) published in GaiaDR3 \citep{GaiaDR3}. This study gives $D_S=3.39^{+5.35}_{-2.29}$~kpc assuming the geometric model and $D_S=4.41^{+5.79}_{-3.07}$~kpc assuming the photogeometric model, which is 4.5 and 3.5 times smaller than our spectroscopic value, respectively. Moreover, \citep{Bailer-Jones2018} inferred an even smaller value of the distance $D_S=0.33^{+0.43}_{-0.26}$~kpc based on GaiaDR2 data \citep{GaiaDR2}. The reason of such a discrepancy in the distance can be explained with the help of astrometric quality flags published also in the Gaia archive and presented in Tab.~\ref{tab:gdrsVals}. First of all, the RUWE parameter in both data releases is higher than 1.4 which is the tolerance limit according to the Gaia Data Processing and Analysis Consortium (DPAC) documentation\footnote{\url{https://gea.esac.esa.int/archive/documentation/GDR2/Gaia_archive/chap_datamodel/sec_dm_main_tables/ssec_dm_ruwe.html}}. This can indicate that the astrometric solution for the source is problematic due to the large scatter in the measurements or the binary nature of the object. Another parameter that provides a consistent measure of the
astrometric goodness-of-fit is \url{astrometric_excess_noise} which is also relatively high, given 3.1~mas in the case of Gaia18ajz. Because we do not see any evidence of the binarity in the light curve as well as in the spectroscopic data, we can suspect that Gaia measurements were affected by either some instrumental effects or centroid shift motion due to the astrometric microlensing. In addition, it is clearly visible that the parallaxes and proper motions, both in the GaiaDR2 and DR3 releases, are characterised by relatively high 1-$\sigma$ error bars. It yields small values of the parameter \url{parallax_over_error}, 5.49 (for GaiaDR2) and 2.81 (for GaiaDR3), and shows the low quality of the Gaia measurements for Gaia18ajz. The large scatter of the astrometric data and hints for problems with obtaining a reliable solution lead to distrust of the Gaia astrometric parallax measurement in this case. Therefore, we decided to use the spectroscopic distance as the main constraint for lens mass and distance in the microlensing model.

\section{Lensing object}
\subsection{Assumptions}
Since there is no available astrometric data for this event, determining the exact values of the distance to the lens and its mass is currently impossible. To uncover the most likely parameters of the lens relying solely on photometric data, we employed the \textit{DarkLensCode} described in detail in Appendix \ref{app:dlc}.

Since the \textit{G} filter is wider than the \textit{V} filter, we assumed the upper bound of extinction to the lens $A_G=7.3\,\text{mag}$, which is the value of extinction to the source star in the \textit{V} filter derived from spectroscopy. Following \cite{Kaczmarek2022}, for the lower bound, we assumed $A_G=0\,\text{mag}$. For the proper motion of the source star, we assumed the values of \Gaia DR3 shown in Table \ref{tab:gdrsVals}, despite of elevated RUWE value in Gaia catalogues. The proper motions reported in GDR2 and GDR3 were consistent with each other to within error bars, while the reported parallaxes were very different. For the source distance, we assumed the value from spectroscopy and, at each iteration of the code, we drew $D_S$ from the normal distribution with the mean $15.26\,\text{kpc}$ and standard deviation of $2.46\,\text{kpc}$. We used two mass functions as lens mass priors. First, \cite{2001Kroupa} (hereafter StellarIMF), describes stellar objects
\begin{equation}
    f(M) \sim 
    \begin{cases}
        M^{-0.3}, & ~~ M \leq 0.08 \msun,\\
        M^{-1.3}, & ~~  0.08 \msun < M \leq 0.5 \msun,\\
        M^{-2.3}, & ~~  0.5 \msun < M < 150 \msun,\\
    \end{cases}
    \label{eq:kroupa}
\end{equation}
and second, \cite{2021MrozBHOGLE} (hereafter DarkIMF), describes solitary dark remnants in the Milky Way
\begin{equation}
    f(M) \sim 
    \begin{cases}
        M^{0.51}, & ~~ M \leq 1.0 \msun,\\
        M^{-0.83}, & ~~  1.0 \msun < M < 100 \msun\\
    \end{cases}
    \label{eq:mroz}
\end{equation}
Summary of the input parameters for the \DLC is shown in the Table \ref{tab:dlcparams}.

\begin{table}
\centering
\caption{\label{tab:dlcparams}\textit{DarkLensCode} input parameters.}
     \centering
        \begin{tabular}{c c }
        \hline
        \noalign{\smallskip}
             Parameter &  Value \\
             \noalign{\smallskip}
        \hline
        \hline
        \noalign{\smallskip}
            \texttt{alpha} & $277.56025^{\circ}$ \\
            \noalign{\smallskip}
            \texttt{delta} & $-8.22021^{\circ}$ \\
            \noalign{\smallskip}
            \texttt{t0par} & 8231 \\
            \noalign{\smallskip}
            \texttt{extinction} & 7.3 [mag]\\
            \noalign{\smallskip}
            \texttt{ds\_median} & 15.26 [kpc]\\
            \noalign{\smallskip}
            \texttt{ds\_err\_neg} & 2.46 [kpc]\\
            \noalign{\smallskip}
            \texttt{ds\_err\_pos} & 2.46 [kpc]\\
            \noalign{\smallskip}
            \texttt{mu\_ra} & -5.37 [mas/yr]\\
            \noalign{\smallskip}
            \texttt{mu\_ra\_sig} & 0.59 [mas/yr]\\
            \noalign{\smallskip}
            \texttt{mu\_dec} & -6.69 [mas/yr]\\
            \noalign{\smallskip}
            \texttt{mu\_dec\_sig} & 0.51 [mas/yr]\\
            \noalign{\smallskip}
            \texttt{mu\_ra\_dec\_corr} & 0.38 \\
            \noalign{\smallskip}
        \hline
        \hline
        \end{tabular}
\end{table}

\subsection{Mass and distance}
\begin{figure}
   \centering
   \includegraphics[width=\hsize]{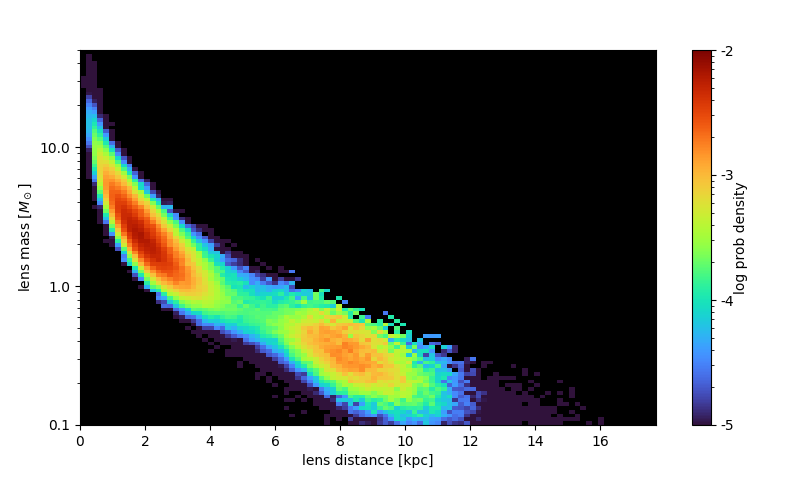}
   \caption{\textit{DarkLensCode} output for the GO+ microlensing model assuming the DarkIMF. Posterior distribution of the distance to the lens and its mass. The colors correspond to the log probability density. The dark color of a bin means that there were no samples present in this bin.}
   \label{fig:mass_dist}
\end{figure}

\begin{figure}
   \centering
   \includegraphics[width=\hsize]{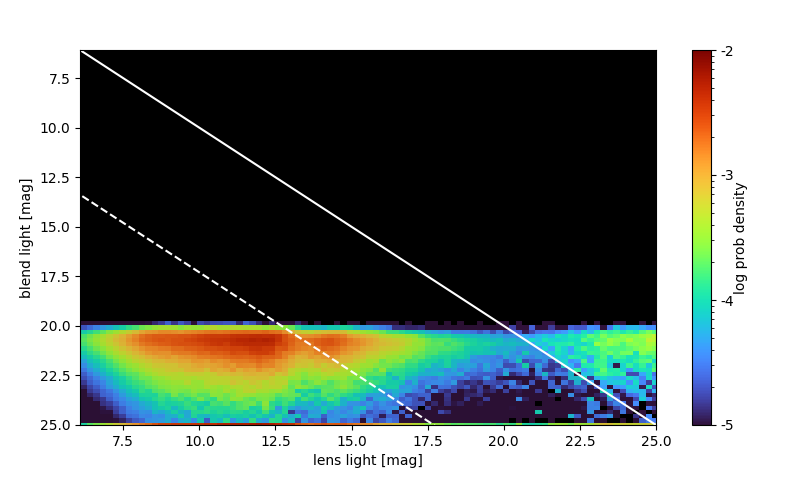}
   \caption{\textit{DarkLensCode} output for the GO+ microlensing model assuming the StellarIMF mass function. Posterior distribution of the blended light magnitude and lens light magnitude assuming that the lens is an MS star. The solid line represents the upper limit of extinction, while the dashed line represents the lower limit of extinction. Lines separate dark lens solutions from the MS star solutions. The colors correspond to the log probability density. The dark color of the bin means that there were no samples present in this bin.}
   \label{fig:blend_lens}
\end{figure}

To estimate the mass and distance of the lens, we employed samples from MCMC models of both the GO+ and GO- models. Figure \ref{fig:mass_dist} shows the two-dimensional posterior distribution of the lens mass and lens distance. Figure \ref{fig:blend_lens} shows the posterior distribution of the blend magnitude and lens magnitude if it were a main sequence star. Both figures come from the GO- photometric model and assume the StellarIMF mass function. 

As can be seen in Figure \ref{fig:mass_dist}, the posterior distribution is bimodal with two distinct solutions. One represents a more massive, close lens while the other is less massive and at a greater distance. We decided to split these solutions at $D_L=5\,\text{kpc}$. Assuming the mass function for luminous sources (StellarIMF), we estimated dark lens probability (DLProb) and probability of each solution (defined as the sum of the weights in each solution divided by the sum of all weights). Since the DLProb of more massive solutions is high, we recalculated all weights using DarkIMF mass function. We report $M_L$, $D_L$, $\theta_E$, and solution probability for both mass functions. Since the calculation of the dark lens probability assumes that the lens is in a MS star, we report the DLProb using only StellarIMF mass function. The results for the GO- photometric model are shown in the table \ref{tab:dlcgfneg} while the results for the GO+ photometric model are in the table \ref{tab:dlcgfpos}.

\begin{table}
\centering
\caption{\label{tab:dlcgfneg}\textit{DarkLensCode} output for GO- solution for each possible scenario. Estimations for the lens' distance and mass, as well as $\theta_E$ and the probability of the lens being a dark remnant for each possible scenario.}
     \centering
        \begin{tabular}{c c c}
        \hline
        \noalign{\smallskip}
             Parameter &  Near & Far \\
             \noalign{\smallskip}
        \hline
        \hline
        \noalign{\smallskip}
            DLProb & $99.8\% - 100.0\%$ & $5.7\% - 93.5\%$ \\
            \noalign{\smallskip}
            \hline
             \noalign{\smallskip}
            \multicolumn{3}{c}{StellarIMF} \\
             \noalign{\smallskip}
            \hline
            \noalign{\smallskip}
            Solution probability & $47\%$ & $53\%$ \\
            \noalign{\smallskip}
            $M_L$ [$M_{\odot}$] & $7.2^{+5.1}_{-2.7}$ &  $0.94^{+0.33}_{-0.26}$ \\
            \noalign{\smallskip}
            $D_L$ [kpc] & $1.88^{+0.91}_{-0.71}$ & $8.4^{+1.0}_{-1.0}$ \\
            \noalign{\smallskip}
            $\theta_E$ [mas] & $5.5^{+3.7}_{-2.0}$& $0.72^{+0.21}_{-0.18}$\\
            \noalign{\smallskip}
            \hline
             \noalign{\smallskip}
            \multicolumn{3}{c}{DarkIMF} \\
             \noalign{\smallskip}
            \hline
            \noalign{\smallskip}
            Solution probability & $96\%$ & $4\%$ \\
            \noalign{\smallskip}
            $M_L$ [$M_{\odot}$] & $11.1^{+10.3}_{-4.7}$ &  $1.11^{+0.38}_{-0.26}$ \\
            \noalign{\smallskip}
            $D_L$ [kpc] & $1.31^{+0.80}_{-0.60}$ & $8.0^{+0.9}_{-1.0}$ \\
            \noalign{\smallskip}
            $\theta_E$ [mas] & $8.2^{+7.5}_{-3.4}$& $0.81^{+0.23}_{-0.18}$\\
            \noalign{\smallskip}
        \hline
        \hline
        \end{tabular}
         \tablefoot{The minimum probability of dark lens occurrence corresponds to the upper limit of extinction (7.3 mag), while the maximum probability of dark lens occurrence corresponds to the lower limit of extinction (0 mag). The probability of each solution is defined as the sum of the weights in each solution divided by the sum of all weights.}
\end{table}

\begin{table}
\centering
\caption{\label{tab:dlcgfpos}\textit{DarkLensCode}  output for GO+ solution for each possible scenario. Estimations for the lens' distance and mass, as well as $\theta_E$ and the probability of the lens being a dark remnant.}
     \centering
        \begin{tabular}{c c c}
        \hline
        \noalign{\smallskip}
             Parameter & Near & Far \\
             \noalign{\smallskip}
        \hline
        \hline
        \noalign{\smallskip}
            DLProb & $85\%- 100.0\%$ & $0.1\% - 32.7\%$ \\
            \noalign{\smallskip}
            \hline
             \noalign{\smallskip}
            \multicolumn{3}{c}{StellarIMF} \\
             \noalign{\smallskip}
            \hline
            \noalign{\smallskip}
            Solution probability & $94\%$ & $6\%$ \\
            \noalign{\smallskip}
            $M_L$ [$M_{\odot}$] & $2.8^{+2.3}_{-1.1}$ &  $0.40^{+0.19}_{-0.14}$ \\
            \noalign{\smallskip}
            $D_L$ [kpc] & $1.70^{+0.79}_{-0.66}$ & $8.0^{+1.1}_{-1.2}$ \\
            \noalign{\smallskip}
            $\theta_E$ [mas] & $4.1^{+2.9}_{-1.5}$& $0.53^{+0.20}_{-0.15}$\\
            \noalign{\smallskip}
            \hline
             \noalign{\smallskip}
            \multicolumn{3}{c}{DarkIMF} \\
             \noalign{\smallskip}
            \hline
            \noalign{\smallskip}
            Solution probability & $99.9\%$ & $0.1\%$ \\
            \noalign{\smallskip}
            $M_L$ [$M_{\odot}$] & $4.9^{+5.4}_{-2.3}$ &  $0.59^{+0.29}_{-0.21}$ \\
            \noalign{\smallskip}
            $D_L$ [kpc] & $1.14^{+0.75}_{-0.57}$ & $7.3^{+1.2}_{-1.6}$ \\
            \noalign{\smallskip}
            $\theta_E$ [mas] & $6.5^{+6.9}_{-2.8}$& $0.69^{+0.30}_{-0.20}$\\
            \noalign{\smallskip}
        \hline
        \hline
        \end{tabular}
        \tablefoot{The minimum probability of dark lens occurrence corresponds to the upper limit of extinction (7.3 mag), while the maximum probability of dark lens occurrence corresponds to the lower limit of extinction (0 mag). The probability of each solution is defined as the sum of the weights in each solution divided by the sum of all weights.}
\end{table}

\section{Discussion}

The Gaia18ajz microlensing event lasted about 1000 days, making it one of the longest microlensing events ever studied. The most probable explanation of the light curve shape is microlensing caused by a single lens on a single source with a visible annual parallax effect. 
The standard Paczy\'nski model without the inclusion of the parallax effect, visible in Figures \ref{fig:lc1}, \ref{fig:lc2}, \ref{fig:lc3} as a dashed line, cannot fully explain the changes in magnitude. The light curve does not show any features due to binarity of the lens, in particular the caustic crossings, hence we can immediately rule out binary lenses with separations of the order of single AUs. The effect of space parallax was not included in the models obtained. Due to significant measurement error caused by the low brightness of the event as well as the small annual parallax effect, we can assume that space parallax has minimal effect on the light curve.

The event was observed by various observatories (Table \ref{tab:photometry}, Figure \ref{fig:data}). However, due to the low brightness of the event, at a baseline below 19 mag in \textit{G} and below 21 mag in \textit{V}, most of the data is characterised by a significant measurement error. Datasets from \Gaia and OGLE have the best quality. Although the data from other observatories have bigger error bars, they are consistent with the data from \Gaia and OGLE. 

High-resolution spectroscopic observations were possible around the maximum of the Gaia18ajz brightness. We used the VLT/X-Shooter instrument at two epochs when the source was the most amplified: $\sim 5$ times in the case of the first spectrum and $\sim 4$ times in the case of the second spectrum. Due to this, we were able to determine the atmospheric parameters of the source, the line-of-sight extinction, and the spectroscopic distance. Based on spectral analysis, the Gaia18ajz source was classified as a reddened K5 supergiant or a bright giant at a large distance of 15.26 kpc. The spectroscopic parameters, especially the distance, constrained the microlensing models and helped in estimates of the lens' distance and mass as well as the probability of the lens being dark remnant.

The event can be explained by two models, one with positive and one with negative impact parameter $u_0$. Models obtained using only data from Gaia (G models), data from Gaia and OGLE (GO models) and data from Gaia and all of the ground-based follow-up (GF models) agree with each other within error bars. Although the differences are not great, the GO+ model has the lowest $\chi^2$ of all models. The discrepancy between both solutions cannot be resolved with the available photometric data. The model with a positive impact parameter is more likely than the model with a negative impact parameter. In case of Gaia + OGLE models, GO+ is 24.68 times more likely than GO-.

Estimates of the distance to the lens and its mass obtained using \DLC indicate two possible scenarios. In the first scenario, the lens is positioned nearer with a greater mass, whereas in the alternate scenario, the lens is situated farther away with a lesser mass. Based on calculations by $\DLC$, the probability that the lens is a dark remnant exceeds $79.6\%$ in the more massive scenario, while in the second scenario, this probability falls below $32.7\%$ for the more probable GO+ model. For the GO- model, in the more massive case, the lens is a dark remnant object in more than $99.5\%$. In the other case, this probability is between $4.8\%$ and $98.0\%$. Estimates of the distance to the lens and its mass are highly dependent on the assumed mass function. 

If we assume that the lens is a star and the mass function is defined by Equation \ref{eq:kroupa} \citep{2001Kroupa} (StellarIMF), for a less massive solution we get $M_L = 0.94^{+0.33}_{-0.26}\,M_{\odot}$ located at $D_L = 8.4^{+1.0}_{-1.0}\,\text{kpc}$ for the GO- model and $M_L = 0.40^{+0.19}_{-0.14}\,M_{\odot}$ at $D_L = 8.0^{+1.1}_{-1.2}\,\text{kpc}$ for the GO+ model. Regardless of extinction, for GO- model the lens is unlikely to be a star and belong to the more massive solution. For more likely GO+ model, if the extinction is close to the maximal value, the GLProb of the model would be close to 85\% which gives 15\% chance of lens being a MS-star with $M_L = 2.8^{+2.3}_{-1.1}\,M_{\odot}$ at $D_L = 1.70^{+0.79}_{-0.66}\,\text{kpc}$. In the other 85\%, the lens would be a dark remnant which contradicts our assumption of the StellarIMF.

If we assume that the lens is a dark remnant, and the mass function is defined by Equation \ref{eq:mroz} \citep{2021MrozBHOGLE} (DarkIMF), for the more massive solution we obtain $M_L = 11.1^{+10.3}_{-4.7}\,M_{\odot}$ located at $D_L = 1.31^{+0.80}_{-0.60}\,\text{kpc}$ for the GO- model and $M_L = 4.9^{+5.4}_{-2.3}\,M_{\odot}$ at $D_L = 1.14^{+0.75}_{-0.57}\,\text{kpc}$ for the GO+ model. In this scenario, the lens is likely a stellar-mass black hole. For a model with a positive impact parameter, the less massive solution is highly improbable. However, if the extinction is in the lower half of the possible range, the GLProb for the GO- model would be in the higher half of the possible range. In that case, the lens would have a mass of $M_L=1.11^{+0.38}_{-0.26}\,M_{\odot}$ at a distance of $D_L = 8.0^{+0.9}_{-1.0}\,\text{kpc}$. This would classify the lens as a massive white dwarf or a light neutron star.

The assumed distance to the source star affects the lens mass and the distance estimates obtained. A small distance, such as $D_S=4.41^{+5.79}_{-3.07}\,\text{kpc}$ from GDR3 would result in a closer lens, which would eliminate the less massive solution of $\DLC$. In such a case, if the lens is an MS star, the lens would be bright enough to be visible, which is in disagreement with the blending parameter. A greater distance to the source star would result in a greater distance to the lens and a lower lens mass.

In the \textit{DarkLensCode} analysis, we assume that the proper motion of \Gaia DR3 represents the proper motion of the source star. The microlensing event occurred after the data span included in GDR2 (2014.5-2016.4), so it could not have influenced the measurement of proper motion in GDR2. Furthermore, the proper motion values from \Gaia DR2 and \Gaia DR3 (2014.5-2017.4) agree with each other within the error bars. Since the proper motion signal is stronger than the parallax signal and grows over time as the star moves, the proper motion should be measured accurately, even if the parallax is not and the RUWE is elevated. In this case, the high RUWE is probably not caused by the microlensing event alone (as discussed further); it could be the result of a single bad measurement, which should not significantly affect the proper motion measurement. This justifies our assumption to use the proper motion from \Gaia as the proper motion of the source star, additionally, given almost negligible blended light. Moreover, our experiments with \textit{DarkLensCode} showed that using random proper motion values from a Gaussian distribution, with medians matching the GDR3 measurements, does not significantly change the result but instead makes the observed degeneracy more diffuse.

\begin{figure}[h]
   \centering
   \includegraphics[width=\hsize]{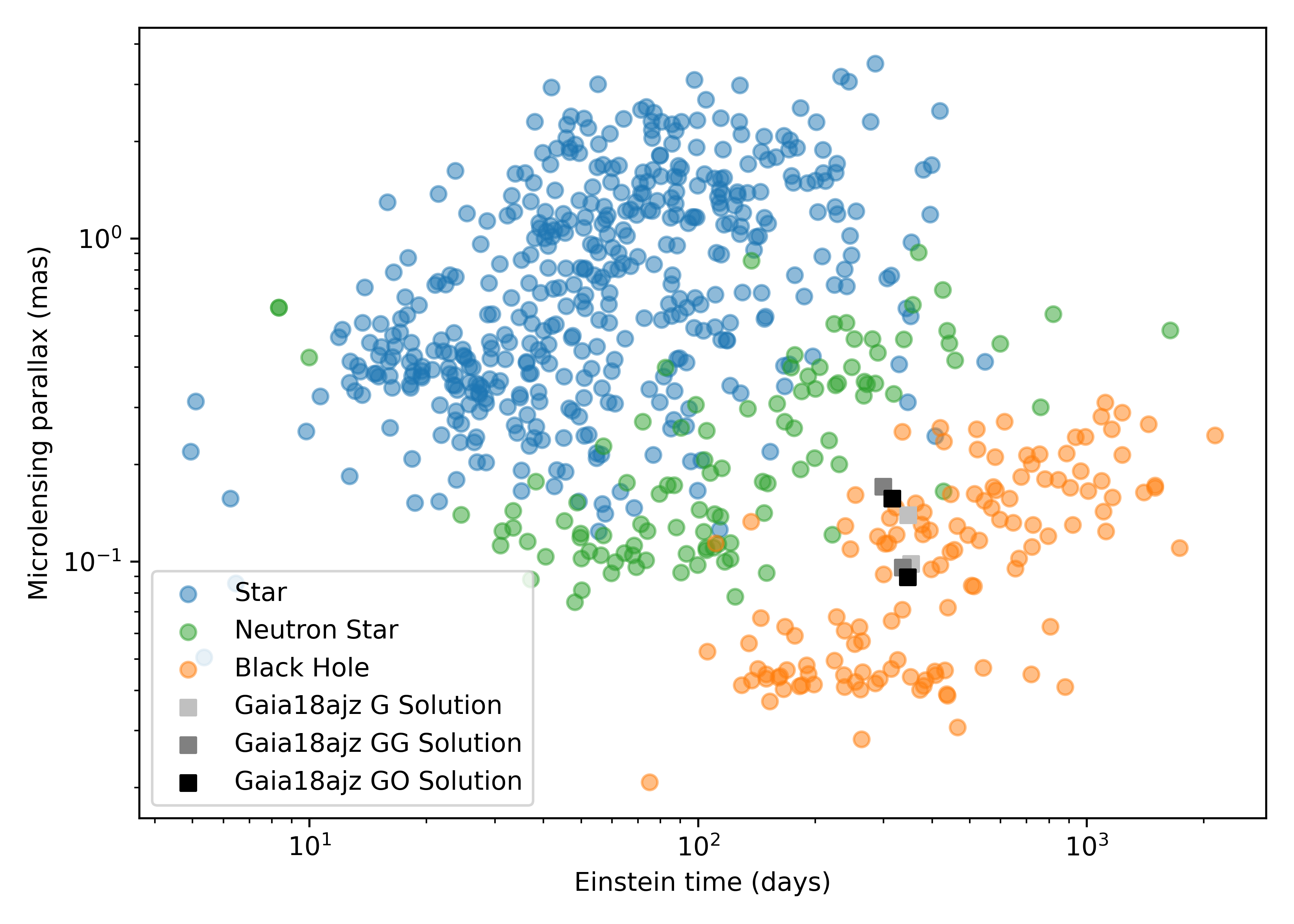}
    \caption{Location of all parallax solutions for Gaia18ajz photometry superimposed on the distribution of $t_E$ and $\pi_E$ parameters expected for microlensing events using the simulated data from \cite{Sweeney2024}. Note that the microlensing events due to stars should be $100\times$ more numerous in this plot, due to the undersampling in the \cite{Sweeney2024} data.}
    \label{fig:te-pie}
\end{figure}

The possibility that Gaia18ajz lens is a stellar-mass black hole is also visible when comparing the measured $t_E$ and $\pi_E$ parameters of all parallax solutions with the distributions of typical timescales and parallaxes of microlensing events. Fig. \ref{fig:te-pie} shows a location of Gaia18ajz solutions on microlensing events of stars, neutron stars, and black holes computed by \cite{Sweeney2024} based on the Galaxy model from \cite{Sweeney2022}. Half of the solutions fall into the region where black holes dominate, with the other half still being in the region of plausible black hole or neutron star events.

The \Gaia astrometric time series is currently unavailable and will not be accessible until the publication of \Gaia Data Release 4 in 2026. However, \Gaia DR2 \citep{GaiaDR2} and \Gaia DR3 \citep{GaiaDR3} reported on the astrometric parameters of the object associated with Gaia18ajz (Source ID=4156664130700362752 in both releases), as shown in Table \ref{tab:gdrsVals}. \Gaia DR3, in particular, used the \Gaia astrometric time series from 25 July 2014 (10:30 UTC) to 28 May 2017 (08:44 UTC) (up to JD = 2457890.86389). Gaia18ajz event was discovered on 14 February 2018, however, the photometric deviation began already before JD=2457800. As shown in \cite{BelokurovEvans2002}, the astrometric microlensing signal becomes significant already around $\sqrt{2}\theta_E$, hence a couple of $t_E$-s before the peak of the event. However, the elevated value of the RUWE parameter \citep{Lindegren2018} indicates that the fit of the 6-parameter astrometric model to the astrometric data has significant residuals. Furthermore, the value of the parallax (3.24$\pm$0.59 mas and 1.52$\pm$0.54 for GDR2 and GDR3, respectively) is far from the expected value for the source distance we found with spectroscopy. Although the astrometric microlensing effect could cause the RUWE to grow from 1.49 in GDR2 to 1.53 in GDR3, it cannot explain all the anomalies in the astrometric fit. Using the Astromet code \footnote{\href{https://github.com/zpenoyre/astromet.py}{https://github.com/zpenoyre/astromet.py}} and applying the method of \cite{Jablonska2022}, we concluded that $\theta_E$ needed to reproduce the parallaxes of GDR2 and GDR3 would cause the separation of the images to be high enough for \Gaia to detect them as separate sources. Based on that, we can assume that some other effect has to be responsible for disagreement in parallax and high RUWE in \Gaia DR2 (and in effect in GDR3).

If it could be confirmed that the lens is a black hole, it would be the second known isolated black hole. The only currently known isolated black hole was discovered by the \cite{Sahu_2022, Lam_2022}. If the mass of the black hole in Gaia18ajz event was equal to $M_L = 4.9^{+5.4}_{-2.3}\,M_{\odot}$ as in the GO+ solution, it would be the lightest isolated black hole currently known. According to \cite{Mroz_2022} the only currently known isolated black hole has a mass of $7.88\pm0.82\,M_{\odot}$. In this case the Gaia18ajz lens is also most likely lighter than previously detected \Gaia black holes GaiaBH1 with a mass of $9.62\pm0.18\,M_{\odot}$ \citep{GaiaBH1, Chakrabarti2023}, GaiaBH2 with a mass of $8.9\pm0.3\,M_{\odot}$ \citep{GaiaBH2, Ataru2022} and GaiaBH3 with $33\,M_{\odot}$ \citep{GaiaBH3}. On the other hand, the GO- solution yields mass of $11.1^{+10.3}_{-4.7}\,M_{\odot}$, hence it would make Gaia18ajz the most massive non-interacting single black hole known so far.

\section{Conclusions}
In this work, we presented the investigation and analysis of a long microlensing event Gaia18ajz located towards the Galactic Disk, discovered by the {\gaia} space satellite. The event exhibited a microlensing parallax effect perturbed by the Earth's orbital motion. The investigation is based on {\gaia} data and ground photometry, as well as spectroscopy follow-up observations.
The event has two best-fitting solutions with a negative and positive impact parameter, $u0$. The model with a positive impact parameter is more likely. The parameters of these solutions are presented in Tables \ref{tab:valSolutionsGaia}, \ref{tab:valSolutionsOGLE}, and \ref{tab:valSolutionsFup}. 

Using \DLC \ref{app:dlc}, we estimated the probability density of the lens mass and distance. We found two possible scenarios, one closer and more massive, and one further and less massive. We present the results in Tables \ref{tab:dlcgfneg} and \ref{tab:dlcgfpos}. We conclude that in the more massive scenario, the lens is a black hole with mass $M_L=11.1^{+10.3}_{-4.7}\,M_{\odot}$ at $D_L=1.31^{+0.80}_{-0.60}\,\text{kpc}$ for the less probable model or $M_L=4.9^{+5.4}_{-2.3}\,M_{\odot}$ at $D_L=1.14^{+0.75}_{-0.57}\,\text{kpc}$ for more probable photometric microlensing solution. For the less massive solution, we obtain a white dwarf or a star with mass $M_L=0.94^{+0.33}_{-0.26}\,M_{\odot}$ at $D_L=8.4^{+1.0}_{-1.0}\,\text{kpc}$ or $M_L=1.11^{+0.38}_{-0.26}\,M_{\odot}$ at $D_L=8.0^{+0.9}_{-1.0}\,\text{kpc}$. 
The final \Gaia Data Release DR4 (2026), with astrometric time series data for the Gaia18ajz source, could address the issue of elevated RUWE and possibly resolve the ambiguity between near- and far-recognition scenarios, as well as confirm which photometric solution is correct.

\begin{acknowledgements}
The authors would like to thank Dr Radek Poleski for discussion and his comments during the creation of this work.
This work has made use of data from the European Space Agency (ESA) mission {\it Gaia} (\url{https://www.cosmos.esa.int/gaia}), processed by the {\it Gaia} Data Processing and Analysis Consortium (DPAC,
\url{https://www.cosmos.esa.int/web/gaia/dpac/consortium}). Funding for the DPAC has been provided by national institutions, in particular the institutions
participating in the {\it Gaia} Multilateral Agreement. 
The project has received funding from the European Union's Horizon 2020 research and innovation programme under grant agreement No. 101004719 (OPTICON RadioNet Pilot, ORP).
BHTOM acknowledges the following people who helped with its development: Patrik Sivak, Kacper Raciborski, Piotr Trzcionkowski, and AKOND company.
This paper made use of the Whole Sky Database (wsdb) created by Sergey Koposov and maintained at the Institute of Astronomy, Cambridge by Sergey Koposov, Vasily Belokurov and Wyn Evans with financial support from the Science \& Technology Facilities Council (STFC) and the European Research Council (ERC), with the use of Q3C software
(http://adsabs.harvard.edu/abs/2006ASPC..351..735K). 
This research has used the VizieR catalogue access tool, CDS, Strasbourg, France.  
We acknowledge also the use of ChatGPT3.5 from OpenAI, which was helpful as a tool to improve the text of some parts of the manuscript,  aiming to mitigate any potential impact of non-native language usage on the overall quality of the publication. In particular, the title of the work, the abstract, and portions of the introduction were generated collaboratively with the aid of ChatGPT. Additionally, the description of the data was automatically generated using the BHTOM ver.2.0 Publication feature.
J.~M.~Carrasco was (partially) supported by the Spanish MICIN/AEI/10.13039/501100011033 and by "ERDF A way of making Europe" by the “European Union” through grant PID2021-122842OB-C21, and the Institute of Cosmos Sciences University of Barcelona (ICCUB, Unidad de Excelencia ’Mar\'{\i}a de Maeztu’) through grant CEX2019-000918-M. The Joan Oró Telescope (TJO) of the Montsec Observatory (OdM) is owned by the Catalan Government and operated by the Institute for Space Studies of Catalonia (IEEC). ZK is a Fellow of the International Max Planck Research School for Astronomy and Cosmic Physics at the University of Heidelberg (IMPRS-HD).
\end{acknowledgements}

\bibliography{bibs}

\begin{thebibliography}{102}
\expandafter\ifx\csname natexlab\endcsname\relax\def\natexlab#1{#1}\fi

\bibitem[{{Abbott} {et~al.}(2016){Abbott}, {Abbott}, {Abbott}, {Abernathy},
  {Acernese}, {Ackley}, {Adams}, {Adams}, {Addesso}, {Adhikari}, \&
  et~al.}]{Abbott2016}
{Abbott}, B.~P., {Abbott}, R., {Abbott}, T.~D., {et~al.} 2016, Physical Review
  Letters, 116, 061102

\bibitem[{Abbott {et~al.}(2019)Abbott, Abbott, Abbott, Acernese, Ackley, Adams,
  Adams, Addesso, Adhikari, Adya, Affeldt, Agarwal, Agathos, Agatsuma,
  Aggarwal, Aguiar, Aiello, Ain, Ajith, Allen, Allen, Allocca, Aloy, Altin,
  Amato, Ananyeva, Anderson, Anderson, Angelova, Antier, Appert, Arai, Araya,
  Areeda, Ar\`ene, Arnaud, Arun, Ascenzi, Ashton, Ast, Aston, Astone, Atallah,
  Aubin, Aufmuth, Aulbert, AultONeal, Austin, Avila-Alvarez, Babak, Bacon,
  Badaracco, Bader, Bae, Baker, Baldaccini, Ballardin, Ballmer, Banagiri,
  Barayoga, Barclay, Barish, Barker, Barkett, Barnum, Barone, Barr, Barsotti,
  Barsuglia, Barta, Bartlett, Bartos, Bassiri, Basti, Batch, Bawaj, Bayley,
  Bazzan, B\'ecsy, Beer, Bejger, Belahcene, Bell, Beniwal, Bensch, Berger,
  Bergmann, Bernuzzi, Bero, Berry, Bersanetti, Bertolini, Betzwieser, Bhandare,
  Bilenko, Bilgili, Billingsley, Billman, Birch, Birney, Birnholtz, Biscans,
  Biscoveanu, Bisht, Bitossi, Bizouard, Blackburn, Blackman, Blair, Blair,
  Blair, Bloemen, Bock, Bode, Boer, Boetzel, Bogaert, Bohe, Bondu, Bonilla,
  Bonnand, Booker, Boom, Booth, Bork, Boschi, Bose, Bossie, Bossilkov, Bosveld,
  Bouffanais, Bozzi, Bradaschia, Brady, Bramley, Branchesi, Brau, Briant,
  Brighenti, Brillet, Brinkmann, Brisson, Brockill, Brooks, Brown, Brunett,
  Buchanan, Buikema, Bulik, Bulten, Buonanno, Buskulic, Buy, Byer, Cabero,
  Cadonati, Cagnoli, Cahillane, Bustillo, Callister, Calloni, Camp, Canepa,
  Canizares, Cannon, Cao, Cao, Capano, Capocasa, Carbognani, Caride, Carney,
  Carullo, Diaz, Casentini, Caudill, Cavagli\`a, Cavalier, Cavalieri, Cella,
  Cepeda, Cerd\'a-Dur\'an, Cerretani, Cesarini, Chaibi, Chamberlin, Chan, Chao,
  Charlton, Chase, Chassande-Mottin, Chatterjee, Chatziioannou, Cheeseboro,
  Chen, Chen, Chen, Cheng, Chia, Chincarini, Chiummo, Chmiel, Cho, Cho, Chow,
  Christensen, Chu, Chua, Chua, Chung, Chung, Ciani, Ciobanu, Ciolfi, Cipriano,
  Cirelli, Cirone, Clara, Clark, Clearwater, Cleva, Cocchieri, Coccia, Cohadon,
  Cohen, Colla, Collette, Collins, Cominsky, Constancio, Conti, Cooper, Corban,
  Corbitt, Cordero-Carri\'on, Corley, Cornish, Corsi, Cortese, Costa, Cotesta,
  Coughlin, Coughlin, Coulon, Countryman, Couvares, Covas, Cowan, Coward,
  Cowart, Coyne, Coyne, Creighton, Creighton, Cripe, Crowder, Cullen, Cumming,
  Cunningham, Cuoco, Canton, D\'alya, Danilishin, D'Antonio, Danzmann,
  Dasgupta, Costa, Dattilo, Dave, Davier, Davis, Daw, Day, DeBra, Deenadayalan,
  Degallaix, De~Laurentis, Del\'eglise, Del~Pozzo, Demos, Denker, Dent,
  De~Pietri, Derby, Dergachev, De~Rosa, De~Rossi, DeSalvo, de~Varona,
  Dhurandhar, D\'{\i}az, Dietrich, Di~Fiore, Di~Giovanni, Di~Girolamo,
  Di~Lieto, Ding, Di~Pace, Di~Palma, Di~Renzo, Dmitriev, Doctor, Dolique,
  Donovan, Dooley, Doravari, Dorrington, \'Alvarez, Downes, Drago,
  Dreissigacker, Driggers, Du, Dudi, Dupej, Dwyer, Easter, Edo, Edwards,
  Effler, Eggenstein, Ehrens, Eichholz, Eikenberry, Eisenmann, Eisenstein,
  Essick, Estelles, Estevez, Etienne, Etzel, Evans, Evans, Fafone, Fair,
  Fairhurst, Fan, Farinon, Farr, Farr, Fauchon-Jones, Favata, Fays, Fee,
  Fehrmann, Feicht, Fejer, Feng, Fernandez-Galiana, Ferrante, Ferreira,
  Ferrini, Fidecaro, Fiori, Fiorucci, Fishbach, Fisher, Fishner, Fitz-Axen,
  Flaminio, Fletcher, Fong, Font, Forsyth, Forsyth, Fournier, Frasca, Frasconi,
  Frei, Freise, Frey, Frey, Fritschel, Frolov, Fulda, Fyffe, Gabbard, Gadre,
  Gaebel, Gair, Gammaitoni, Ganija, Gaonkar, Garcia, Garc\'{\i}a-Quir\'os,
  Garufi, Gateley, Gaudio, Gaur, Gayathri, Gemme, Genin, Gennai, George,
  George, Gergely, Germain, Ghonge, Ghosh, Ghosh, Ghosh, Giacomazzo, Giaime,
  Giardina, Giazotto, Gill, Giordano, Glover, Goetz, Goetz, Goncharov,
  Gonz\'alez, Castro, Gopakumar, Gorodetsky, Gossan, Gosselin, Gouaty, Grado,
  Graef, Granata, Grant, Gras, Gray, Greco, Green, Green, Gretarsson, Groot,
  Grote, Grunewald, Gruning, Guidi, Gulati, Guo, Gupta, Gupta, Gushwa,
  Gustafson, Gustafson, Halim, Hall, Hall, Hamilton, Hamilton, Hammond, Haney,
  Hanke, Hanks, Hanna, Hannam, Hannuksela, Hanson, Hardwick, Harms, Harry,
  Harry, Hart, Haster, Haughian, Healy, Heidmann, Heintze, Heitmann, Hello,
  Hemming, Hendry, Heng, Hennig, Heptonstall, Hernandez, Heurs, Hild, Hinderer,
  Hoak, Hochheim, Hofman, Holland, Holt, Holz, Hopkins, Horst, Hough, Houston,
  Howell, Hreibi, Huerta, Huet, Hughey, Hulko, Husa, Huttner, Huynh-Dinh, Iess,
  Indik, Ingram, Inta, Intini, Isa, Isac, Isi, Iyer, Izumi, Jacqmin, Jani,
  Jaranowski, Johnson, Johnson, Jones, Jones, Jonker, Ju, Junker, Kalaghatgi,
  Kalogera, Kamai, Kandhasamy, Kang, Kanner, Kapadia, Karki, Karvinen,
  Kasprzack, Kastaun, Katolik, Katsanevas, Katsavounidis, Katzman, Kaufer,
  Kawabe, Keerthana, K\'ef\'elian, Keitel, Kemball, Kennedy, Key, Khalili,
  Khamesra, Khan, Khan, Khan, Khan, Khazanov, Kijbunchoo, Kim, Kim, Kim, Kim,
  Kim, Kim, King, King, Kinley-Hanlon, Kirchhoff, Kissel, Kleybolte, Klimenko,
  Knowles, Koch, Koehlenbeck, Koley, Kondrashov, Kontos, Korobko, Korth,
  Kowalska, Kozak, Kr\"amer, Kringel, Krishnan, Kr\'olak, Kuehn, Kumar, Kumar,
  Kumar, Kuo, Kutynia, Kwang, Lackey, Lai, Landry, Landry, Lang, Lange, Lantz,
  Lanza, Lartaux-Vollard, Lasky, Laxen, Lazzarini, Lazzaro, Leaci, Leavey, Lee,
  Lee, Lee, Lee, Lee, Lehmann, Lenon, Leonardi, Leroy, Letendre, Levin, Li, Li,
  Li, Linker, Littenberg, Liu, Liu, Lo, Lockerbie, London, Longo, Lorenzini,
  Loriette, Lormand, Losurdo, Lough, Lousto, Lovelace, L\"uck, Lumaca,
  Lundgren, Lynch, Ma, Macas, Macfoy, Machenschalk, MacInnis, Macleod,
  Hernandez, Maga\~na Sandoval, Zertuche, Magee, Majorana, Maksimovic, Man,
  Mandic, Mangano, Mansell, Manske, Mantovani, Marchesoni, Marion, M\'arka,
  M\'arka, Markakis, Markosyan, Markowitz, Maros, Marquina, Martelli,
  Martellini, Martin, Martin, Martynov, Mason, Massera, Masserot, Massinger,
  Masso-Reid, Mastrogiovanni, Matas, Matichard, Matone, Mavalvala, Mazumder,
  McCann, McCarthy, McClelland, McCormick, McCuller, McGuire, McIver, McManus,
  McRae, McWilliams, Meacher, Meadors, Mehmet, Meidam, Mejuto-Villa, Melatos,
  Mendell, Mendoza-Gandara, Mercer, Mereni, Merilh, Merzougui, Meshkov,
  Messenger, Messick, Metzdorff, Meyers, Miao, Michel, Middleton, Mikhailov,
  Milano, Miller, Miller, Miller, Miller, Millhouse, Mills, Milovich-Goff,
  Minazzoli, Minenkov, Ming, Mishra, Mitra, Mitrofanov, Mitselmakher,
  Mittleman, Moffa, Mogushi, Mohan, Mohapatra, Montani, Moore, Moraru, Moreno,
  Morisaki, Mours, Mow-Lowry, Mueller, Muir, Mukherjee, Mukherjee, Mukherjee,
  Mukund, Mullavey, Munch, Mu\~niz, Muratore, Murray, Nagar, Napier,
  Nardecchia, Naticchioni, Nayak, Neilson, Nelemans, Nelson, Nery, Neunzert,
  Nevin, Newport, Ng, Ng, Nguyen, Nguyen, Nichols, Nielsen, Nissanke, Nitz,
  Nocera, Nolting, North, Nuttall, Obergaulinger, Oberling, O'Brien, O'Dea,
  Ogin, Oh, Oh, Ohme, Ohta, Okada, Oliver, Oppermann, Oram, O'Reilly, Ormiston,
  Ortega, O'Shaughnessy, Ossokine, Ottaway, Overmier, Owen, Pace, Pagano, Page,
  Page, Pai, Pai, Palamos, Palashov, Palomba, Pal-Singh, Pan, Pan, Pang, Pang,
  Pankow, Pannarale, Pant, Paoletti, Paoli, Papa, Parida, Parker, Pascucci,
  Pasqualetti, Passaquieti, Passuello, Patil, Patricelli, Pearlstone, Pedersen,
  Pedraza, Pedurand, Pekowsky, Pele, Penn, Perez, Perreca, Perri, Pfeiffer,
  Phelps, Phukon, Piccinni, Pichot, Piergiovanni, Pierro, Pillant, Pinard,
  Pinto, Pirello, Pitkin, Poggiani, Popolizio, Porter, Possenti, Post, Powell,
  Prasad, Pratt, Pratten, Predoi, Prestegard, Principe, Privitera, Prodi,
  Prokhorov, Puncken, Punturo, Puppo, P\"urrer, Qi, Quetschke, Quintero,
  Quitzow-James, Raab, Rabeling, Radkins, Raffai, Raja, Rajan, Rajbhandari,
  Rakhmanov, Ramirez, Ramos-Buades, Rana, Rapagnani, Raymond, Razzano, Read,
  Regimbau, Rei, Reid, Reitze, Ren, Ricci, Ricker, Riemenschneider, Riles,
  Rizzo, Robertson, Robie, Robinet, Robson, Rocchi, Rolland, Rollins, Roma,
  Romano, Romel, Romie, Rosi\ifmmode~\acute{n}\else \'{n}\fi{}ska, Ross, Rowan,
  R\"udiger, Ruggi, Rutins, Ryan, Sachdev, Sadecki, Sakellariadou, Salconi,
  Saleem, Salemi, Samajdar, Sammut, Sampson, Sanchez, Sanchez, Sanchis-Gual,
  Sandberg, Sanders, Sarin, Sassolas, Sathyaprakash, Saulson, Sauter, Savage,
  Sawadsky, Schale, Scheel, Scheuer, Schmidt, Schnabel, Schofield, Sch\"onbeck,
  Schreiber, Schuette, Schulte, Schutz, Schwalbe, Scott, Scott, Seidel,
  Sellers, Sengupta, Sentenac, Sequino, Sergeev, Setyawati, Shaddock, Shaffer,
  Shah, Shahriar, Shaner, Shao, Shapiro, Shawhan, Shen, Shoemaker, Shoemaker,
  Siellez, Siemens, Sieniawska, Sigg, Silva, Singer, Singh, Singhal, Sintes,
  Slagmolen, Slaven-Blair, Smith, Smith, Smith, Somala, Son, Sorazu,
  Sorrentino, Souradeep, Spencer, Srivastava, Staats, Steinke, Steinlechner,
  Steinlechner, Steinmeyer, Steltner, Stevenson, Stocks, Stone, Stops, Strain,
  Stratta, Strigin, Strunk, Sturani, Stuver, Summerscales, Sun, Sunil, Suresh,
  Sutton, Swinkels, Szczepa\ifmmode~\acute{n}\else \'{n}\fi{}czyk, Tacca, Tait,
  Talbot, Talukder, Tanner, T\'apai, Taracchini, Tasson, Taylor, Taylor,
  Tewari, Theeg, Thies, Thomas, Thomas, Thomas, Thorne, Thrane, Tiwari, Tiwari,
  Tokmakov, Toland, Tonelli, Tornasi, Torres-Forn\'e, Torrie, T\"oyr\"a,
  Travasso, Traylor, Trinastic, Tringali, Trozzo, Tsang, Tse, Tso, Tsuna,
  Tsukada, Tuyenbayev, Ueno, Ugolini, Urban, Usman, Vahlbruch, Vajente, Valdes,
  van Bakel, van Beuzekom, van~den Brand, Van Den~Broeck, Vander-Hyde, van~der
  Schaaf, van Heijningen, van Veggel, Vardaro, Varma, Vass, Vas\'uth, Vecchio,
  Vedovato, Veitch, Veitch, Venkateswara, Venugopalan, Verkindt, Vetrano,
  Vicer\'e, Viets, Vinciguerra, Vine, Vinet, Vitale, Vo, Vocca, Vorvick,
  Vyatchanin, Wade, Wade, Wade, Walet, Walker, Wallace, Walsh, Wang, Wang,
  Wang, Wang, Wang, Ward, Warner, Was, Watchi, Weaver, Wei, Weinert, Weinstein,
  Weiss, Wellmann, Wen, Wessel, We\ss{}els, Westerweck, Wette, Whelan, Whiting,
  Whittle, Wilken, Williams, Williams, Williamson, Willis, Willke, Wimmer,
  Winkler, Wipf, Wittel, Woan, Woehler, Wofford, Wong, Worden, Wright, Wu,
  Wysocki, Xiao, Yam, Yamamoto, Yancey, Yang, Yap, Yazback, Yu, Yu, Yvert,
  Zadro\ifmmode~\dot{z}\else \.{z}\fi{}ny, Zanolin, Zelenova, Zendri, Zevin,
  Zhang, Zhang, Zhang, Zhang, Zhang, Zhao, Zhou, Zhou, Zhu, Zhu, Zimmerman,
  Zlochower, Zucker, \& Zweizig}]{Abbott2019}
Abbott, B.~P., Abbott, R., Abbott, T.~D., {et~al.} 2019, Phys. Rev. X, 9,
  011001

\bibitem[{{Abbott} {et~al.}(2017){Abbott}, {Abbott}, {Abbott}, {Acernese},
  {Ackley}, {Adams}, {Adams}, {Addesso}, {Adhikari}, {Adya}, \&
  et~al.}]{Abbott2017GW50MSUN}
{Abbott}, B.~P., {Abbott}, R., {Abbott}, T.~D., {et~al.} 2017, Physical Review
  Letters, 118, 221101

\bibitem[{{Andrews} \& {Kalogera}(2022)}]{Andrews2022}
{Andrews}, J.~J. \& {Kalogera}, V. 2022, \apj, 930, 159

\bibitem[{Askar {et~al.}(2024)Askar, Baldassare, \& Mezcua}]{Askar2023IMBH}
Askar, A., Baldassare, V.~F., \& Mezcua, M. 2024, Intermediate-Mass Black Holes
  in Star Clusters and Dwarf Galaxies

\bibitem[{{Bachelet} {et~al.}(2022){Bachelet}, {Zieli{\'n}ski}, {Gromadzki},
  {Gezer}, {Rybicki}, {Kruszy{\'n}ska}, {Ihanec}, {Wyrzykowski}, {Street},
  {Tsapras}, {Hundertmark}, {Cassan}, {Harbeck}, \& {Rabus}}]{2022Bachelet}
{Bachelet}, E., {Zieli{\'n}ski}, P., {Gromadzki}, M., {et~al.} 2022, \aap, 657,
  A17

\bibitem[{{Bahramian} {et~al.}(2017){Bahramian}, {Heinke}, {Tudor},
  {Miller-Jones}, {Bogdanov}, {Maccarone}, {Knigge}, {Sivakoff}, {Chomiuk},
  {Strader}, {Garcia}, \& {Kallman}}]{2017MNRAS.467.2199B}
{Bahramian}, A., {Heinke}, C.~O., {Tudor}, V., {et~al.} 2017, \mnras, 467, 2199

\bibitem[{{Bailer-Jones} {et~al.}(2021){Bailer-Jones}, {Rybizki}, {Fouesneau},
  {Demleitner}, \& {Andrae}}]{2021Bailer-Jones}
{Bailer-Jones}, C.~A.~L., {Rybizki}, J., {Fouesneau}, M., {Demleitner}, M., \&
  {Andrae}, R. 2021, \aj, 161, 147

\bibitem[{{Bailer-Jones} {et~al.}(2018){Bailer-Jones}, {Rybizki}, {Fouesneau},
  {Mantelet}, \& {Andrae}}]{Bailer-Jones2018}
{Bailer-Jones}, C.~A.~L., {Rybizki}, J., {Fouesneau}, M., {Mantelet}, G., \&
  {Andrae}, R. 2018, \aj, 156, 58

\bibitem[{{Batista} {et~al.}(2011){Batista}, {Gould}, {Dieters}, {Dong},
  {Bond}, {Beaulieu}, {Maoz}, {Monard}, {Christie}, {McCormick}, {Albrow},
  {Horne}, {Tsapras}, {Burgdorf}, {Calchi Novati}, {Skottfelt}, {Caldwell},
  {Koz{\l}owski}, {Kubas}, {Gaudi}, {Han}, {Bennett}, {An}, {MOA
  Collaboration}, {Abe}, {Botzler}, {Douchin}, {Freeman}, {Fukui}, {Furusawa},
  {Hearnshaw}, {Hosaka}, {Itow}, {Kamiya}, {Kilmartin}, {Korpela}, {Lin},
  {Ling}, {Makita}, {Masuda}, {Matsubara}, {Miyake}, {Muraki}, {Nagaya},
  {Nishimoto}, {Ohnishi}, {Okumura}, {Perrott}, {Rattenbury}, {Saito},
  {Sullivan}, {Sumi}, {Sweatman}, {Tristram}, {von Seggern}, {Yock}, {PLANET
  Collaboration}, {Brillant}, {Calitz}, {Cassan}, {Cole}, {Cook}, {Coutures},
  {Dominis Prester}, {Donatowicz}, {Greenhill}, {Hoffman}, {Jablonski}, {Kane},
  {Kains}, {Marquette}, {Martin}, {Martioli}, {Meintjes}, {Menzies},
  {Pedretti}, {Pollard}, {Sahu}, {Vinter}, {Wambsganss}, {Watson}, {Williams},
  {Zub}, {FUN Collaboration}, {Allen}, {Bolt}, {Bos}, {DePoy}, {Drummond},
  {Eastman}, {Gal-Yam}, {Gorbikov}, {Higgins}, {Janczak}, {Kaspi}, {Lee},
  {Mallia}, {Maury}, {Monard}, {Moorhouse}, {Morgan}, {Natusch}, {Ofek},
  {Park}, {Pogge}, {Polishook}, {Santallo}, {Shporer}, {Spector}, {Thornley},
  {Yee}, {MiNDSTEp Consortium}, {Bozza}, {Browne}, {Dominik}, {Dreizler},
  {Finet}, {Glitrup}, {Grundahl}, {Harps{\o}e}, {Hessman}, {Hinse},
  {Hundertmark}, {J{\o}rgensen}, {Liebig}, {Maier}, {Mancini}, {Mathiasen},
  {Rahvar}, {Ricci}, {Scarpetta}, {Southworth}, {Surdej}, {Zimmer}, {RoboNet
  Collaboration}, {Allan}, {Bramich}, {Snodgrass}, {Steele}, \&
  {Street}}]{2011Batista}
{Batista}, V., {Gould}, A., {Dieters}, S., {et~al.} 2011, \aap, 529, A102

\bibitem[{{Belczynski} {et~al.}(2020){Belczynski}, {Hirschi}, {Kaiser}, {Liu},
  {Casares}, {Lu}, {O'Shaughnessy}, {Heger}, {Justham}, \&
  {Soria}}]{Belczynski2020}
{Belczynski}, K., {Hirschi}, R., {Kaiser}, E.~A., {et~al.} 2020, \apj, 890, 113

\bibitem[{{Belokurov} \& {Evans}(2002)}]{BelokurovEvans2002}
{Belokurov}, V.~A. \& {Evans}, N.~W. 2002, \mnras, 331, 649

\bibitem[{{Blanco-Cuaresma}(2019)}]{BlancoCuaresma2019}
{Blanco-Cuaresma}, S. 2019, \mnras, 486, 2075

\bibitem[{{Blanco-Cuaresma} {et~al.}(2014){Blanco-Cuaresma}, {Soubiran},
  {Heiter}, \& {Jofr{\'e}}}]{BlancoCuaresma2014}
{Blanco-Cuaresma}, S., {Soubiran}, C., {Heiter}, U., \& {Jofr{\'e}}, P. 2014,
  \aap, 569, A111

\bibitem[{{Bressan} {et~al.}(2012){Bressan}, {Marigo}, {Girardi}, {Salasnich},
  {Dal Cero}, {Rubele}, \& {Nanni}}]{Bressan2012}
{Bressan}, A., {Marigo}, P., {Girardi}, L., {et~al.} 2012, \mnras, 427, 127

\bibitem[{{Buchner} {et~al.}(2014){Buchner}, {Georgakakis}, {Nandra}, {Hsu},
  {Rangel}, {Brightman}, {Merloni}, {Salvato}, {Donley}, \&
  {Kocevski}}]{2014BuchnerPyMultiNest}
{Buchner}, J., {Georgakakis}, A., {Nandra}, K., {et~al.} 2014, \aap, 564, A125

\bibitem[{{Casares} {et~al.}(1992){Casares}, {Charles}, \&
  {Naylor}}]{1992Natur.355..614C}
{Casares}, J., {Charles}, P.~A., \& {Naylor}, T. 1992, \nat, 355, 614

\bibitem[{{Cassan}(2023)}]{Cassan2023}
{Cassan}, A. 2023, \aap, 676, A110

\bibitem[{{Chakrabarti} {et~al.}(2023){Chakrabarti}, {Simon}, {Craig},
  {Reggiani}, {Brandt}, {Guhathakurta}, {Dalba}, {Kirby}, {Chang}, {Hey},
  {Savino}, {Geha}, \& {Thompson}}]{Chakrabarti2023}
{Chakrabarti}, S., {Simon}, J.~D., {Craig}, P.~A., {et~al.} 2023, \aj, 166, 6

\bibitem[{{Corral-Santana} {et~al.}(2016){Corral-Santana}, {Casares},
  {Mu{\~n}oz-Darias}, {Bauer}, {Mart{\'{\i}}nez-Pais}, \&
  {Russell}}]{Corral-Santana2016}
{Corral-Santana}, J.~M., {Casares}, J., {Mu{\~n}oz-Darias}, T., {et~al.} 2016,
  \aap, 587, A61

\bibitem[{{Dominik} \& {Sahu}(2000)}]{DominikSahu2000}
{Dominik}, M. \& {Sahu}, K.~C. 2000, The Astrophysical Journal, 534, 213

\bibitem[{{Dong} {et~al.}(2019){Dong}, {M{\'e}rand}, {Delplancke-Str{\"o}bele},
  {Gould}, \& {Zang}}]{Dong2019}
{Dong}, S., {M{\'e}rand}, A., {Delplancke-Str{\"o}bele}, F., {Gould}, A., \&
  {Zang}, W. 2019, The Messenger, 178, 45

\bibitem[{{Einstein}(1936)}]{Einstein1936}
{Einstein}, A. 1936, Science, 84, 506

\bibitem[{El-Badry {et~al.}(2023)El-Badry, Rix, Cendes, Rodriguez, Conroy,
  Quataert, Hawkins, Zari, Hobson, Breivik, Rau, Berger, Shahaf, Seeburger,
  Burdge, Latham, Buchhave, Bieryla, Bashi, Mazeh, \& Faigler}]{GaiaBH2}
El-Badry, K., Rix, H., Cendes, Y., {et~al.} 2023, Monthly Notices of the Royal
  Astronomical Society, 521, 4323, publisher Copyright: {\textcopyright} 2023
  The Author(s) Published by Oxford University Press on behalf of Royal
  Astronomical Society.

\bibitem[{El-Badry {et~al.}(2022)El-Badry, Rix, Quataert, Howard, Isaacson,
  Fuller, Hawkins, Breivik, Wong, Rodriguez, Conroy, Shahaf, Mazeh, Arenou,
  Burdge, Bashi, Faigler, Weisz, Seeburger, Almada Monter, \& Wojno}]{GaiaBH1}
El-Badry, K., Rix, H.-W., Quataert, E., {et~al.} 2022, Monthly Notices of the
  Royal Astronomical Society, 518, 1057

\bibitem[{{Feroz} {et~al.}(2009){Feroz}, {Hobson}, \&
  {Bridges}}]{2009FerozMultiNest}
{Feroz}, F., {Hobson}, M.~P., \& {Bridges}, M. 2009, \mnras, 398, 1601

\bibitem[{Foreman-Mackey(2016)}]{corner}
Foreman-Mackey, D. 2016, Journal of Open Source Software, 1, 24

\bibitem[{{Foreman-Mackey} {et~al.}(2013){Foreman-Mackey}, {Hogg}, {Lang}, \&
  {Goodman}}]{2013Emcee}
{Foreman-Mackey}, D., {Hogg}, D.~W., {Lang}, D., \& {Goodman}, J. 2013, \pasp,
  125, 306

\bibitem[{{Fukui} {et~al.}(2019){Fukui}, {Suzuki}, {Koshimoto}, {Bachelet},
  {Vanmunster}, {Storey}, {Maehara}, {Yanagisawa}, {Yamada}, {Yonehara},
  {Hirano}, {Bennett}, {Bozza}, {Mawet}, {Penny}, {Awiphan}, {Oksanen},
  {Heintz}, {Oberst}, {B{\'e}jar}, {Casasayas-Barris}, {Chen}, {Crouzet},
  {Hidalgo}, {Klagyivik}, {Murgas}, {Narita}, {Palle}, {Parviainen},
  {Watanabe}, {Kusakabe}, {Mori}, {Terada}, {de Leon}, {Hernandez}, {Luque},
  {Monelli}, {Monta{\~n}es-Rodriguez}, {Prieto-Arranz}, {Murata}, {Shugarov},
  {Kubota}, {Otsuki}, {Shionoya}, {Nishiumi}, {Nishide}, {Fukagawa}, {Onodera},
  {Villanueva}, {Street}, {Tsapras}, {Hundertmark}, {Kuzuhara}, {Fujita},
  {Beichman}, {Beaulieu}, {Alonso}, {Reichart}, {Kawai}, \&
  {Tamura}}]{2019Fukui}
{Fukui}, A., {Suzuki}, D., {Koshimoto}, N., {et~al.} 2019, \aj, 158, 206

\bibitem[{{Gaia Collaboration}(2020)}]{GaiaEDR3Cat}
{Gaia Collaboration}. 2020, VizieR Online Data Catalog, I/350

\bibitem[{{Gaia Collaboration} {et~al.}(2018){Gaia Collaboration}, {Brown},
  {Vallenari}, {Prusti}, {de Bruijne}, {Babusiaux}, {Bailer-Jones}, {Biermann},
  {Evans}, {Eyer}, \& et~al.}]{GaiaDR2}
{Gaia Collaboration}, {Brown}, A.~G.~A., {Vallenari}, A., {et~al.} 2018, \aap,
  616, A1

\bibitem[{{Gaia Collaboration} {et~al.}(2024){Gaia Collaboration}, {Panuzzo,
  P.}, {Mazeh, T.}, {Arenou, F.}, {Holl, B.}, {Caffau, E.}, {Jorissen, A.},
  {Babusiaux, C.}, {Gavras, P.}, {Sahlmann, J.}, {Bastian, U.}, {Wyrzykowski,
  Ł.}, {Eyer, L.}, {Leclerc, N.}, {Bauchet, N.}, {Bombrun, A.}, {Mowlavi, N.},
  {Seabroke, G. M.}, {Teyssier, D.}, {Balbinot, E.}, {Helmi, A.}, {Brown, A. G.
  A.}, {Vallenari, A.}, {Prusti, T.}, {de Bruijne, J. H. J.}, {Barbier, A.},
  {Biermann, M.}, {Creevey, O. L.}, {Ducourant, C.}, {Evans, D. W.}, {Guerra,
  R.}, {Hutton, A.}, {Jordi, C.}, {Klioner, S. A.}, {Lammers, U.}, {Lindegren,
  L.}, {Luri, X.}, {Mignard, F.}, {Nicolas, C.}, {Randich, S.}, {Sartoretti,
  P.}, {Smiljanic, R.}, {Tanga, P.}, {Walton, N. A.}, {Aerts, C.},
  {Bailer-Jones, C. A. L.}, {Cropper, M.}, {Drimmel, R.}, {Jansen, F.}, {Katz,
  D.}, {Lattanzi, M. G.}, {Soubiran, C.}, {Thévenin, F.}, {van Leeuwen, F.},
  {Andrae, R.}, {Audard, M.}, {Bakker, J.}, {Blomme, R.}, {Castañeda, J.}, {De
  Angeli, F.}, {Fabricius, C.}, {Fouesneau, M.}, {Frémat, Y.}, {Galluccio,
  L.}, {Guerrier, A.}, {Heiter, U.}, {Masana, E.}, {Messineo, R.},
  {Nienartowicz, K.}, {Pailler, F.}, {Riclet, F.}, {Roux, W.}, {Sordo, R.},
  {Gracia-Abril, G.}, {Portell, J.}, {Altmann, M.}, {Benson, K.}, {Berthier,
  J.}, {Burgess, P. W.}, {Busonero, D.}, {Busso, G.}, {Cacciari, C.},
  {Cánovas, H.}, {Carrasco, J. M.}, {Carry, B.}, {Cellino, A.}, {Cheek, N.},
  {Clementini, G.}, {Damerdji, Y.}, {Davidson, M.}, {de Teodoro, P.},
  {Delchambre, L.}, {Dell’Oro, A.}, {Fraile Garcia, E.}, {Garabato, D.},
  {García-Lario, P.}, {Haigron, R.}, {Hambly, N. C.}, {Harrison, D. L.},
  {Hatzidimitriou, D.}, {Hernández, J.}, {Hestroffer, D.}, {Hodgkin, S. T.},
  {Jamal, S.}, {Jevardat de Fombelle, G.}, {Jordan, S.}, {Krone-Martins, A.},
  {Lanzafame, A. C.}, {Löffler, W.}, {Lorca, A.}, {Marchal, O.}, {Marrese, P.
  M.}, {Moitinho, A.}, {Muinonen, K.}, {Nuñez Campos, M.}, {Oreshina-Slezak,
  I.}, {Osborne, P.}, {Pancino, E.}, {Pauwels, T.}, {Recio-Blanco, A.},
  {Riello, M.}, {Rimoldini, L.}, {Robin, A. C.}, {Roegiers, T.}, {Sarro, L.
  M.}, {Schultheis, M.}, {Smith, M.}, {Sozzetti, A.}, {Utrilla, E.}, {van
  Leeuwen, M.}, {Weingrill, K.}, {Abbas, U.}, {Ábrahám, P.}, {Abreu Aramburu,
  A.}, {Ahmed, S.}, {Altavilla, G.}, {Álvarez, M. A.}, {Anders, F.},
  {Anderson, R. I.}, {Anglada Varela, E.}, {Antoja, T.}, {Baig, S.}, {Baines,
  D.}, {Baker, S. G.}, {Balaguer-Núñez, L.}, {Balog, Z.}, {Barache, C.},
  {Barros, M.}, {Barstow, M. A.}, {Bartolomé, S.}, {Bashi, D.}, {Bassilana,
  J.-L.}, {Baudeau, N.}, {Becciani, U.}, {Bedin, L. R.}, {Bellas-Velidis, I.},
  {Bellazzini, M.}, {Beordo, W.}, {Bernet, M.}, {Bertolotto, C.}, {Bertone,
  S.}, {Bianchi, L.}, {Binnenfeld, A.}, {Blanco-Cuaresma, S.}, {Bland-Hawthorn,
  J.}, {Blazere, A.}, {Boch, T.}, {Bossini, D.}, {Bouquillon, S.}, {Bragaglia,
  A.}, {Braine, J.}, {Bratsolis, E.}, {Breedt, E.}, {Bressan, A.}, {Brouillet,
  N.}, {Brugaletta, E.}, {Bucciarelli, B.}, {Butkevich, A. G.}, {Buzzi, R.},
  {Camut, A.}, {Cancelliere, R.}, {Cantat-Gaudin, T.}, {Capilla Guilarte, D.},
  {Carballo, R.}, {Carlucci, T.}, {Carnerero, M. I.}, {Carretero, J.}, {Carton,
  S.}, {Casamiquela, L.}, {Casey, A.}, {Castellani, M.}, {Castro-Ginard, A.},
  {Ceraj, L.}, {Cesare, V.}, {Charlot, P.}, {Chaudet, C.}, {Chemin, L.},
  {Chiavassa, A.}, {Chornay, N.}, {Chosson, D.}, {Cooper, W. J.}, {Cornez, T.},
  {Cowell, S.}, {Crosta, M.}, {Crowley, C.}, {Cruz Reyes, M.}, {Dafonte, C.},
  {Dal Ponte, M.}, {David, M.}, {de Laverny, P.}, {De Luise, F.}, {De March,
  R.}, {de Torres, A.}, {del Peloso, E. F.}, {Delbo, M.}, {Delgado, A.},
  {Delisle, J.-B.}, {Demouchy, C.}, {Denis, E.}, {Dharmawardena, T. E.}, {Di
  Giacomo, F.}, {Diener, C.}, {Distefano, E.}, {Dolding, C.}, {Dsilva, K.},
  {Enke, H.}, {Fabre, C.}, {Fabrizio, M.}, {Faigler, S.}, {Fatović, M.},
  {Fedorets, G.}, {Fernández-Hernández, J.}, {Fernique, P.}, {Figueras, F.},
  {Fouron, C.}, {Fragkoudi, F.}, {Gai, M.}, {Galinier, M.}, {Garcia-Serrano,
  A.}, {García-Torres, M.}, {Garofalo, A.}, {Gerlach, E.}, {Geyer, R.},
  {Giacobbe, P.}, {Gilmore, G.}, {Girona, S.}, {Giuffrida, G.}, {Gomboc, A.},
  {Gomez, A.}, {González-Santamaría, I.}, {Gosset, E.}, {Granvik, M.},
  {Gregori Barrera, V.}, {Gutiérrez-Sánchez, R.}, {Haywood, M.}, {Helmer,
  A.}, {Hidalgo, S. L.}, {Hilger, T.}, {Hobbs, D.}, {Hottier, C.}, {Huckle, H.
  E.}, {Jiménez-Arranz, Ó.}, {Juaristi Campillo, J.}, {Kaczmarek, Z.},
  {Kervella, P.}, {Khanna, S.}, {Kontizas, M.}, {Kordopatis, G.}, {Korn, A.
  J.}, {Kóspál, Á}, {Kostrzewa-Rutkowska, Z.}, {Kruszyńska, K.}, {Kun, M.},
  {Lambert, S.}, {Lanza, A. F.}, {Lebreton, Y.}, {Lebzelter, T.}, {Leccia, S.},
  {Lecoutre, G.}, {Liao, S.}, {Liberato, L.}, {Licata, E.}, {Livanou, E.},
  {Lobel, A.}, {López-Miralles, J.}, {Loup, C.}, {Madarász, M.}, {Mahy, L.},
  {Mann, R. G.}, {Manteiga, M.}, {Marcellino, C. P.}, {Marchant, J. M.},
  {Marconi, M.}, {Marín Pina, D.}, {Marinoni, S.}, {Marshall, D. J.}, {Martín
  Lozano, J.}, {Martin Polo, L.}, {Martín-Fleitas, J. M.}, {Marton, G.},
  {Mascarenhas, D.}, {Masip, A.}, {Mastrobuono-Battisti, A.}, {McMillan, P.
  J.}, {Meichsner, J.}, {Merc, J.}, {Messina, S.}, {Millar, N. R.}, {Mints,
  A.}, {Mohamed, D.}, {Molina, D.}, {Molinaro, R.}, {Molnár, L.}, {Monguió,
  M.}, {Montegriffo, P.}, {Monti, L.}, {Mora, A.}, {Morbidelli, R.}, {Morris,
  D.}, {Mudimadugula, R.}, {Muraveva, T.}, {Musella, I.}, {Nagy, Z.},
  {Nardetto, N.}, {Navarrete, C.}, {Oh, S.}, {Ordenovic, C.}, {Orenstein, O.},
  {Pagani, C.}, {Pagano, I.}, {Palaversa, L.}, {Palicio, P. A.},
  {Pallas-Quintela, L.}, {Pawlak, M.}, {Penttilä, A.}, {Pesciullesi, P.},
  {Pinamonti, M.}, {Plachy, E.}, {Planquart, L.}, {Plum, G.}, {Poggio, E.},
  {Pourbaix, D.}, {Price-Whelan, A. M.}, {Pulone, L.}, {Rabin, V.}, {Rainer,
  M.}, {Raiteri, C. M.}, {Ramos, P.}, {Ramos-Lerate, M.}, {Ratajczak, M.}, {Re
  Fiorentin, P.}, {Regibo, S.}, {Reylé, C.}, {Ripepi, V.}, {Riva, A.}, {Rix,
  H.-W.}, {Rixon, G.}, {Robert, G.}, {Robichon, N.}, {Robin, C.},
  {Romero-Gómez, M.}, {Rowell, N.}, {Ruz Mieres, D.}, {Rybicki, K. A.},
  {Sadowski, G.}, {Sagristà Sellés, A.}, {Sanna, N.}, {Santoveña, R.},
  {Sarasso, M.}, {Sarmiento, M. H.}, {Sarrate Riera, C.}, {Sciacca, E.},
  {Ségransan, D.}, {Semczuk, M.}, {Shahaf, S.}, {Siebert, A.}, {Slezak, E.},
  {Smart, R. L.}, {Snaith, O. N.}, {Solano, E.}, {Solitro, F.}, {Souami, D.},
  {Souchay, J.}, {Spitoni, E.}, {Spoto, F.}, {Squillante, L. A.}, {Steele, I.
  A.}, {Steidelmüller, H.}, {Surdej, J.}, {Szabados, L.}, {Taris, F.},
  {Taylor, M. B.}, {Teixeira, R.}, {Tepper-Garcia, T.}, {Thuillot, W.},
  {Tolomei, L.}, {Tonello, N.}, {Torra, F.}, {Torralba Elipe, G.}, {Trabucchi,
  M.}, {Trentin, E.}, {Tsantaki, M.}, {Turon, C.}, {Ulla, A.}, {Unger, N.},
  {Valtchanov, I.}, {Vanel, O.}, {Vecchiato, A.}, {Vicente, D.}, {Villar, E.},
  {Weiler, M.}, {Zhao, H.}, {Zorec, J.}, {Zucker, S.}, {Župić, A.}, \&
  {Zwitter, T.}}]{GaiaBH3}
{Gaia Collaboration}, {Panuzzo, P.}, {Mazeh, T.}, {et~al.} 2024, \aap, 686, L2

\bibitem[{{Gaia Collaboration} {et~al.}(2016){Gaia Collaboration}, {Prusti},
  {de Bruijne}, {Brown}, {Vallenari}, {Babusiaux}, {Bailer-Jones}, {Bastian},
  {Biermann}, {Evans}, \& et~al.}]{Gaia}
{Gaia Collaboration}, {Prusti}, T., {de Bruijne}, J.~H.~J., {et~al.} 2016,
  \aap, 595, A1

\bibitem[{{Gaia Collaboration} {et~al.}(2023){Gaia Collaboration}, {Vallenari},
  {Brown}, {Prusti}, {de Bruijne}, {Arenou}, {Babusiaux}, {Biermann},
  {Creevey}, {Ducourant}, {Evans}, {Eyer}, {Guerra}, {Hutton}, {Jordi},
  {Klioner}, {Lammers}, {Lindegren}, {Luri}, {Mignard}, {Panem}, {Pourbaix},
  {Randich}, {Sartoretti}, {Soubiran}, {Tanga}, {Walton}, {Bailer-Jones},
  {Bastian}, {Drimmel}, {Jansen}, {Katz}, {Lattanzi}, {van Leeuwen}, {Bakker},
  {Cacciari}, {Casta{\~n}eda}, {De Angeli}, {Fabricius}, {Fouesneau},
  {Fr{\'e}mat}, {Galluccio}, {Guerrier}, {Heiter}, {Masana}, {Messineo},
  {Mowlavi}, {Nicolas}, {Nienartowicz}, {Pailler}, {Panuzzo}, {Riclet}, {Roux},
  {Seabroke}, {Sordo}, {Th{\'e}venin}, {Gracia-Abril}, {Portell}, {Teyssier},
  {Altmann}, {Andrae}, {Audard}, {Bellas-Velidis}, {Benson}, {Berthier},
  {Blomme}, {Burgess}, {Busonero}, {Busso}, {C{\'a}novas}, {Carry}, {Cellino},
  {Cheek}, {Clementini}, {Damerdji}, {Davidson}, {de Teodoro}, {Nu{\~n}ez
  Campos}, {Delchambre}, {Dell'Oro}, {Esquej}, {Fern{\'a}ndez-Hern{\'a}ndez},
  {Fraile}, {Garabato}, {Garc{\'\i}a-Lario}, {Gosset}, {Haigron}, {Halbwachs},
  {Hambly}, {Harrison}, {Hern{\'a}ndez}, {Hestroffer}, {Hodgkin}, {Holl},
  {Jan{\ss}en}, {Jevardat de Fombelle}, {Jordan}, {Krone-Martins}, {Lanzafame},
  {L{\"o}ffler}, {Marchal}, {Marrese}, {Moitinho}, {Muinonen}, {Osborne},
  {Pancino}, {Pauwels}, {Recio-Blanco}, {Reyl{\'e}}, {Riello}, {Rimoldini},
  {Roegiers}, {Rybizki}, {Sarro}, {Siopis}, {Smith}, {Sozzetti}, {Utrilla},
  {van Leeuwen}, {Abbas}, {{\'A}brah{\'a}m}, {Abreu Aramburu}, {Aerts},
  {Aguado}, {Ajaj}, {Aldea-Montero}, {Altavilla}, {{\'A}lvarez}, {Alves},
  {Anders}, {Anderson}, {Anglada Varela}, {Antoja}, {Baines}, {Baker},
  {Balaguer-N{\'u}{\~n}ez}, {Balbinot}, {Balog}, {Barache}, {Barbato},
  {Barros}, {Barstow}, {Bartolom{\'e}}, {Bassilana}, {Bauchet}, {Becciani},
  {Bellazzini}, {Berihuete}, {Bernet}, {Bertone}, {Bianchi}, {Binnenfeld},
  {Blanco-Cuaresma}, {Blazere}, {Boch}, {Bombrun}, {Bossini}, {Bouquillon},
  {Bragaglia}, {Bramante}, {Breedt}, {Bressan}, {Brouillet}, {Brugaletta},
  {Bucciarelli}, {Burlacu}, {Butkevich}, {Buzzi}, {Caffau}, {Cancelliere},
  {Cantat-Gaudin}, {Carballo}, {Carlucci}, {Carnerero}, {Carrasco},
  {Casamiquela}, {Castellani}, {Castro-Ginard}, {Chaoul}, {Charlot}, {Chemin},
  {Chiaramida}, {Chiavassa}, {Chornay}, {Comoretto}, {Contursi}, {Cooper},
  {Cornez}, {Cowell}, {Crifo}, {Cropper}, {Crosta}, {Crowley}, {Dafonte},
  {Dapergolas}, {David}, {David}, {de Laverny}, {De Luise}, {De March}, {De
  Ridder}, {de Souza}, {de Torres}, {del Peloso}, {del Pozo}, {Delbo},
  {Delgado}, {Delisle}, {Demouchy}, {Dharmawardena}, {Di Matteo}, {Diakite},
  {Diener}, {Distefano}, {Dolding}, {Edvardsson}, {Enke}, {Fabre}, {Fabrizio},
  {Faigler}, {Fedorets}, {Fernique}, {Fienga}, {Figueras}, {Fournier},
  {Fouron}, {Fragkoudi}, {Gai}, {Garcia-Gutierrez}, {Garcia-Reinaldos},
  {Garc{\'\i}a-Torres}, {Garofalo}, {Gavel}, {Gavras}, {Gerlach}, {Geyer},
  {Giacobbe}, {Gilmore}, {Girona}, {Giuffrida}, {Gomel}, {Gomez},
  {Gonz{\'a}lez-N{\'u}{\~n}ez}, {Gonz{\'a}lez-Santamar{\'\i}a},
  {Gonz{\'a}lez-Vidal}, {Granvik}, {Guillout}, {Guiraud},
  {Guti{\'e}rrez-S{\'a}nchez}, {Guy}, {Hatzidimitriou}, {Hauser}, {Haywood},
  {Helmer}, {Helmi}, {Sarmiento}, {Hidalgo}, {Hilger}, {H{\l}adczuk}, {Hobbs},
  {Holland}, {Huckle}, {Jardine}, {Jasniewicz}, {Jean-Antoine Piccolo},
  {Jim{\'e}nez-Arranz}, {Jorissen}, {Juaristi Campillo}, {Julbe}, {Karbevska},
  {Kervella}, {Khanna}, {Kontizas}, {Kordopatis}, {Korn}, {K{\'o}sp{\'a}l},
  {Kostrzewa-Rutkowska}, {Kruszy{\'n}ska}, {Kun}, {Laizeau}, {Lambert},
  {Lanza}, {Lasne}, {Le Campion}, {Lebreton}, {Lebzelter}, {Leccia}, {Leclerc},
  {Lecoeur-Taibi}, {Liao}, {Licata}, {Lindstr{\o}m}, {Lister}, {Livanou},
  {Lobel}, {Lorca}, {Loup}, {Madrero Pardo}, {Magdaleno Romeo}, {Managau},
  {Mann}, {Manteiga}, {Marchant}, {Marconi}, {Marcos}, {Marcos Santos},
  {Mar{\'\i}n Pina}, {Marinoni}, {Marocco}, {Marshall}, {Martin Polo},
  {Mart{\'\i}n-Fleitas}, {Marton}, {Mary}, {Masip}, {Massari},
  {Mastrobuono-Battisti}, {Mazeh}, {McMillan}, {Messina}, {Michalik}, {Millar},
  {Mints}, {Molina}, {Molinaro}, {Moln{\'a}r}, {Monari}, {Mongui{\'o}},
  {Montegriffo}, {Montero}, {Mor}, {Mora}, {Morbidelli}, {Morel}, {Morris},
  {Muraveva}, {Murphy}, {Musella}, {Nagy}, {Noval}, {Oca{\~n}a}, {Ogden},
  {Ordenovic}, {Osinde}, {Pagani}, {Pagano}, {Palaversa}, {Palicio},
  {Pallas-Quintela}, {Panahi}, {Payne-Wardenaar}, {Pe{\~n}alosa Esteller},
  {Penttil{\"a}}, {Pichon}, {Piersimoni}, {Pineau}, {Plachy}, {Plum}, {Poggio},
  {Pr{\v{s}}a}, {Pulone}, {Racero}, {Ragaini}, {Rainer}, {Raiteri}, {Rambaux},
  {Ramos}, {Ramos-Lerate}, {Re Fiorentin}, {Regibo}, {Richards}, {Rios Diaz},
  {Ripepi}, {Riva}, {Rix}, {Rixon}, {Robichon}, {Robin}, {Robin}, {Roelens},
  {Rogues}, {Rohrbasser}, {Romero-G{\'o}mez}, {Rowell}, {Royer}, {Ruz Mieres},
  {Rybicki}, {Sadowski}, {S{\'a}ez N{\'u}{\~n}ez}, {Sagrist{\`a} Sell{\'e}s},
  {Sahlmann}, {Salguero}, {Samaras}, {Sanchez Gimenez}, {Sanna},
  {Santove{\~n}a}, {Sarasso}, {Schultheis}, {Sciacca}, {Segol}, {Segovia},
  {S{\'e}gransan}, {Semeux}, {Shahaf}, {Siddiqui}, {Siebert}, {Siltala},
  {Silvelo}, {Slezak}, {Slezak}, {Smart}, {Snaith}, {Solano}, {Solitro},
  {Souami}, {Souchay}, {Spagna}, {Spina}, {Spoto}, {Steele},
  {Steidelm{\"u}ller}, {Stephenson}, {S{\"u}veges}, {Surdej}, {Szabados},
  {Szegedi-Elek}, {Taris}, {Taylor}, {Teixeira}, {Tolomei}, {Tonello}, {Torra},
  {Torra}, {Torralba Elipe}, {Trabucchi}, {Tsounis}, {Turon}, {Ulla}, {Unger},
  {Vaillant}, {van Dillen}, {van Reeven}, {Vanel}, {Vecchiato}, {Viala},
  {Vicente}, {Voutsinas}, {Weiler}, {Wevers}, {Wyrzykowski}, {Yoldas}, {Yvard},
  {Zhao}, {Zorec}, {Zucker}, \& {Zwitter}}]{GaiaDR3}
{Gaia Collaboration}, {Vallenari}, A., {Brown}, A.~G.~A., {et~al.} 2023, \aap,
  674, A1

\bibitem[{{Giersz} {et~al.}(2022){Giersz}, {Askar}, {Klencki}, \&
  {Morawski}}]{Giersz2022}
{Giersz}, M., {Askar}, A., {Klencki}, J., \& {Morawski}, J. 2022, in Fifteenth
  Marcel Grossmann Meeting on General Relativity, ed. E.~S. {Battistelli},
  R.~T. {Jantzen}, \& R.~{Ruffini}, 1618--1627

\bibitem[{{Gobat}(2022)}]{assymetric_sampling}
{Gobat}, C. 2022, {Asymmetric Uncertainty}: Handling nonstandard numerical
  uncertainties, Astrophysics Source Code Library

\bibitem[{{Gomel} {et~al.}(2023){Gomel}, {Mazeh}, {Faigler}, {Bashi}, {Eyer},
  {Rimoldini}, {Audard}, {Mowlavi}, {Holl}, {Jevardat}, {Nienartowicz},
  {Lecoeur}, \&
  {Wyrzykowski}}]{Gomel2023-Gaia-DR3-ellipsoidal-compact-binaries}
{Gomel}, R., {Mazeh}, T., {Faigler}, S., {et~al.} 2023, \aap, 674, A19

\bibitem[{{Gould}(2000)}]{Gould2000b}
{Gould}, A. 2000, \apj, 542, 785

\bibitem[{{Gould}(2004)}]{Gould2004}
{Gould}, A. 2004, \apj, 606, 319

\bibitem[{{Grevesse} {et~al.}(2007){Grevesse}, {Asplund}, \&
  {Sauval}}]{Grevesse2007}
{Grevesse}, N., {Asplund}, M., \& {Sauval}, A.~J. 2007, \ssr, 130, 105

\bibitem[{{Gustafsson} {et~al.}(2008){Gustafsson}, {Edvardsson}, {Eriksson},
  {J{\o}rgensen}, {Nordlund}, \& {Plez}}]{Gustafsson2008}
{Gustafsson}, B., {Edvardsson}, B., {Eriksson}, K., {et~al.} 2008, \aap, 486,
  951

\bibitem[{{Han} \& {Gould}(2003)}]{2003HanGould}
{Han}, C. \& {Gould}, A. 2003, \apj, 592, 172

\bibitem[{{Heiter} {et~al.}(2021){Heiter}, {Lind}, {Bergemann}, {Asplund},
  {Mikolaitis}, {Barklem}, {Masseron}, {de Laverny}, {Magrini}, {Edvardsson},
  {J{\"o}nsson}, {Pickering}, {Ryde}, {Bayo Ar{\'a}n}, {Bensby}, {Casey},
  {Feltzing}, {Jofr{\'e}}, {Korn}, {Pancino}, {Damiani}, {Lanzafame}, {Lardo},
  {Monaco}, {Morbidelli}, {Smiljanic}, {Worley}, {Zaggia}, {Randich}, \&
  {Gilmore}}]{Heiter2021}
{Heiter}, U., {Lind}, K., {Bergemann}, M., {et~al.} 2021, \aap, 645, A106

\bibitem[{{Higson} {et~al.}(2019){Higson}, {Handley}, {Hobson}, \&
  {Lasenby}}]{2019Higson}
{Higson}, E., {Handley}, W., {Hobson}, M., \& {Lasenby}, A. 2019, Statistics
  and Computing, 29, 891

\bibitem[{{Hodgkin} {et~al.}(2021){Hodgkin}, {Harrison}, {Breedt}, {Wevers},
  {Rixon}, {Delgado}, {Yoldas}, {Kostrzewa-Rutkowska}, {Wyrzykowski}, {van
  Leeuwen}, {Blagorodnova}, {Campbell}, {Eappachen}, {Fraser}, {Ihanec},
  {Koposov}, {Kruszy{\'n}ska}, {Marton}, {Rybicki}, {Brown}, {Burgess},
  {Busso}, {Cowell}, {De Angeli}, {Diener}, {Evans}, {Gilmore}, {Holland},
  {Jonker}, {van Leeuwen}, {Mignard}, {Osborne}, {Portell}, {Prusti},
  {Richards}, {Riello}, {Seabroke}, {Walton}, {{\'A}brah{\'a}m}, {Altavilla},
  {Baker}, {Bastian}, {O'Brien}, {de Bruijne}, {Butterley}, {Carrasco},
  {Casta{\~n}eda}, {Clark}, {Clementini}, {Copperwheat}, {Cropper},
  {Damljanovic}, {Davidson}, {Davis}, {Dennefeld}, {Dhillon}, {Dolding},
  {Dominik}, {Esquej}, {Eyer}, {Fabricius}, {Fridman}, {Froebrich}, {Garralda},
  {Gomboc}, {Gonz{\'a}lez-Vidal}, {Guerra}, {Hambly}, {Hardy}, {Holl},
  {Hourihane}, {Japelj}, {Kann}, {Kiss}, {Knigge}, {Kolb}, {Komossa},
  {K{\'o}sp{\'a}l}, {Kov{\'a}cs}, {Kun}, {Leto}, {Lewis}, {Littlefair},
  {Mahabal}, {Mundell}, {Nagy}, {Padeletti}, {Palaversa}, {Pigulski},
  {Pretorius}, {van Reeven}, {Ribeiro}, {Roelens}, {Rowell}, {Schartel},
  {Scholz}, {Schwope}, {Sip{\H{o}}cz}, {Smartt}, {Smith}, {Serraller},
  {Steeghs}, {Sullivan}, {Szabados}, {Szegedi-Elek}, {Tisserand}, {Tomasella},
  {van Velzen}, {Whitelock}, {Wilson}, \& {Young}}]{Hodgkin_2021}
{Hodgkin}, S.~T., {Harrison}, D.~L., {Breedt}, E., {et~al.} 2021, \aap, 652,
  A76

\bibitem[{{Hodgkin} {et~al.}(2013){Hodgkin}, {Wyrzykowski}, {Blagorodnova}, \&
  {Koposov}}]{Hodgkin_2013}
{Hodgkin}, S.~T., {Wyrzykowski}, L., {Blagorodnova}, N., \& {Koposov}, S. 2013,
  Philosophical Transactions of the Royal Society of London Series A, 371,
  20120239

\bibitem[{{Horvath} {et~al.}(2023){Horvath}, {Bernardo}, {Bachega}, {de
  S{\'a}}, {Rocha}, \& {Moraes}}]{Horvath2023}
{Horvath}, J.~E., {Bernardo}, A. L.~C., {Bachega}, R. R.~A., {et~al.} 2023,
  Astronomische Nachrichten, 344, e20220106

\bibitem[{{Jab{\l}o{\'n}ska} {et~al.}(2022){Jab{\l}o{\'n}ska}, {Wyrzykowski},
  {Rybicki}, {Kruszy{\'n}ska}, {Kaczmarek}, \& {Penoyre}}]{Jablonska2022}
{Jab{\l}o{\'n}ska}, M., {Wyrzykowski}, {\L}., {Rybicki}, K.~A., {et~al.} 2022,
  \aap, 666, L16

\bibitem[{{Jordi} {et~al.}(2010){Jordi}, {Gebran}, {Carrasco}, {de Bruijne},
  {Voss}, {Fabricius}, {Knude}, {Vallenari}, {Kohley}, \&
  {Mora}}]{2010JordiGaiaPhot}
{Jordi}, C., {Gebran}, M., {Carrasco}, J.~M., {et~al.} 2010, \aap, 523, A48

\bibitem[{{Kaczmarek} {et~al.}(2022){Kaczmarek}, {McGill}, {Evans}, {Smith},
  {Wyrzykowski}, {Howil}, \& {Jab{\l}o{\'n}ska}}]{Kaczmarek2022}
{Kaczmarek}, Z., {McGill}, P., {Evans}, N.~W., {et~al.} 2022, \mnras, 514, 4845

\bibitem[{{Kroupa}(2001)}]{2001Kroupa}
{Kroupa}, P. 2001, \mnras, 322, 231

\bibitem[{{Kruszynska} {et~al.}(2018){Kruszynska}, {Gromadzki}, {Wyrzykowski},
  {Zielinski}, {Zielinski}, {Rybicki}, {Mroz}, {Hamanowicz}, \&
  {Hodgkin}}]{Kruszynska2018ATel}
{Kruszynska}, K., {Gromadzki}, M., {Wyrzykowski}, L., {et~al.} 2018, The
  Astronomer's Telegram, 11634, 1

\bibitem[{{Kruszy{\'n}ska} {et~al.}(2024){Kruszy{\'n}ska}, {Wyrzykowski},
  {Rybicki}, {Howil}, {Jab{\l}o{\'n}ska}, {Kaczmarek}, {Ihanec},
  {Maskoli{\={u}}nas}, {Bronikowski}, \& {Pylypenko}}]{Kruszynska2024}
{Kruszy{\'n}ska}, K., {Wyrzykowski}, {\L}., {Rybicki}, K.~A., {et~al.} 2024,
  arXiv e-prints, arXiv:2401.13759

\bibitem[{{Kruszy{\'n}ska} {et~al.}(2022){Kruszy{\'n}ska}, {Wyrzykowski},
  {Rybicki}, {Maskoli{\={u}}nas}, {Bachelet}, {Rattenbury}, {Mr{\'o}z},
  {Zieli{\'n}ski}, {Howil}, {Kaczmarek}, {Hodgkin}, {Ihanec}, {Gezer},
  {Gromadzki}, {Miko{\l}ajczyk}, {Stankevi{\v{c}}i{\={u}}t{\.{e}}},
  {{\v{C}}epas}, {Pak{\v{s}}tien{\.{e}}}, {{\v{S}}i{\v{s}}kauskait{\.{e}}},
  {Zdanavi{\v{c}}ius}, {Bozza}, {Dominik}, {Figuera Jaimes}, {Fukui},
  {Hundertmark}, {Narita}, {Street}, {Tsapras}, {Bronikowski},
  {Jab{\l}o{\'n}ska}, {Jab{\l}onowska}, \&
  {Zi{\'o}{\l}kowska}}]{Kruszynska2022}
{Kruszy{\'n}ska}, K., {Wyrzykowski}, {\L}., {Rybicki}, K.~A., {et~al.} 2022,
  \aap, 662, A59

\bibitem[{{Kurucz}(1993)}]{Kurucz1993}
{Kurucz}, R.~L. 1993, {SYNTHE spectrum synthesis programs and line data}

\bibitem[{{Lam} \& {Lu}(2023)}]{Lam2023}
{Lam}, C.~Y. \& {Lu}, J.~R. 2023, \apj, 955, 116

\bibitem[{Lam {et~al.}(2022)Lam, Lu, Udalski, Bond, Bennett, Skowron, Mróz,
  Poleski, Sumi, Szymański, Kozłowski, Pietrukowicz, Soszyński, Ulaczyk,
  Łukasz Wyrzykowski, Miyazaki, Suzuki, Koshimoto, Rattenbury, Hosek, Abe,
  Barry, Bhattacharya, Fukui, Fujii, Hirao, Itow, Kirikawa, Kondo, Matsubara,
  Matsumoto, Muraki, Olmschenk, Ranc, Okamura, Satoh, Silva, Toda, Tristram,
  Vandorou, Yama, Abrams, Agarwal, Rose, \& Terry}]{Lam_2022}
Lam, C.~Y., Lu, J.~R., Udalski, A., {et~al.} 2022, The Astrophysical Journal
  Letters, 933, L23

\bibitem[{{Leveque} {et~al.}(2023){Leveque}, {Giersz}, {Askar}, {Arca-Sedda},
  \& {Olejak}}]{Leveque2023}
{Leveque}, A., {Giersz}, M., {Askar}, A., {Arca-Sedda}, M., \& {Olejak}, A.
  2023, \mnras, 520, 2593

\bibitem[{{Lindegren} {et~al.}(2018){Lindegren}, {Hern{\'a}ndez}, {Bombrun},
  {Klioner}, {Bastian}, {Ramos-Lerate}, {de Torres}, {Steidelm{\"u}ller},
  {Stephenson}, {Hobbs}, {Lammers}, {Biermann}, {Geyer}, {Hilger}, {Michalik},
  {Stampa}, {McMillan}, {Casta{\~n}eda}, {Clotet}, {Comoretto}, {Davidson},
  {Fabricius}, {Gracia}, {Hambly}, {Hutton}, {Mora}, {Portell}, {van Leeuwen},
  {Abbas}, {Abreu}, {Altmann}, {Andrei}, {Anglada}, {Balaguer-N{\'u}{\~n}ez},
  {Barache}, {Becciani}, {Bertone}, {Bianchi}, {Bouquillon}, {Bourda},
  {Br{\"u}semeister}, {Bucciarelli}, {Busonero}, {Buzzi}, {Cancelliere},
  {Carlucci}, {Charlot}, {Cheek}, {Crosta}, {Crowley}, {de Bruijne}, {de
  Felice}, {Drimmel}, {Esquej}, {Fienga}, {Fraile}, {Gai}, {Garralda},
  {Gonz{\'a}lez-Vidal}, {Guerra}, {Hauser}, {Hofmann}, {Holl}, {Jordan},
  {Lattanzi}, {Lenhardt}, {Liao}, {Licata}, {Lister}, {L{\"o}ffler},
  {Marchant}, {Martin-Fleitas}, {Messineo}, {Mignard}, {Morbidelli}, {Poggio},
  {Riva}, {Rowell}, {Salguero}, {Sarasso}, {Sciacca}, {Siddiqui}, {Smart},
  {Spagna}, {Steele}, {Taris}, {Torra}, {van Elteren}, {van Reeven}, \&
  {Vecchiato}}]{Lindegren2018}
{Lindegren}, L., {Hern{\'a}ndez}, J., {Bombrun}, A., {et~al.} 2018, \aap, 616,
  A2

\bibitem[{{Mao} {et~al.}(2002){Mao}, {Smith}, {Wo{\'z}niak}, {Udalski},
  {Szyma{\'n}ski}, {Kubiak}, {Pietrzy{\'n}ski}, {Soszy{\'n}ski}, \&
  {{\.Z}ebru{\'n}}}]{Mao2002}
{Mao}, S., {Smith}, M.~C., {Wo{\'z}niak}, P., {et~al.} 2002, \mnras, 329, 349

\bibitem[{{Maskoli{\={u}}nas} {et~al.}(2023){Maskoli{\={u}}nas}, {Wyrzykowski},
  {Howil}, {Rybicki}, {Zieli{\'n}ski}, {Kaczmarek}, {Kruszy{\'n}ska},
  {Jab{\l}o{\'n}ska}, {Zdanavi{\v{c}}ius}, {Pak{\v{s}}tien{\.{e}}},
  {{\v{C}}epas}, {Miko{\l}ajczyk}, {Janulis}, {Gromadzki}, {Ihanec},
  {Adomavi{\v{c}}ien{\.{e}}}, {{\v{S}}i{\v{s}}kauskait{\.{e}}}, {Bronikowski},
  {Sivak}, {Stankevi{\v{c}}i{\={u}}t{\.{e}}}, {Sitek}, {Ratajczak},
  {Pylypenko}, {Gezer}, {Awiphan}, {Bachelet}, {B{\k{a}}kowska}, {Boyle},
  {Bozza}, {Brincat}, {Burgaz}, {Butterley}, {Carrasco}, {Cassan}, {Cusano},
  {Damljanovic}, {Davidson}, {Dhillon}, {Dominik}, {Dubois}, {Esenoglu},
  {Figuera Jaimes}, {Fukui}, {Galdies}, {Garofalo}, {Godunova}, {G{\"u}ver},
  {Heidt}, {Hundertmark}, {Izviekova}, {Joachimczyk}, {Kami{\'n}ska},
  {Kami{\'n}ski}, {Kaptan}, {Kvernadze}, {Kvaratskhelia}, {Littlefair},
  {Michniewicz}, {Nakhatutai}, {Og{\l}oza}, {Ohsawa}, {Olszewska},
  {Poli{\'n}ska}, {Popowicz}, {Qvam}, {Radziwonowicz}, {Reichart},
  {S{\l}owikowska}, {Simon}, {Sonbas}, {Stojanovic}, {Tsapras}, {Vanaverbeke},
  {Wambsganss}, {Wilson}, {{\.Z}ejmo}, \& {Zola}}]{gaia19dke}
{Maskoli{\={u}}nas}, M., {Wyrzykowski}, {\L}., {Howil}, K., {et~al.} 2023,
  arXiv e-prints, arXiv:2309.03324

\bibitem[{{McGill} {et~al.}(2023){McGill}, {Anderson}, {Casertano}, {Sahu},
  {Bergeron}, {Blouin}, {Dufour}, {Smith}, {Evans}, {Belokurov}, {Smart},
  {Bellini}, {Calamida}, {Dominik}, {Kains}, {Kl{\"u}ter}, {Nielsen}, \&
  {Wambsganss}}]{McGill2023}
{McGill}, P., {Anderson}, J., {Casertano}, S., {et~al.} 2023, \mnras, 520, 259

\bibitem[{{Montegriffo} {et~al.}(2022){Montegriffo}, {De Angeli}, {Andrae},
  {Riello}, {Pancino}, {Sanna}, {Bellazzini}, {Evans}, {Carrasco}, {Sordo},
  {Busso}, {Cacciari}, {Jordi}, {van Leeuwen}, {Vallenari}, {Altavilla},
  {Barstow}, {Brown}, {Burgess}, {Castellani}, {Cowell}, {Davidson}, {De
  Luise}, {Delchambre}, {Diener}, {Fabricius}, {Fremat}, {Fouesneau},
  {Gilmore}, {Giuffrida}, {Hambly}, {Harrison}, {Hidalgo}, {Hodgkin},
  {Holland}, {Marinoni}, {Osborne}, {Pagani}, {Palaversa}, {Piersimoni},
  {Pulone}, {Ragaini}, {Rainer}, {Richards}, {Rowell}, {Ruz-Mieres}, {Sarro},
  {Walton}, \& {Yoldas}}]{Montegriffo2022}
{Montegriffo}, P., {De Angeli}, F., {Andrae}, R., {et~al.} 2022, arXiv
  e-prints, arXiv:2206.06205

\bibitem[{{Mr{\'o}z} {et~al.}(2019){Mr{\'o}z}, {Udalski}, {Skowron}, {Skowron},
  {Soszy{\'n}ski}, {Pietrukowicz}, {Szyma{\'n}ski}, {Poleski}, {Koz{\l}owski},
  \& {Ulaczyk}}]{2019Mroz}
{Mr{\'o}z}, P., {Udalski}, A., {Skowron}, D.~M., {et~al.} 2019, \apjl, 870, L10

\bibitem[{{Mr{\'o}z} {et~al.}(2020){Mr{\'o}z}, {Udalski}, {Szyma{\'n}ski},
  {Soszy{\'n}ski}, {Pietrukowicz}, {Koz{\l}owski}, {Skowron}, {Poleski},
  {Ulaczyk}, {Gromadzki}, {Rybicki}, {Iwanek}, \& {Wrona}}]{Mroz2020}
{Mr{\'o}z}, P., {Udalski}, A., {Szyma{\'n}ski}, M.~K., {et~al.} 2020, \apjs,
  249, 16

\bibitem[{{Mr{\'o}z} {et~al.}(2021){Mr{\'o}z}, {Udalski}, {Wyrzykowski},
  {Skowron}, {Poleski}, {Szymanski}, {Soszynski}, \&
  {Ulaczyk}}]{2021MrozBHOGLE}
{Mr{\'o}z}, P., {Udalski}, A., {Wyrzykowski}, L., {et~al.} 2021, arXiv
  e-prints, arXiv:2107.13697

\bibitem[{{Mr{\'o}z} \& {Wyrzykowski}(2021)}]{MrozWyrzykowski2021}
{Mr{\'o}z}, P. \& {Wyrzykowski}, {\L}. 2021, \actaa, 71, 89

\bibitem[{Mróz {et~al.}(2022)Mróz, Udalski, \& Gould}]{Mroz_2022}
Mróz, P., Udalski, A., \& Gould, A. 2022, The Astrophysical Journal Letters,
  937, L24

\bibitem[{{Paczynski}(1986)}]{Paczynski1986}
{Paczynski}, B. 1986, \apj, 304, 1

\bibitem[{{Pecaut} \& {Mamajek}(2013)}]{PecautMamajek2013}
{Pecaut}, M.~J. \& {Mamajek}, E.~E. 2013, \apjs, 208, 9

\bibitem[{{Pejcha} \& {Thompson}(2015)}]{Pejcha2015}
{Pejcha}, O. \& {Thompson}, T.~A. 2015, \apj, 801, 90

\bibitem[{{Poleski}(2013)}]{2013Poleski}
{Poleski}, R. 2013, arXiv e-prints, arXiv:1306.2945

\bibitem[{{Poleski} \& {Yee}(2019)}]{MulensModel}
{Poleski}, R. \& {Yee}, J.~C. 2019, Astronomy and Computing, 26, 35

\bibitem[{{Reid} {et~al.}(2009){Reid}, {Menten}, {Zheng}, {Brunthaler},
  {Moscadelli}, {Xu}, {Zhang}, {Sato}, {Honma}, {Hirota}, {Hachisuka}, {Choi},
  {Moellenbrock}, \& {Bartkiewicz}}]{2009Reid}
{Reid}, M.~J., {Menten}, K.~M., {Zheng}, X.~W., {et~al.} 2009, \apj, 700, 137

\bibitem[{{Rybicki} {et~al.}(2018){Rybicki}, {Wyrzykowski}, {Klencki}, {de
  Bruijne}, {Belczy{\'n}ski}, \& {Chru{\'s}li{\'n}ska}}]{Rybicki2018}
{Rybicki}, K.~A., {Wyrzykowski}, {\L}., {Klencki}, J., {et~al.} 2018, \mnras,
  476, 2013

\bibitem[{{Sahu} {et~al.}(2017){Sahu}, {Anderson}, {Casertano}, {Bond},
  {Bergeron}, {Nelan}, {Pueyo}, {Brown}, {Bellini}, {Levay}, {Sokol},
  {Dominik}, {Calamida}, {Kains}, \& {Livio}}]{SahuWD2017}
{Sahu}, K.~C., {Anderson}, J., {Casertano}, S., {et~al.} 2017, Science, 356,
  1046

\bibitem[{Sahu {et~al.}(2022)Sahu, Anderson, Casertano, Bond, Udalski, Dominik,
  Calamida, Bellini, Brown, Rejkuba, Bajaj, Kains, Ferguson, Fryer, Yock,
  Mróz, Kozłowski, Pietrukowicz, Poleski, Skowron, Soszyński, Szymański,
  Ulaczyk, Łukasz Wyrzykowski, Collaboration), Barry, Bennett, Bond, Hirao,
  Silva, Kondo, Koshimoto, Ranc, Rattenbury, Sumi, Suzuki, Tristram, Vandorou,
  Collaboration), Beaulieu, Marquette, Cole, Fouqué, Hill, Dieters, Coutures,
  Dominis-Prester, Bennett, Bachelet, Menzies, Albrow, Pollard, Collaboration),
  Gould, Yee, Allen, Almeida, Christie, Drummond, Gal-Yam, Gorbikov, Jablonski,
  Lee, Maoz, Manulis, McCormick, Natusch, Pogge, Shvartzvald, Collaboration),
  Jørgensen, Alsubai, Andersen, Bozza, Novati, Burgdorf, Hinse, Hundertmark,
  Husser, Kerins, Longa-Peña, Mancini, Penny, Rahvar, Ricci, Sajadian,
  Skottfelt, Snodgrass, Southworth, Tregloan-Reed, Wambsganss, Wertz,
  Consortium), Tsapras, Street, Bramich, Horne, Steele, \&
  Collaboration)}]{Sahu_2022}
Sahu, K.~C., Anderson, J., Casertano, S., {et~al.} 2022, The Astrophysical
  Journal, 933, 83

\bibitem[{{Schlafly} \& {Finkbeiner}(2011{\natexlab{a}})}]{Schlafly2011}
{Schlafly}, E.~F. \& {Finkbeiner}, D.~P. 2011{\natexlab{a}}, \apj, 737, 103

\bibitem[{{Schlafly} \& {Finkbeiner}(2011{\natexlab{b}})}]{2011SchlaflyRedd}
{Schlafly}, E.~F. \& {Finkbeiner}, D.~P. 2011{\natexlab{b}}, \apj, 737, 103

\bibitem[{{Schlegel} {et~al.}(1998){Schlegel}, {Finkbeiner}, \&
  {Davis}}]{Schlegel98}
{Schlegel}, D.~J., {Finkbeiner}, D.~P., \& {Davis}, M. 1998, \apj, 500, 525

\bibitem[{{Sch{\"o}nrich} {et~al.}(2010){Sch{\"o}nrich}, {Binney}, \&
  {Dehnen}}]{2010Schoenrich}
{Sch{\"o}nrich}, R., {Binney}, J., \& {Dehnen}, W. 2010, \mnras, 403, 1829

\bibitem[{{Shahaf} {et~al.}(2023){Shahaf}, {Bashi}, {Mazeh}, {Faigler},
  {Arenou}, {El-Badry}, \& {Rix}}]{Shahaf2023}
{Shahaf}, S., {Bashi}, D., {Mazeh}, T., {et~al.} 2023, \mnras, 518, 2991

\bibitem[{{Skilling}(2004)}]{2004Skilling}
{Skilling}, J. 2004, in American Institute of Physics Conference Series, Vol.
  735, Bayesian Inference and Maximum Entropy Methods in Science and
  Engineering: 24th International Workshop on Bayesian Inference and Maximum
  Entropy Methods in Science and Engineering, ed. R.~{Fischer}, R.~{Preuss}, \&
  U.~V. {Toussaint}, 395--405

\bibitem[{{Skowron} {et~al.}(2011{\natexlab{a}}){Skowron}, {Udalski}, {Gould},
  {Dong}, {Monard}, {Han}, {Nelson}, {McCormick}, {Moorhouse}, {Thornley},
  {Maury}, {Bramich}, {Greenhill}, {Koz{\l}owski}, {Bond}, {Poleski},
  {Wyrzykowski}, {Ulaczyk}, {Kubiak}, {Szyma{\'n}ski}, {Pietrzy{\'n}ski},
  {Soszy{\'n}ski}, {OGLE Collaboration}, {Gaudi}, {Yee}, {Hung}, {Pogge},
  {DePoy}, {Lee}, {Park}, {Allen}, {Mallia}, {Drummond}, {Bolt},
  {{\ensuremath{\mu}}FUN Collaboration}, {Allan}, {Browne}, {Clay}, {Dominik},
  {Fraser}, {Horne}, {Kains}, {Mottram}, {Snodgrass}, {Steele}, {Street},
  {Tsapras}, {RoboNet Collaboration}, {Abe}, {Bennett}, {Botzler}, {Douchin},
  {Freeman}, {Fukui}, {Furusawa}, {Hayashi}, {Hearnshaw}, {Hosaka}, {Itow},
  {Kamiya}, {Kilmartin}, {Korpela}, {Lin}, {Ling}, {Makita}, {Masuda},
  {Matsubara}, {Muraki}, {Nagayama}, {Miyake}, {Nishimoto}, {Ohnishi},
  {Perrott}, {Rattenbury}, {Saito}, {Skuljan}, {Sullivan}, {Sumi}, {Suzuki},
  {Sweatman}, {Tristram}, {Wada}, {Yock}, {MOA Collaboration}, {Beaulieu},
  {Fouqu{\'e}}, {Albrow}, {Batista}, {Brillant}, {Caldwell}, {Cassan}, {Cole},
  {Cook}, {Coutures}, {Dieters}, {Dominis Prester}, {Donatowicz}, {Kane},
  {Kubas}, {Marquette}, {Martin}, {Menzies}, {Sahu}, {Wambsganss}, {Williams},
  {Zub}, \& {PLANET Collaboration}}]{2011Skowron}
{Skowron}, J., {Udalski}, A., {Gould}, A., {et~al.} 2011{\natexlab{a}}, \apj,
  738, 87

\bibitem[{{Skowron} {et~al.}(2011{\natexlab{b}}){Skowron}, {Udalski}, {Gould},
  {Dong}, {Monard}, {Han}, {Nelson}, {McCormick}, {Moorhouse}, {Thornley},
  {Maury}, {Bramich}, {Greenhill}, {Koz{\l}owski}, {Bond}, {Poleski},
  {Wyrzykowski}, {Ulaczyk}, {Kubiak}, {Szyma{\'n}ski}, {Pietrzy{\'n}ski},
  {Soszy{\'n}ski}, {OGLE Collaboration}, {Gaudi}, {Yee}, {Hung}, {Pogge},
  {DePoy}, {Lee}, {Park}, {Allen}, {Mallia}, {Drummond}, {Bolt}, {{$\mu$}FUN
  Collaboration}, {Allan}, {Browne}, {Clay}, {Dominik}, {Fraser}, {Horne},
  {Kains}, {Mottram}, {Snodgrass}, {Steele}, {Street}, {Tsapras}, {RoboNet
  Collaboration}, {Abe}, {Bennett}, {Botzler}, {Douchin}, {Freeman}, {Fukui},
  {Furusawa}, {Hayashi}, {Hearnshaw}, {Hosaka}, {Itow}, {Kamiya}, {Kilmartin},
  {Korpela}, {Lin}, {Ling}, {Makita}, {Masuda}, {Matsubara}, {Muraki},
  {Nagayama}, {Miyake}, {Nishimoto}, {Ohnishi}, {Perrott}, {Rattenbury},
  {Saito}, {Skuljan}, {Sullivan}, {Sumi}, {Suzuki}, {Sweatman}, {Tristram},
  {Wada}, {Yock}, {MOA Collaboration}, {Beaulieu}, {Fouqu{\'e}}, {Albrow},
  {Batista}, {Brillant}, {Caldwell}, {Cassan}, {Cole}, {Cook}, {Coutures},
  {Dieters}, {Dominis Prester}, {Donatowicz}, {Kane}, {Kubas}, {Marquette},
  {Martin}, {Menzies}, {Sahu}, {Wambsganss}, {Williams}, {Zub}, \& {PLANET
  Collaboration}}]{Skowron2011}
{Skowron}, J., {Udalski}, A., {Gould}, A., {et~al.} 2011{\natexlab{b}}, \apj,
  738, 87

\bibitem[{{Speagle}(2020)}]{2020Speagle}
{Speagle}, J.~S. 2020, \mnras, 493, 3132

\bibitem[{{Stanek} {et~al.}(1994){Stanek}, {Mateo}, {Udalski}, {Szymanski},
  {Kaluzny}, \& {Kubiak}}]{1994Stanek}
{Stanek}, K.~Z., {Mateo}, M., {Udalski}, A., {et~al.} 1994, \apjl, 429, L73

\bibitem[{{Sweeney} {et~al.}(2024){Sweeney}, {Tuthill}, {Krone-Martins},
  {M{\'e}rand}, {Scalzo}, \& {Martinod}}]{Sweeney2024}
{Sweeney}, D., {Tuthill}, P., {Krone-Martins}, A., {et~al.} 2024, arXiv
  e-prints, arXiv:2403.14612

\bibitem[{{Sweeney} {et~al.}(2022){Sweeney}, {Tuthill}, {Sharma}, \&
  {Hirai}}]{Sweeney2022}
{Sweeney}, D., {Tuthill}, P., {Sharma}, S., \& {Hirai}, R. 2022, \mnras, 516,
  4971

\bibitem[{{Tanikawa} {et~al.}(2023){Tanikawa}, {Hattori}, {Kawanaka},
  {Kinugawa}, {Shikauchi}, \& {Tsuna}}]{Ataru2022}
{Tanikawa}, A., {Hattori}, K., {Kawanaka}, N., {et~al.} 2023, \apj, 946, 79

\bibitem[{{Udalski} {et~al.}(2015){Udalski}, {Szyma{\'n}ski}, \&
  {Szyma{\'n}ski}}]{Udalski2015}
{Udalski}, A., {Szyma{\'n}ski}, M.~K., \& {Szyma{\'n}ski}, G. 2015, \actaa, 65,
  1

\bibitem[{{Vernet} {et~al.}(2011){Vernet}, {Dekker}, {D'Odorico}, {Kaper},
  {Kjaergaard}, {Hammer}, {Randich}, {Zerbi}, {Groot}, {Hjorth}, {Guinouard},
  {Navarro}, {Adolfse}, {Albers}, {Amans}, {Andersen}, {Andersen}, {Binetruy},
  {Bristow}, {Castillo}, {Chemla}, {Christensen}, {Conconi}, {Conzelmann},
  {Dam}, {de Caprio}, {de Ugarte Postigo}, {Delabre}, {di Marcantonio},
  {Downing}, {Elswijk}, {Finger}, {Fischer}, {Flores}, {Fran{\c{c}}ois},
  {Goldoni}, {Guglielmi}, {Haigron}, {Hanenburg}, {Hendriks}, {Horrobin},
  {Horville}, {Jessen}, {Kerber}, {Kern}, {Kiekebusch}, {Kleszcz}, {Klougart},
  {Kragt}, {Larsen}, {Lizon}, {Lucuix}, {Mainieri}, {Manuputy}, {Martayan},
  {Mason}, {Mazzoleni}, {Michaelsen}, {Modigliani}, {Moehler}, {M{\o}ller},
  {Norup S{\o}rensen}, {N{\o}rregaard}, {P{\'e}roux}, {Patat}, {Pena}, {Pragt},
  {Reinero}, {Rigal}, {Riva}, {Roelfsema}, {Royer}, {Sacco}, {Santin},
  {Schoenmaker}, {Spano}, {Sweers}, {Ter Horst}, {Tintori}, {Tromp}, {van
  Dael}, {van der Vliet}, {Venema}, {Vidali}, {Vinther}, {Vola}, {Winters},
  {Wistisen}, {Wulterkens}, \& {Zacchei}}]{Vernet2011}
{Vernet}, J., {Dekker}, H., {D'Odorico}, S., {et~al.} 2011, \aap, 536, A105

\bibitem[{{Vigna-G{\'o}mez} \& {Ramirez-Ruiz}(2023)}]{Vigna-Gomez2023}
{Vigna-G{\'o}mez}, A. \& {Ramirez-Ruiz}, E. 2023, \apjl, 946, L2

\bibitem[{{Wang} \& {Chen}(2019{\natexlab{a}})}]{Wang2019}
{Wang}, S. \& {Chen}, X. 2019{\natexlab{a}}, \apj, 877, 116

\bibitem[{{Wang} \& {Chen}(2019{\natexlab{b}})}]{2019WangChenExt}
{Wang}, S. \& {Chen}, X. 2019{\natexlab{b}}, \apj, 877, 116

\bibitem[{{Wiktorowicz} {et~al.}(2019){Wiktorowicz}, {Wyrzykowski},
  {Chruslinska}, {Klencki}, {Rybicki}, \& {Belczynski}}]{Wiktorowicz2019}
{Wiktorowicz}, G., {Wyrzykowski}, {\L}., {Chruslinska}, M., {et~al.} 2019,
  \apj, 885, 1

\bibitem[{{Wyrzykowski} \& {Hodgkin}(2012)}]{Wyrzykowski2012}
{Wyrzykowski}, {\L}. \& {Hodgkin}, S. 2012, in IAU Symposium, Vol. 285, IAU
  Symposium, ed. E.~{Griffin}, R.~{Hanisch}, \& R.~{Seaman}, 425--428

\bibitem[{{Wyrzykowski} {et~al.}(2016){Wyrzykowski}, {Kostrzewa-Rutkowska},
  {Skowron}, {Rybicki}, {Mr{\'o}z}, {Koz{\l}owski}, {Udalski}, {Szyma{\'n}ski},
  {Pietrzy{\'n}ski}, {Soszy{\'n}ski}, {Ulaczyk}, {Pietrukowicz}, {Poleski},
  {Pawlak}, {I{\l}kiewicz}, \& {Rattenbury}}]{Wyrzykowski2016}
{Wyrzykowski}, {\L}., {Kostrzewa-Rutkowska}, Z., {Skowron}, J., {et~al.} 2016,
  \mnras, 458, 3012

\bibitem[{{Wyrzykowski} {et~al.}(2023){Wyrzykowski}, {Kruszy{\'n}ska},
  {Rybicki}, {Holl}, {Lec{\oe}ur-Ta{\"\i}bi}, {Mowlavi}, {Nienartowicz},
  {Jevardat de Fombelle}, {Rimoldini}, {Audard}, {Garcia-Lario}, {Gavras},
  {Evans}, {Hodgkin}, \& {Eyer}}]{Wyrzykowski2023}
{Wyrzykowski}, {\L}., {Kruszy{\'n}ska}, K., {Rybicki}, K.~A., {et~al.} 2023,
  \aap, 674, A23

\bibitem[{{Wyrzykowski} \& {Mandel}(2020)}]{Wyrzykowski2020}
{Wyrzykowski}, {\L}. \& {Mandel}, I. 2020, \aap, 636, A20

\bibitem[{{Zieli{\'n}ski} {et~al.}(2020){Zieli{\'n}ski}, {Wyrzykowski},
  {Miko{\l}ajczyk}, {Rybicki}, \& {Ko{\l}aczkowski}}]{Zielinski2020}
{Zieli{\'n}ski}, P., {Wyrzykowski}, {\l}., {Miko{\l}ajczyk}, P., {Rybicki}, K.,
  \& {Ko{\l}aczkowski}, Z. 2020, in XXXIX Polish Astronomical Society Meeting,
  ed. K.~{Ma{\l}ek}, M.~{Poli{\'n}ska}, A.~{Majczyna}, G.~{Stachowski},
  R.~{Poleski}, {\L}.~{Wyrzykowski}, \& A.~{R{\'o}{\.z}a{\'n}ska}, Vol.~10,
  190--193

\bibitem[{{Zieli{\'n}ski} {et~al.}(2019){Zieli{\'n}ski}, {Wyrzykowski},
  {Rybicki}, {Ko{\l}aczkowski}, {Bru{\'s}}, \&
  {Miko{\l}ajczyk}}]{Zielinski2019}
{Zieli{\'n}ski}, P., {Wyrzykowski}, {\L}., {Rybicki}, K., {et~al.} 2019,
  Contributions of the Astronomical Observatory Skalnate Pleso, 49, 125

\end{thebibliography}
\bibliographystyle{aa}

\begin{appendix}
\section{\textit{DarkLensCode}}\label{app:dlc}
The \DLC (DLC) is an open-source software \footnote{\url{https://github.com/BHTOM-Team/DarkLensCode/}} used to find the posterior distribution of lens distance and lens mass, using probability density of the photometric model parameters, and the Galactic model. The final estimates are the median values of the mass and distance posteriors obtained.  

This method was first introduced in \cite{2011Skowron}, then explored in \cite{Wyrzykowski2016}, and then refined by \cite{Wyrzykowski2020},  \cite{MrozWyrzykowski2021}, \cite{Kaczmarek2022} and \cite{gaia19dke}.
In \cite{Kruszynska2022}, we expanded the expected proper motion calculation method to provide estimates for events outside the Galactic bulge, following \cite{2009Reid}.
In \cite{Kruszynska2024}, we added an option for broken power-law mass functions, allowing more flexibility in the mass prior.

\begin{figure*}
\centering
\includegraphics[width=\hsize]{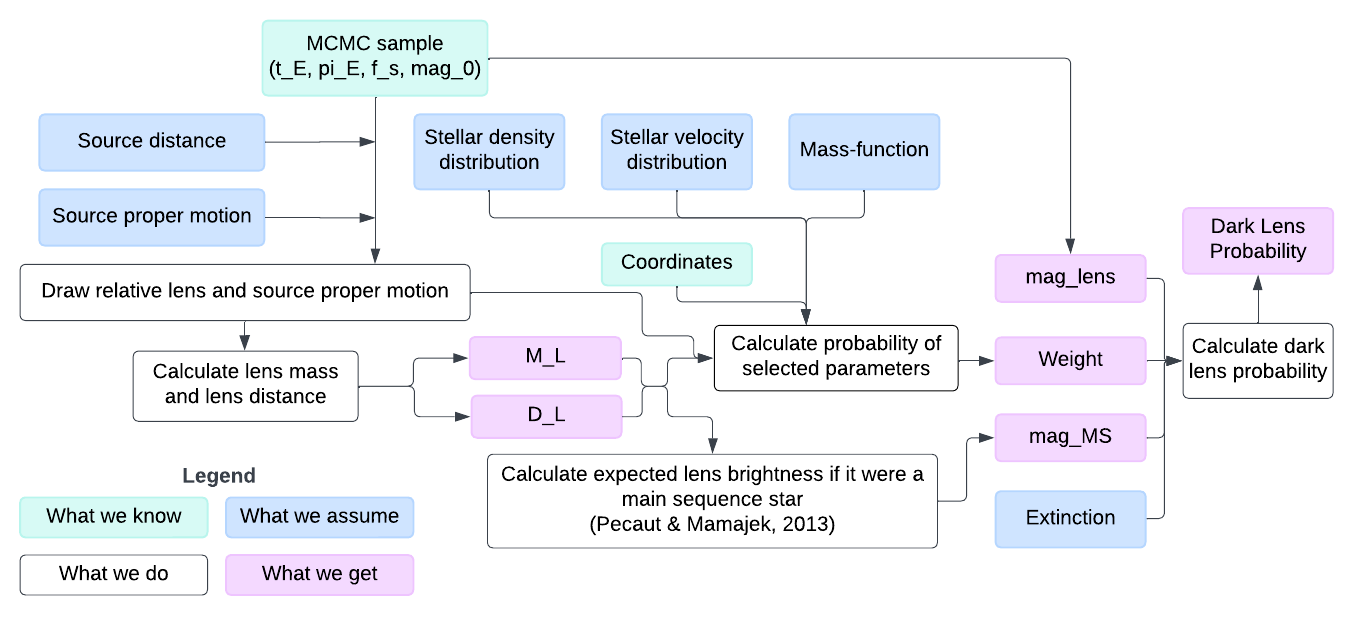}
  \caption{Schematic representation of steps done while estimating mass and distance of the lens with \DLC. $M_\mathrm{L}$ and $D_\mathrm{L}$ are lens mass and distance respectively, $m_\mathrm{lens}$ is lens's observed magnitude, and $m_\mathrm{MS}$ is the observed magnitude of a MS star of $M_\mathrm{L}$ mass at a $D_\mathrm{L}$ distance.}
     \label{fig:DLC_schema}
\end{figure*}

The DLC uses several steps to find the posterior distributions of the lens mass and distance that we present in sections \ref{sec:step1}-\ref{sec:step4}. We presented these steps in Figure \ref{fig:DLC_schema}. In sections \ref{sec:input} and \ref{sec:output} we describe \textbf{ what are the input and output data}.

The DLC can be applied to microlensing events, where the microlensing parallax effect was detected or constrained. The microlensing mass $M_\mathrm{L}$ and distance (parallax) $\pi_\mathrm{L}$ of the lens can be then computed following \cite{Gould2000b}:
\begin{equation}
    M_\mathrm{L} = \frac{\theta_\mathrm{E}}{\kappa \pi_\mathrm{E}}, \pi_\mathrm{L} = \frac{\theta_\mathrm{E}}{\pi_\mathrm{E}} + \pi_\mathrm{S}, 
\end{equation}
where $\theta_\mathrm{E}$ is the angular Einstein radius, $\pi_\mathrm{E}$ is the microlensing parallax, $\pi_\mathrm{S}$ is the parallax of the source, and $\kappa = 8.144 \frac{\mathrm{mas}}{\msun}$ is a constant.

\subsection{Note on the distance to the source}
We obtain the microlensing parallax $\pi_\mathrm{E}$ thanks to the photometric microlensing model.
However, obtaining the distance to the source (parallax $\pi_\mathrm{S}$) is necessary and is not always straightforward, especially in the case of severely blended events.
In the case of Galactic bulge microlensing events, it is common practice to utilise a colour-magnitude diagram (CMD) to estimate the source distance.
In particular, for the sources associated with the Red Clump Giants, is is typically assumed that they are at around 8~kpc from the observer, like in \citealt{1994Stanek, Wyrzykowski2016}.
Another way to obtain the distances, especially for events located away from the bulge, is to utilise \gaia's distances obtained from astrometric time-series \citep{Gaia}.

However, a word of caution is necessary when using \gaia parallaxes for microlensing events. Catalogues released so far, \gaia~DR2 \citep{GaiaDR2} and \gaia~DR3 \citep{GaiaDR3}, covered years 2014-2016 and 2014-2017, respectively. 
If the microlensing event in question occurred somewhere within these time windows, its astrometric measurements used by \gaia to derive geometric parallax and proper motions could be affected by astrometric microlensing \citep{DominikSahu2000, BelokurovEvans2002, Rybicki2018}, as already shown in \cite{Jablonska2022}.
Moreover, blending, very common in microlensing events since they occur in the densest parts of the sky, will jeopardise the \gaia measurement of source parallax and its proper motion, yielding only a light-weighted average parallax and proper motion of the source and the blend (note, \gaia can disentangle sources as close as around 200~mas, for \gaia~DR3, which will improve in further data releases, however, in microlensing events the source and luminous lens/blend typically are separated at distances smaller than Gaia's resolution).

Spectroscopic distances are yet another method of determining the source distance in a microlensing event. When data is timely taken close to the peak of the event, the light of the source is often amplified a couple of times, hence allowing separation of source contribution over the blend, \citep[e.g.][]{2019Fukui, 2022Bachelet}. 

In cases where the distance of the source can not be determined, DLC offers an option to draw the source distance from a distribution of stars according to the Galaxy model, however, this severely blurs the resulting distributions for lens distance and lens dark nature.

\subsection{Input}\label{sec:input}
Typically, the photometric data of a microlensing event are modelled with tools like Monte Carlo Markov Chain (for example with \texttt{emcee} Python package, \citep{2013Emcee}), Nested Sampling \citep{2004Skilling} (for example, \texttt{pyMultiNest} package \citep{2009FerozMultiNest, 2014BuchnerPyMultiNest}) or Dynamic Nested Sampling \cite{2019Higson} (for example \texttt{dynesty} package \citep{2020Speagle}). 
The main input to DLC is the posterior distribution of all microlensing parameters in a parallax microlensing model in a geocentric frame (in the case of this work, one mode of the posterior distribution). 
Moreover, it also requires the posterior distribution for the blending parameter, defined as $f_s=\frac{F_S}{F_S+F_B}$ or $f_b=\frac{F_B}{F_S+F_B}$, where $F_S$ and $F_B$ are fluxes of the source and the blend, respectively.

Next, it requires source information, such as its distance and proper motion vector in the equatorial coordinate system. 
If any or both of these quantities are not known, the DLC will use the distribution of these parameters obtained from the Galaxy model evaluated in the direction towards the source. 

Finally, the DLC requires an extinction estimate in the same filter as the blending parameter.
This value can be obtained from the \texttt{gaia\_source} {\gaia}DR~2 or {\gaia}DR~3 catalogues \citep{GaiaDR2, GaiaEDR3Cat}, or from spectra.
When possible, we recommend to use the {\gaia}DR3 (\texttt{ag\_gspphot} value). 
Otherwise, the reddening maps such as \cite{2011SchlaflyRedd} can be used.
If the blending parameter is provided in a filter for which the reddening maps do not provide the extinction value, we use the relation of $E(B-V)$ with the extinction in the desired filter from \cite{2019WangChenExt}.
As a sanity check, we compare the obtained value to extinction in $u'$, $g'$, $r'$, and $i'$ filters from \cite{2011SchlaflyRedd}.

The DLC also has several flags and parameters that allow user customisation for their desired result.
These are described in the README document available with the package\footnote{\url{https://github.com/BHTOM-Team/DarkLensCode/blob/main/README.md}}.

\subsection{Step 1 - Finding the mass and distance of the lens}\label{sec:step1}
In this and the three following sections, we explain the steps made by the DLC to obtain the posterior distribution of the lens mass and distance.

First, we randomly select one solution from the posterior distribution of the microlensing model, to establish the Einstein timescale of the event $\tE$, the length and direction of the parallax vector $\piE$, and the observed brightness of the blend $G_\mathrm{blend}$.

Next, we select a random value of the relative proper motion from a flat distribution between 0 and 30 $\mathrm{mas\,yr}^{-1}$ and as well as a distance to the source $D_\mathrm{S}$, based on the location of the event on the sky. 
If the source distance $D_\mathrm{S}$ is not provided, it is also drawn from a flat distribution, and later on weighted by the Galactic model.
Otherwise, if the user provides an estimate obtained by other means, we sample the distance using the \texttt{asymmetric\_uncertainty} package \citep{assymetric_sampling}. In the case of symmetric error, it is equivalent to a Gaussian distribution.
These values allow us to find the angular Einstein radius $\thetaE$, and, in turn, the mass $M_\mathrm{L}$ and distance to the lens $D_\mathrm{L}$. 

Next, we find the observed magnitude $G_\mathrm{MS}$ of a main-sequence (MS) star with $M_\mathrm{L}$ mass at a $D_\mathrm{L}$ distance.
We calculate two values of the observed magnitude: $G_\mathrm{MS,0}$ without applying any extinction and $G_\mathrm{MS,A}$ assuming that the extinction to the lens is equal to the extinction in that direction $A_G$.
For the MS brightness, we use \cite{PecautMamajek2013} tables provided by the authors on Eric Mamajek's website\footnote{\url{http://www.pas.rochester.edu/~emamajek/}}.

Then, we apply priors in the form of weight to the drawn values.
We used three priors: the relative proper motion of the lens and source, the distance to the lens, and the mass of the lens. 
When we analyse the relative proper motion and the distance, we consider two scenarios: the lens residing in the Galactic disc or Galactic bulge.
For the source, we assume that if the event's Galactic longitude was within 10 degrees from the Galactic centre, the source is located in the Galactic bulge.
Otherwise, we conclude that it is located in the Galactic disc.
Our final mass and distance estimate is a weighted median using multiplied probabilities as weights $f_\mu$ assuming the proper motion distribution, $\nu_D$ assuming the distance probability density distribution, and $f_M$ assuming the mass distribution.
All weights were combined using this relation \citep{2003HanGould, 2011Batista}:
\begin{equation} \label{eq:DLCweight}
    \begin{split}
        w_\mathrm{d} = & \frac{\mathrm{au}\,M_\mathrm{L} D^4_\mathrm{L} \mu^4_\mathrm{LS} \tE}{4\piE} \nu_{D,\mathrm{d}} f_{\mu, \mathrm{d}} f_M,\\
        w_\mathrm{b} = & \frac{\mathrm{au}\,M_\mathrm{L} D^4_\mathrm{L} \mu^4_\mathrm{LS} \tE}{4 \piE} \nu_{D,\mathrm{b}} f_{\mu, \mathrm{b}} f_M,
    \end{split}
\end{equation}
where $w_\mathrm{d}$ is the weight for the lens located in the Galactic disc, and $w_\mathrm{b}$ is the weight located in the Galactic bulge.
The final weight applied to the obtained lens mass and distance is the larger of the two values  $w_\mathrm{d}$ and $w_\mathrm{b}$.
We derive these weights following the steps outlined below.

\subsection{Step 2 - The relative proper motion prior}\label{sec:step2}

For the relative proper motion prior, we assume that it is a normal distribution $N(\mu_\mathrm{exp, LS}, \sigma_\mu)$.
We use the following procedure to find the expected value $\mu_\mathrm{exp, LS}$ and the standard deviation $\sigma_\mu$.
First, we assume that the velocity of the lens in the Milky Way is described by a normal distribution with $N(v, \sigma_v)$, where $v$ is the expected velocity for the population under consideration, and $\sigma_v$ is its standard deviation.
The standard \textbf{deviation} values are as follows: $(\sigma_{v,l}, \sigma_{v,b}) = (100, 100)\,\mathrm{km}\,\mathrm{s}^{-1}$ for lens within the Galactic bulge, and $(\sigma_{v,l}, \sigma_{v,b}) = (30, 20)\,\mathrm{km}\,\mathrm{s}^{-1}$ for lens within the Galactic disc, where $l$ is the Galactic latitude, and $b$ the Galactic longitude.
The mean velocities for distributions are calculated using the Galactic model.
First, we find the velocities ($U$, $V$, $W$) in the Cartesian Galactic coordinates using the following relation \citep{2009Reid}:
\begin{equation}
    \begin{split}
    ~~~~~~ & U = V_r(R) \sin\beta, \\
    & V = V_r(R) \cos\beta - V_\mathrm{rot} - V_\odot, \\
    ~~~~~ & W = -W_\odot,
    \end{split}
\end{equation}
where $V_r(R) = V_\mathrm{rot} - 1.34 (R-R_\odot)$ is the rotational velocity of the object, $V_\mathrm{rot} = 222.3\,\mathrm{km}\,\mathrm{s}^{-1}$ is the rotational velocity of the Sun, $R$ is the radial distance to the object from the Galactic centre projected on the Galactic plane in kpc, $R_\odot = 8.1$~kpc, $\beta$ is the angle between the observer, the Galactic centre and the observed object, and $(U_\odot, V_\odot, W_\odot) = (11.1, 12.2, 7.3)\,\mathrm{km}\,\mathrm{s}^{-1}$ are the velocities of the Sun in the Cartesian Galactic coordinates \citep{2010Schoenrich}. 
We transforme the velocity to the Galactic coordinate system following \cite{2009Reid} and \cite{2019Mroz}.
If the lens belongs to the Galactic disc:
\begin{equation}
    \begin{split}
    v_{l,d} = & V\cos l - U\sin l, \\
    v_{b,d} = & W\cos l - (U\cos l + V\sin l \sin b).
    \end{split}
\end{equation}
If the lens belongs to the Galactic bulge, then we use the simplified equations (because $(\beta, l, b) \approx (0, 0, 0)$):
\begin{equation}
    \begin{split}
    v_{l,b} = & - W_\odot, \\
    v_{b,b} = & - V_\odot.
    \end{split}
\end{equation}
We transform these velocities into the lens's expected proper motion and its standard deviation using the well-known relation: $\mu_\mathrm{L} = v_\mathrm{L} / (4.74 D_\mathrm{L})$.
Finally, the prior of the relative proper motion between the lens and the source is a normal distribution with the expected value of $\mu_\mathrm{L} - \mu_\mathrm{S}$ and the standard deviation of $(\sigma_{\mu, \mathrm{L}}^2 + \sigma_{\mu, \mathrm{S}}^2)^{1/2}$.

If a proper motion is available for this event in \gaia Data Release main catalogue, we assume it was measured for the source and use it to establish the expected relative proper motion.
Otherwise, we use the Galactic model to calculate the expected source's proper motion $\mu_\mathrm{S}$. 
If the source's Galactic latitude is within 10~degrees of the Galactic Center and its distance from the Galactic Center is smaller than 2.4~kpc, we assume that its proper motion is equal to $(\mu_{s,l}, mu_{s,b}) = (-6.12, -0.19)$~mas~$\mathrm{yr}^{-1}$ with uncertainty of $\sigma_{\mu,l} = \sigma_{\mu,b} = 2.64$~mas~$\mathrm{yr}^{-1}$ \cite{2021MrozBHOGLE}.
In any other case, we use the same procedure, which was used to find the lens's proper motion.

The standard deviation of the relative proper motion is calculated using the standard deviation of the lens's and source's proper motion. 
If the source's proper motion is known from \gaia, we use its error as standard deviation.
We use equations presented in \cite{2013Poleski} to transform the proper motion from equatorial coordinates to the Galactic coordinate system.
We calculate the probabilities $f_{\mu,\mathrm{d}}$ and $f_{\mu,\mathrm{b}}$ of the relative proper motion drawn from the flat distribution at the beginning of the \DLC procedure occurring in the Milky Way using the prior for the relative proper motion described above for each scenario of the lens location. 

\subsection{Step 3 - The distance prior}\label{sec:step3}

We weigh the drawn distances for the source and lens using the following distributions described in \cite{2003HanGould} and \cite{2011Batista}.
For an object located in the Galactic disc, the probability of having a distance $R$ from the Galactic centre measured on the Galactic plane and height $z$ from the Galactic plane is:
\begin{equation}
    \begin{split}
        \nu_\mathrm{d}(R,z) = & 1.07 \msun \mathrm{pc}^{-3} \exp(-R/H) \\
         & [(1-B)\exp(|z|h_1) + B\exp(-|z|h_2)],
    \end{split}
\end{equation}
where $H = 2.75$~kpc, $h_1 = 0.156$~kpc, $h_2 = 0.439$~kpc, and $B = 0.381$.
An object located in the Galactic bulge would have a probability equal to:
\begin{equation}
    \nu_\mathrm{b} = 1.23 \msun \mathrm{pc}^{-3} \exp(-0.5 r^2_s),
\end{equation}
where $r^4_s = ((x'/x_0)^2 + (y'/y_0)^2)^2 + (z'/z_0)^4$.
$(x', y', z')$ are measured in a Cartesian coordinate system with its centre in the Galactic centre, the $x'$ axis is rotated by 20~degrees from the line connecting the Galactic centre and the Sun in the direction of the positive Galactic longitude, and the $z'$ axis is directed towards the Galactic North Pole.
The scale parameters are $(x_0, y_0, z_0) = (1.58, 0.62, 0.43)$~kpc.
If the distance $R$ is larger than 2.4~kpc, the $\nu_\mathrm{b}$ is multiplied by:
\begin{equation}
    n_\mathrm{b} = \exp(-0.5 (\frac{r_r - 2.4}{0.5})^2),
\end{equation}
where $r_r = (x'/x_0)^2 + (y'/y_0)^2$, which smooths the transition from $\nu_\mathrm{b}$ to  $\nu_\mathrm{d}$.

\subsection{Step 4 - The mass prior}\label{sec:step4}

The mass $M_\mathrm{L}$ can be weighted using different scenarios, in particular, a mass function from \cite{2001Kroupa} (StellarIMF), which describes MS stars, and a mass function \cite{2021MrozBHOGLE} (DarkIMF), which describes solitary dark remnants, or a simple mass function of $f(M) \sim M^{-1}$ when completely unknown.
The DLC can estimate the mass of the lens for only one type of mass function and the probability $p_M$ is calculated using only one type of mass function $f(M_\mathrm{L})$.
The probability density function is then multiplied by a Jacobian \citep{Skowron2011}: $D_L^4\mu_{rel}^{4}t_E/\pi_E$.

\begin{figure*}
   \centering
   \includegraphics[width=0.45\hsize]{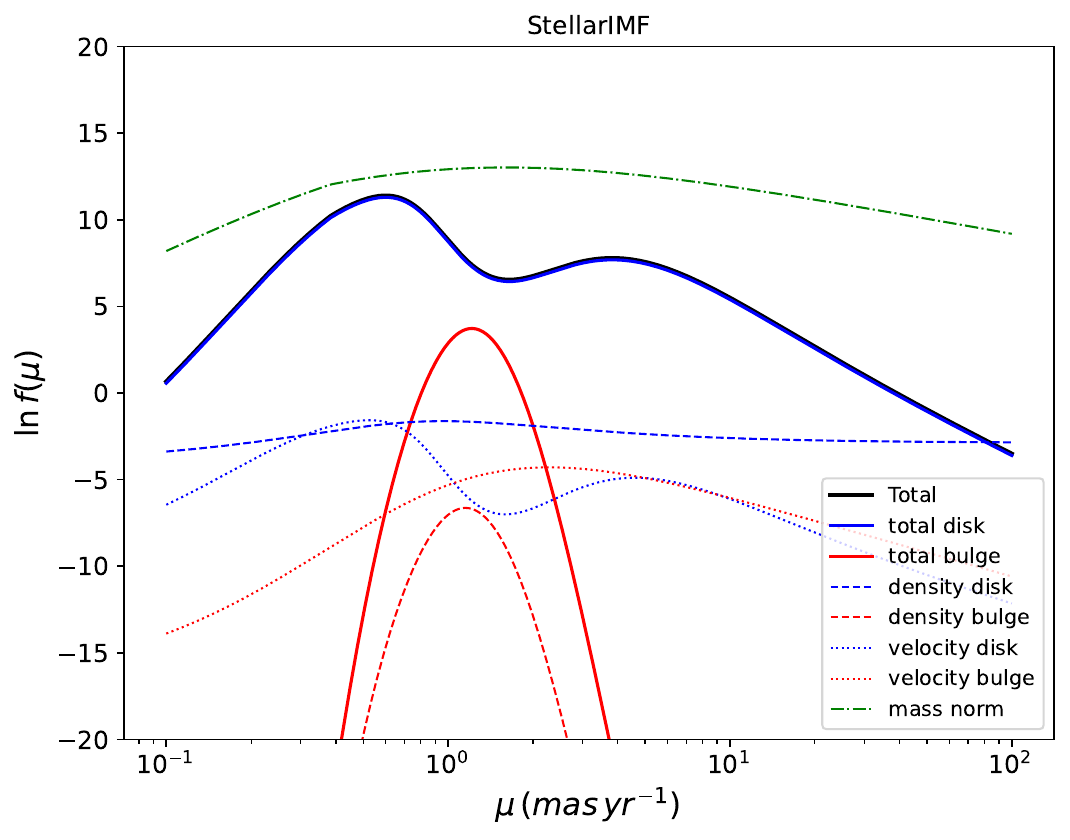}
   \includegraphics[width=0.45\hsize]{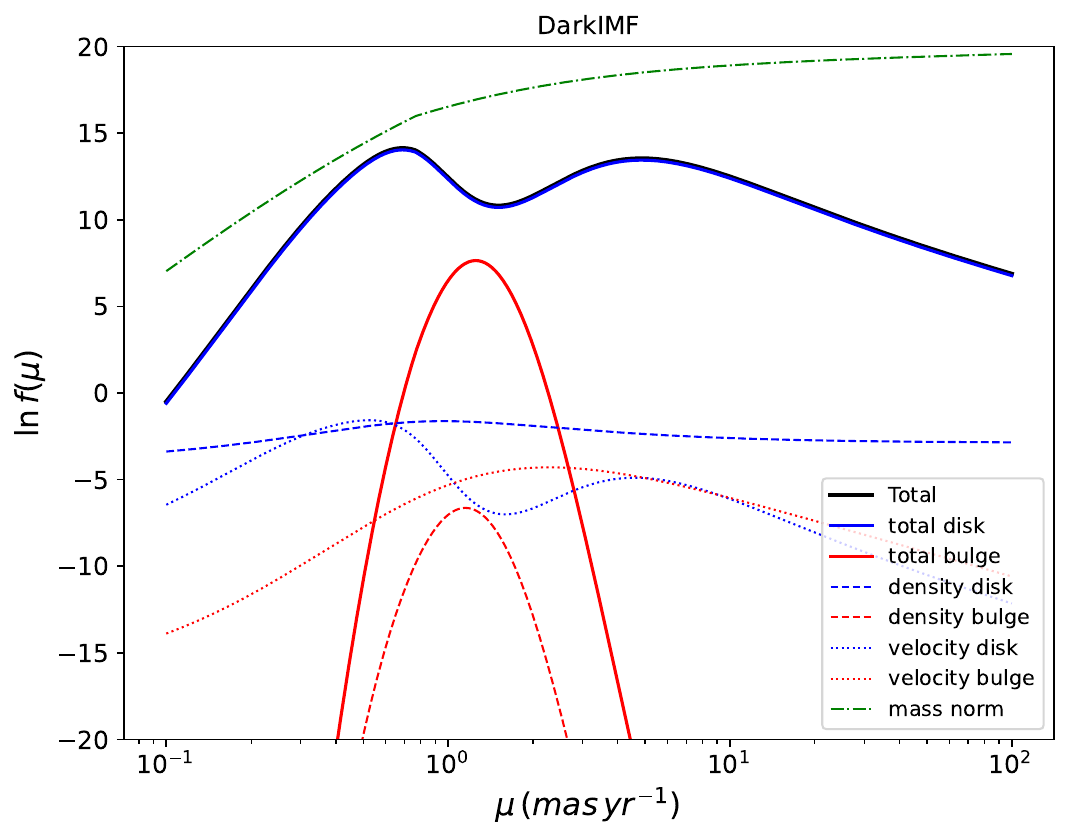}
   \caption{Density probabilities as a function of heliocentric relative proper motion for Gaia18ajz event for two mass functions, StellarIMF and DarkIMF, and their components from disk and bulge. Thin dashed lines show the density of lenses along the line of sight (distance prior), while dotted lines show the velocity (relative proper motion) priors. Dot-dashed line shows the mass function combined with the Jacobian.}
   \label{fig:mass_func_Gaia18ajz}
\end{figure*}

\subsection{Output}\label{sec:output}
The DLC produces two files after it completes its run.
The first one contains the posterior distribution of all estimated parameters:
the weight coming as a larger of the two values calculated using the \ref{eq:DLCweight} equation, the mass of the lens $M_\mathrm{L}$ in Solar masses, the distance to the lens in kpc, the magnitude of the blend coming from the photometric microlensing model, the magnitude of an MS star with the $M_\mathrm{L}$ mass including the full value of the extinction, the magnitude of an MS star with the $M_\mathrm{L}$ mass without extinction, the magnitude of the source, the time of the peak of magnitude from the microlensing model, the Einstein timescale in days, the impact parameter $u_0$, the northern and eastern component of the microlensing parallax, the baseline magnitude, the blending parameter and the relative proper motion in $\mathrm{mas}\,\mathrm{yr}^{-1}$.
The second file contains the summary with the median values if mass and distance of the lens, as well as the probability that the lens is a dark object with and without the inclusion of extinction.

\end{appendix}

\end{document}